\documentclass[useAMS,usenatbib]{mn2e}
\usepackage{graphicx}
\usepackage{amsmath}
\usepackage{bm}

\title[Large-Scale Polarized Foreground Component Separation for {\it Planck}]{Large-Scale Polarized Foreground Component Separation for Planck}
\author[C. Armitage-Caplan et al.]
{Charmaine Armitage-Caplan,$^1$\thanks{armitage-caplan@physics.ox.ac.uk}
Joanna Dunkley,$^1$
Hans Kristian Eriksen,$^2$
\newauthor
Clive Dickinson,$^3$\\
$^1$Sub-department of Astrophysics, University of Oxford, Denys Wilkinson Building, Keble Road, Oxford OX1 3RH, UK\\
$^2$Institute of Theoretical Astrophysics, University of Oslo, P.O. Box 1029 Blindern, N-0315 Oslo, Norway\\
$^3$Jodrell Bank Centre for Astrophysics, Alan Turing Building, School of Physics \& Astronomy, \\The University of Manchester, Oxford Road, Manchester M13 9PL, UK
}

\date{}

\begin{document}

\label{firstpage}

\maketitle

\begin{abstract}
We use Bayesian component estimation methods to examine the prospects for large-scale polarized map and cosmological parameter estimation with simulated {\it Planck} data assuming simplified white noise properties.  The sky signal is parametrized as the sum of the CMB, synchrotron emission, and thermal dust emission.  The synchrotron and dust components are modelled as power-laws, with a spatially varying spectral index for synchrotron and a uniform index for dust.  Using the Gibbs sampling technique, we estimate the linear polarisation Q and U posterior amplitudes of the CMB, synchrotron and dust maps as well as the two spectral indices in $\sim4^{\circ}$ pixels.  We use the recovered CMB map and its covariance in an exact pixel likelihood algorithm to estimate the optical depth to reionization $\tau$, the tensor-to-scalar ratio $r$, and to construct conditional likelihood slices for $C_{\ell}^{EE}$ and $C_{\ell}^{BB}$.  Given our foreground model, we find $\sigma(\tau)\approx0.004$ for $\tau=0.1$, $\sigma(r)\approx0.03$ for a model with $r=0.1$, and a 95\% upper limit of $r<0.02$ for $r=0.0$.  

\end{abstract}
\nokeywords
\section{Introduction}

The {\it Planck} satellite mission, launched in May 2009, is measuring the polarization signal
of the CMB in seven channels over the frequency range 30-353 GHz.  It is now in normal operation and performing as expected \citep{Planck06,Bersanelli10,Mandolesi10,Rosset10,Tauber11}.
With a sensitivity of $\Delta T/T \sim 3 \times 10^{-6}$ in polarization and an angular resolution down to $5'$ \citep{Planck06}, {\it Planck} will produce polarization
data which offer a multitude of opportunities including: possible recovery of inflationary B-modes at large scales, tighter constraints on the 
parameters describing the epoch of reionization, and greater understanding of the polarized nature
of Galactic foregrounds (particularly dust, which dominates at the high frequency channels of {\it Planck}).  

In this paper, we adapt existing foreground separation tools to explore
the prospects of estimating low-resolution polarized CMB maps from the {\it Planck} mission.  We separate the polarization from the temperature signal and focus on the problem of estimating the Q and U component maps.  From these estimated CMB maps and their covariance, we then use a low-$\ell$ pixel likelihood estimator to get estimates on the optical depth to reionization, $\tau$, and the tensor-to-scalar ratio, $r$, and to construct low-$\ell$ conditional likelihood slices of $C_{\ell}^{EE}$ and $C_{\ell}^{BB}$.

For an all-sky experiment like {\it Planck}, component separation of the polarization signal will be much more difficult than for the temperature counterpart, in part, because the ratio of the foreground signal to CMB signal is higher.
There is a large body of work on CMB foreground subtraction for temperature
measurements (see reviews by \cite{DC09} and \cite{Leach08}), and a growing body of work on
the problem of polarized foreground subtraction, including applications to the {\it WMAP} data and {\it Planck} simulated data.  The five-year analysis of the {\it WMAP} data included results from two approaches to component separation: template cleaning \citep{Gold09} and a Bayesian parameter estimation method \citep{Dunkley-WMAP} which we use in this paper.  For {\it Planck}, \cite{EGP09} compare an internal linear combination technique with their template-fitting scheme to assess the impact of foregrounds on B-mode detection at low $\ell$s.
\cite{Betoule09} provide constraints on $r$ with {\it Planck} data and future CMB experiments using a component separation method based on SMICA
and \cite{Ricciardi10} describe a correlated component analysis for {\it Planck}.  
Finally, forecasts on polarization foreground removal for a future CMBPol mission have been presented in \cite{Dunkley-cmbpol}.

In \S\ref{sec:bayes_est}, we provide details of our Bayesian parameter technique, Gibbs sampling, and our model of the data.  In \S\ref{sec:psm}, we describe the experimental specifications of {\it Planck} and our simulated maps.
We present the estimated maps from the parametric technique, the estimated $\tau$ and $r$ likelihoods, and the low-$\ell$ conditional likelihood slices on $C_{\ell}^{EE}$ and $C_{\ell}^{BB}$ in \S\ref{sec:results}.

\section{Methods}
\label{sec:methods}

In the Bayesian parameter estimation method of foreground separation, the emission models of the CMB and foregrounds are parametrized based on our understanding of their frequency dependence.  We then use a sampling method to estimate the marginalized CMB Q and U maps (and additionally the marginalized foreground maps) in every pixel.  In this analysis, we use HEALPix \citep{Healpix} $N_{side} = 16$ maps containing $N_p = 3072$ pixels.    We use two different implementations of Gibbs sampling to estimate the maps and compare their results in the case of a simplified diagonal noise model.  The first is a code called {\it Commander} which we review in \S\ref{sec:commander} and refer the reader to \cite{E06} and \cite{E08} for more details.  The second is a code called {\it Galclean} which we review in \S\ref{sec:galclean} and refer the reader to \cite{Dunkley-WMAP} for more details.

\subsection{Bayesian Estimation of sky maps}
\label{sec:bayes_est}

Consider the simple case in which the data model is given by
\begin{equation}
\mathbf{d} = \mathbf{s} + \mathbf{n}
\end{equation}
where $\mathbf{d}$ are the observed data, consisting of Q and U polarization maps observed by {\it Planck}, $\mathbf{s}$ is the sky signal, and $\mathbf{n}$ is the instrument noise.  We wish to estimate the sky signal $\mathbf{s}$ which is achieved by computing the posterior distribution $P(\mathbf{s}|\mathbf{d})$.  By Bayes' theorem, this distribution can be written as
\begin{equation}
P(\mathbf{s}|\mathbf{d}) \propto P(\mathbf{d}|\mathbf{s}) P(\mathbf{s})
\end{equation}
with prior distribution $P(\mathbf{s})$ and Gaussian likelihood
\begin{equation}
-2 \mathrm{ln} P(\mathbf{d}|\mathbf{s}) = \chi^2 + c,
\end{equation}
with
\begin{equation}
\chi^2 = (\mathbf{d}-\mathbf{s})^t\mathbf{N}^{-1}(\mathbf{d}-\mathbf{s})
\end{equation}
and a normalization term c.  It is straightforward to generalize to multi-frequency data.  In the case that the noise $\mathbf{N}$ is uncorrelated between channels, the likelihood can be written as
\begin{equation}
\chi^2 = \sum_{\nu}[\mathbf{d}_{\nu}-\mathbf{s}_{\nu}]^t\mathbf{N}_{\nu}^{-1}[\mathbf{d}_{\nu}-\mathbf{s}_{\nu}]
\label{eq:likelihood}
\end{equation}
where $\mathbf{d}_{\nu}$ is the observed sky map at frequency $\nu$ and $\mathbf{N}_{\nu}$ is its covariance matrix.

We define the parametric model for the total sky signal in antenna temperature for our three-component model ($k=1$ for CMB, $k=2$ for synchrotron emission, and $k=3$ for thermal dust emission) as
\begin{equation}
\label{eq:model}
\mathbf{s}_{\nu} = \sum\limits_k \bm\alpha_k(\nu;\beta_k)\mathbf{A}_k
\end{equation}
where $\bm{A}_k$ are amplitude vectors of length $2N_p$ and $\bm\alpha_k(\nu;\beta_k)$ are diagonal coefficient matrices of side $2N_p$ at each frequency.  Given that the CMB radiation is blackbody and assuming that the spectral index of the Galactic components do not vary over our frequency range, the coefficients are given by
\begin{align}
\bm\alpha_1(\nu,\beta_1) & = \bm\alpha_1(\nu) = f(\nu)\mathbf{ I}\\
\bm\alpha_2(\nu,\beta_2) & = \rm{diag}[(\nu/\nu_{30})^{\bm\beta_2}]  \\
\label{eq:alpha3}
\bm\alpha_3(\nu,\beta_3) & = \rm{diag}[(\nu/\nu_{353})^{\bm\beta_3}].
\end{align}
Here we have defined the function $f(\nu)$ which converts the CMB signal $\mathrm{I}$ from thermodynamic to antenna temperature, and the two spectral index vectors $\beta_2$ and $\beta_3$ for synchrotron and dust, respectively.  We also set the pivot frequencies to 30 GHz and 353 GHz.

Though the spectral indices for Q and U in a given pixel are expected to be similar (following from the assumption that the polarization angle does not change with frequency), we allow the option for the indices to be sampled independently for Q and for U.  Thus, our model is completely described by $6N_p$ amplitude parameters $\mathbf{A} = (\mathbf{A}_{1},\mathbf{A}_2,\mathbf{A}_3)$ and $4N_p$ spectral index parameters $\bm{\beta} = (\bm{\beta}^Q_2,\bm{\beta}^Q_3,\bm{\beta}^U_2,\bm{\beta}^U_3)$.    We impose a flat prior on amplitude-type parameters and Gaussian priors on the spectral index parameters of $\beta_2 = -3.0 \pm 0.3$ for synchrotron and $\beta_3 = 1.5 \pm 0.5$ for dust.  The priors we have chosen have central value and standard deviation at approximately the average and range of values typically observed and simulated (see, for example, \cite{Fraisse08,Dunkley-cmbpol} for further discussion).

With our data model and priors defined, our aim is to estimate the joint CMB-foreground posterior $P(\mathbf{A},\bm{\beta}|\mathbf{d})$ from which we can then obtain the marginalized distribution for the CMB amplitude vector as
\begin{equation}
p(\mathbf{A}_{1},\mathbf{d}) = \int p(\mathbf{A},\bm\beta|\mathbf{d})d\mathbf{A}_2 d\mathbf{A}_3 d\bm{\beta}
\end{equation}
and similarly for the other model parameters.  

For the multivariate problem that we are considering, Gibbs sampling draws from the joint distribution by sampling each parameter conditionally as follows
\begin{align}
\mathbf{A}^{i+1} & \leftarrow P(\mathbf{A}|\bm\beta,\mathbf{d})\\
\label{eq:beta_samp}
\bm\beta^{i+1} & \leftarrow P(\bm\beta|\mathbf{A},\mathbf{d})
\end{align}
In the next two sections, we compare and contrast the different methods that {\it Commander} and {\it Galclean} implement to sample the amplitude-type and spectral index parameters.

\subsection{Gibbs sampling with {\it Commander}}
\label{sec:commander}

{\it Commander} is a flexible code for joint component separation and CMB power spectrum
estimation \citep{J04,W04,E04,L06}.  {\it Commander} has typically been used in its full implementation, in which the parametric model of the total sky signal is sampled jointly with the CMB power spectrum.  We describe here how to use {\it Commander} to do sampling of the sky signal only.  The theory of Gibbs sampling allows the joint density $P(\mathbf{s},C_{\ell}|\mathbf{d})$ to be sampled by alternately sampling from the two conditional densities
\begin{align} 
   \mathbf{s}^{i+1} & \leftarrow P(\mathbf{s}|C_{\ell}^i,\mathbf{d}) \\
      C_{\ell}^{i+1} & \leftarrow P(C_{\ell}|\mathbf{s}^{i+1},\mathbf{d}).
\end{align}
The conditional sky signal distribution can be written as
\begin{align}
\label{eq:cond_dist}
P(\mathbf{s}|C_{\ell},\mathbf{d}) & \propto P(\mathbf{d}|\mathbf{s},C_{\ell})P(\mathbf{s}|C_{\ell})\\
& \propto e^{-\frac{1}{2}(\mathbf{d}-\mathbf{s})^t\mathbf{N}^{-1}(\mathbf{d}-\mathbf{s})}e^{-\frac{1}{2}\mathbf{s}^t\mathbf{S}^{-1}\mathbf{s}}
\end{align}
where $\mathbf{S}$ and $\mathbf{N}$ are the signal and noise covariance matrices.

Since we are only interested in doing only the component separation part, this is akin to ignoring the the $P(\mathbf{s}|C_{\ell})$ term in the above algorithm, giving the following distribution 
\begin{equation}
P(\mathbf{s}|\mathbf{d}) \propto e^{-\frac{1}{2}(\mathbf{d}-\mathbf{s})^t\mathbf{N}^{-1}(\mathbf{d}-\mathbf{s})}.
\end{equation}
For our purposes of low-$\ell$ component separation, $\tau$ and $r$ estimation, and evaluation of low-$\ell$ $C_{\ell}^{EE}$ and $C_{\ell}^{BB}$ conditional slices, it is arguably optimal to ``switch off" the $C_{\ell}$ sampling step in {\it Commander}.  This is because the $C_{\ell}$ sampling step is time-consuming  and (in the case that the CMB map has approximately Gaussian uncertainties after marginalizing over foregrounds) equivalent estimates on the spectra can be obtained quickly by using an exact pixel likelihood given the estimated CMB map and CMB covariance.  

Alternatively, and without dropping the $P(\mathbf{s}|C_{\ell})$ term in Eq.~\ref{eq:cond_dist}, we can say that we are conditioning on $C_{\ell}=0$ for the correlated CMB component, while simultaneously allowing for an uncorrelated CMB component with a separate value in each pixel.  The net result is that the CMB amplitudes are sampled in the same way as the foreground amplitudes, and the $C_{\ell}$ sampling step has been omitted entirely.

\subsubsection{Sampling the amplitudes}
\label{sec:amp_samp}

The conditional distribution $P(\mathbf{A}|\bm\beta,\mathbf{d})$ for fixed $\bm\beta$ is a $6 N_p$-dimensional Gaussian from which it is straightforward to sample.  
First, we define the data model as
\begin{equation}
\mathbf{d}_{\nu} = \sum\limits_k \bm\alpha_{k}(\nu;\beta_k) \mathbf{A}_k + \mathbf{n}_{\nu}.
\end{equation}
The conditional distribution is
\begin{align} 
   P(\mathbf{A}|\bm\beta,\mathbf{d}) & \propto e^{-\frac{1}{2}\sum\limits_{\nu}[\mathbf{d}_{\nu}-\sum\limits_k \bm\alpha_k(\nu;\beta_k)\mathbf{A}_k]^T \mathbf{N}^{-1}_{\nu}[\mathbf{d}_{\nu}-\sum\limits_k \bm\alpha_k(\nu;\beta_k)\mathbf{A}_k]}\\
   & \propto e^{-\frac{1}{2}(\mathbf{A}-\hat{\mathbf{A}})^T \mathbf{F}^{-1}(\mathbf{A}-\hat{\mathbf{A}})}
\end{align}
where $\hat{\mathbf{A}}$ is the Wiener-filter mean given by $\hat{\mathbf{A}} = \mathbf{F}\mathbf{x}$ with elements 
\begin{align}
\label{eq:Fisher}
\mathbf{F}^{-1} & = \sum\limits_{\nu}\bm\alpha_k^T(\nu;\beta_k) \mathbf{N}_{\nu}^{-1} \bm\alpha_{k'}(\nu;\beta_k)\\
\label{eq:x}
\mathbf{x} & = \sum\limits_{\nu} \bm\alpha_k(\nu;\beta_k) \mathbf{N}_{\nu}^{-1} \mathbf{d}_{\nu}.
\end{align}

The sampling algorithm that {\it Commander} employs solves the set of linear equations
\begin{equation}
\label{eq:linearset}
\mathbf{F}^{-1} \mathbf{A} = \mathbf{b}
\end{equation}
where $\mathbf{b}$ is the Wiener-filter mean plus random fluctuations (given by white noise maps $\mathbf w_{\nu}$)
\begin{equation}
\mathbf b =  \sum\limits_{\nu} \bm\alpha_k(\nu;\beta_k) \mathbf N_{\nu}^{-1} \mathbf d_{\nu} + \sum\limits_{\nu} \bm\alpha_k(\nu;\beta_k) \mathbf N_{\nu}^{-1/2}\mathbf w_{\nu}.
\end{equation}
The solution vector $\mathbf A$ has mean $\hat{\mathbf A}$ and covariance matrix $\mathbf F$, and the next amplitude sample is given by $\mathbf A_{i+1} = \mathbf A$.

\subsubsection{Sampling the spectral indices}
\label{sec:ind_samp}

For fixed amplitude, the spectral index sampler in {\it Commander} is a standard inversion sampler.  This algorithm first maps out the conditional probability distribution $P(\bm\beta|\mathbf A,\mathbf d)$ by evaluating the likelihood (given by Eq.~\ref{eq:likelihood}) at 500 points between the upper and lower $5\sigma$ limits and then computes the corresponding cumulative probability distribution $F(\bm\beta|\mathbf A) = \int_{-\infty}^{\beta} P(\bm\beta'|\mathbf A,\mathbf d)d\bm\beta'$.  Next, a random number $u$ is drawn from the uniform distribution $U[0,1]$.  Thus, the sample from $P(\bm\beta|\mathbf A,\mathbf d)$ is given by $F(\bm\beta|\mathbf A) = u$.

In Commander, Eq.~\ref{eq:beta_samp} can be iterated more than once in each main Gibbs loop as an inexpensive way to reduce correlations between consecutive samples.  We allow three spectral index iterations for each main Gibbs iteration.

\subsection{Gibbs sampling with {\it Galclean}}
\label{sec:galclean}
{\it Galclean}, the Gibbs sampling algorithm described in \cite{Dunkley-WMAP}, solves the same set of equations outlined in \S\ref{sec:bayes_est} and defines the same parametric model given in Eqs.~\ref{eq:model}-\ref{eq:alpha3}.  The method that {\it Galclean} uses to do the amplitude sampling, described below in \S\ref{sec:gal_ampsamp} is similar to that of {\it Commander}, with a different technique for drawing a sample.  The main difference between {\it Galclean} and {\it Commander} is in the spectral index sampling, where {\it Galclean} implements a Metropolis-Hastings algorithm as outlined in \S\ref{sec:gal_specsamp}.  Given enough time to converge, both methods should give the same estimates.

\subsubsection{Amplitude sampling}
\label{sec:gal_ampsamp}
{\it Galclean} solves the set of linear equations given by  $\hat{\mathbf A} = \mathbf F \mathbf x$, where $\mathbf F$ and $\mathbf x$ are given by Eqs.~\ref{eq:Fisher} and \ref{eq:x} , for the Wiener-filtered mean $\hat{\mathbf A}$.  The {\it Galclean} code then draws a Gaussian sample $\mathbf G$ and constructs the next amplitude sample as
\begin{equation}
\mathbf A_{i+1} =\hat{\mathbf A} + \mathbf L^{-1} \mathbf G
\end{equation}
where $\mathbf L$ is the lower cholesky decomposition of the Fisher matrix $\mathbf F = \mathbf L\mathbf L^T$.

\subsubsection{Spectral index sampling}
\label{sec:gal_specsamp}
{\it Galclean} uses the Metropolis-Hastings algorithm (see, for example, \cite{Knox01,Lewis02,Dunkley05}) to sample the index vector.  Briefly, a trial step $\beta_{trial}$ is drawn from a Gaussian distribution of width $\sigma_{trial}$ and centred on the current $\beta$ vector.  The current and trial vectors are used to construct the current and trial posteriors
\begin{equation}
-2 \mathrm{ln} p(\bm\beta|\mathbf A,\mathbf d) = \chi^2
\end{equation}
where $\chi^2$ is given by Eq.~\ref{eq:likelihood}, which includes the full noise correlation matrix.
Then the ratio of the trial to current posterior, $r$, determines whether to move to the trial position (with probability $r$), or to stay at the original position (with probability 1-$r$).

\subsection{Processing the sampled distribution}

We examine the results from the {\it Commander} and {\it Galclean} Gibbs sampling chains by forming the mean map and covariance matrix.  The mean map can be found from the marginalized distribution as
\begin{equation}
\langle \mathbf A_k\rangle = \int p(\mathbf A_k|\mathbf d)\mathbf A_k d\mathbf{A}_k = \frac{1}{n_G}\sum\limits_{i=1}^{n_G} \mathbf A_k^i
\end{equation}
where we sum over the $n_G$ elements in the Gibbs chain for the $k$th component of the model.  We find that a typical Gibbs chain takes roughly 100 iterations to ``burn-in" (or converge to the equilibrium distribution) when the spectral indices are initialized at random values drawn from the Gaussian prior.  Therefore, we remove the first 100 elements in each Gibbs chain before processing the samples.  It should be noted that the marginalized distribution $p(\mathbf A_k|\mathbf d)$ may not be an exact Gaussian, but we assume it to be to good enough approximation.

Similarly, the covariance matrix for $\mathbf A_k$, is estimated by summing over chain elements, as shown here for the general noise case, for the covariance between pixel $x$ and $y$ for component $k$
\begin{align}
C_{xy,k} & = \langle A_{x,k}A_{y,k}\rangle - \langle A_{x,k}\rangle\langle A_{y,k}\rangle\\
	& = \frac{1}{n_G}\sum\limits_{i=1}^{n_G}(A_{x,k}^i-\langle A_{x,k}\rangle ) (A_{y,k'}^i - \langle A_{y,k} \rangle ).
\end{align}
For diagonal noise, the covariance in pixel $x$ for component $k'$ reduces to
\begin{equation}
C_{x,k} = \frac{1}{n_G}\sum\limits_{i=1}^{n_G}(A_{x,k}^i-\langle A_{x,k}\rangle ) (A_{x,k}^i - \langle A_{x,k} \rangle ).
\end{equation}
Note that while we found $n_G=10,000$ samples to be enough to ensure convergence of the chains for the case of diagonal noise, we expect to need an order of magnitude more samples to construct the full covariance matrix in the case of correlated noise.

\subsection{Likelihood estimation}
\label{sec:likelihood}
The product of a Bayesian parametric map estimation method is both a CMB map and a covariance matrix (which can be estimated from the marginalized posterior distribution) and together these products can be used to place constraints on cosmological parameters.  We compute the likelihood of the estimated maps, given a theoretical angular power spectrum, using the method described in \cite{Page07}.  

Summarized here, the standard likelihood is given by
\begin{equation}
L(\mathbf{m}|S) \propto \frac{\exp[-\frac{1}{2}\mathbf{m}^t(S+N)^{-1}\mathbf{m}]}{|S+N|^{1/2}}
\end{equation}
where $\mathbf{m}$ is the data vector containing the temperature, $\mathbf{T}$, and polarization maps, $\mathbf{Q}$ and $\mathbf{U}$, $n_p$ is the number of pixels in each map, and $S$ and $N$ are the signal and noise covariance matrices.  Under the assumption that noise in the temperature can be ignored -- which holds at low multipoles where the signal dominates -- the standard likelihood can be simplified to 
\begin{equation}\begin{split}
L(\mathbf{m}|S) & \propto \frac{\exp[-\frac{1}{2}\mathbf{\tilde{m}}^t(\tilde{S}_P+N_P)^{-1}\mathbf{\tilde{m}}]}{|\tilde{S}_P+N_P|^{1/2}} \\
& \frac{\exp[-\frac{1}{2}\mathbf{T}^tS^{-1}_T \mathbf{T}]}{|S_T|^{1/2}}
\end{split}\end{equation}
where $S_T$ is the temperature signal matrix, $\mathbf{\tilde{m}} = (\tilde{Q}_P,\tilde{U}_P)$ is the new polarization data vector given by
\begin{eqnarray}
\tilde{Q}_P \equiv  Q_P - \frac{1}{2}\sum_{\ell=2}^{23}\frac{S_{\ell}^{TE}}{S_{\ell}^{TT}}\sum_{m=-\ell}^{\ell}T_{\ell m}(_{+2}Y_{\ell m,p} + _{-2}Y_{\ell m,p}^{\ast})    \\
\tilde{U}_P \equiv  U_P - \frac{i}{2}\sum_{\ell=2}^{23}\frac{S_{\ell}^{TE}}{S_{\ell}^{TT}}\sum_{m=-\ell}^{\ell}T_{\ell m}(_{+2}Y_{\ell m,p} + _{-2}Y_{\ell m,p}^{\ast})  
\end{eqnarray}
and $\tilde{S}_P$ is the signal matrix for the new polarization vector.
This new form of the likelihood allows us to factorize it into the likelihood of temperature and polarization, with information about their cross-correlation preserved.

The two cosmological parameters most closely linked with the large scale CMB polarization signal are the optical depth to reionization, $\tau$, and the tensor-to-scalar ratio, $r$.  The signature of reionization is at $\ell<10$ in $C_{\ell}^{EE}$ where the amplitude of the reionization signal is proportional to $\tau^2$.  The tensor-to-scalar ratio $r$ directly scales the $C_{\ell}^{BB}$ power spectrum and is best probed at low $\ell$'s before $C_{\ell}^{BB}$ due to lensing dominates.

By varying only the optical depth to reionization $\tau$ and the power spectrum amplitude (such that the temperature anisotropy power at $\ell=220$ is held constant), we can calculate the likelihood at each value of $\tau$.  This allows us to estimate limits on the optical depth to reionization including foreground uncertainty.  Similarly, we can vary only the tensor-to-scalar ratio $r$, fixing all other parameters at concordance values, and calculate the likelihood at each value of $r$.

The standard likelihood estimator that we use here assumes that the estimated CMB map has approximately Gaussian uncertainties after marginalizing over foregrounds.  In practice, marginalizing over foregrounds may induce non-Gaussianities and there are several ways of addressing this issue.  One option is to modify the standard likelihood to include a non-Gaussian term.  Alternatively, the $C_{\ell}$'s can be sampled jointly with the maps in a full Bayesian framework.  This type of scheme is implemented in the {\it Commander} code, in which the problem of sampling from the joint density $P(\mathbf{s},C_{\ell}|\mathbf{d})$ is reduced to that of sampling from the two conditional densities $P(\mathbf{s}|C_{\ell},\mathbf{d})$ and $P(C_{\ell}|\mathbf{s},\mathbf{d})$.  We have already described the sampling algorithm for the first conditional distribution $P(\mathbf{s}|C_{\ell},\mathbf{d})$ in \S\ref{sec:amp_samp}.  The second conditional distribution, $P(C_{\ell}|\mathbf{s},\mathbf{d})$, reduces to $P(C_{\ell}|\mathbf{s})$ since the data does not further constrain the power spectrum if we already know the CMB sky signal.  Then the distribution reads
\begin{equation}
P(C_{\ell}|\mathbf{s}) \propto \frac{e^{-\frac{1}{2}\mathbf{s}_{\ell}^t\mathbf{S}_{\ell}^{-1}\mathbf{s}_{\ell}}}{\sqrt{|\mathbf{S}_{\ell}|}} = 
\frac{e^{-\frac{2\ell+1}{2}\frac{\sigma_{\ell}}{C_{\ell}}}}{C_{\ell}^{\frac{2\ell+1}{2}}}
\end{equation}
for which there is a simple textbook sampling algorithm detailed in \cite{E04}.

In \S \ref{sec:cl_comparison}, we will compare foreground-marginalized $C_{\ell}$ estimates from our standard pixel-likelihood code with those from the {\it Commander} Gibbs sampler.

\subsection{Comparison to template cleaning}
\label{sec:template_cleaning}
Template cleaning is an alternative method of component separation that assumes that the sky at any frequency can be modeled as a linear sum of fixed spatial templates.  In the regime of perfect templates and no spatial spectral index variations, template fitting would give the optimal foreground subtraction and marginalization.  We do a comparison of our results to a simple template cleaning method which is implemented in {\it Commander}.  The data model is given by
\begin{equation}
\mathbf{d}_{\nu} = \mathbf A + \sum\limits_{j=1}^{N} b_j f_j(\nu) \mathbf{T}_j
\end{equation}
where the first term on the right-hand side is the CMB sky signal, the second term represents the sum over $N$ templates $\mathbf{T}_j$ with fixed frequency scaling $f_j(\nu)$ and overall amplitude $b_j$.  This implementation of template cleaning assumes no spatial variation in the frequency scaling and is therefore limited in usefulness to cases where the spectral index variations in the data are small.  However, it is a fast method that has been successfully used for the analysis of {\it WMAP} data, for example.

We use the {\it WMAP} 23 GHz map for the low-frequency synchrotron template and the {\it Planck} simulated 353 GHz map as the high-frequency dust template.  This leaves six remaining channels of {\it Planck} (30 - 217 GHz) to be used as data for the template fitting.  These templates are fitted to the data, the best fit coefficients for each component are found and the templates are subtracted from the map using these coefficients in order to obtain a clean CMB map at each frequency.  These can then be optimally combined through inverse variance weighting, and the likelihood computed using the method in \S\ref{sec:likelihood}.  This method assumes that the templates have no associated uncertainties; methods to propagate template uncertainty have been considered in e.g.\, \cite{EGP09}.

\section{Simulated maps}
\label{sec:psm}

We use a software package called the Planck Sky Model (PSM, version 1.6.6) developed by the {\it Planck} Working Group 2 to generate our simulated CMB and synchrotron maps.  We generate maps at the seven polarized {\it Planck} frequency channels (30, 44, 70, 100, 143, 217, and 353 GHz).  In our analysis, we do not apply beams or smoothing to the data and we leave the sky unmasked in the Gibbs sampling step; these issues should be included in a more realistic analysis.  

We use the PSM ${\tt gaussian\_cosmo}$ option to generate realizations of the CMB.  The PSM simulates the CMB by feeding a set of standard $\Lambda$CDM cosmological
parameters to CAMB which produces a set of corresponding $C_{\ell}$s.  A Gaussian random CMB 
temperature and polarization field is then drawn according to this spectra.  We use PSM model 6 for the synchrotron emission.  The synchrotron Q and U emission maps are given as an extrapolation in frequency of the polarized 23 GHz {\it WMAP} map:
\begin{equation}
Q_{\nu}(p) = Q_{23}(p)\left(\frac{\nu}{23}\right)^{\beta_s(p)}
\end{equation}
\begin{equation}
U_{\nu}(p) = U_{23}(p)\left(\frac{\nu}{23}\right)^{\beta_s(p)}
\end{equation}

The model for the spectral index is taken to be model 4 of \cite{MAMD08}, given by
\begin{equation}
\beta_s = \frac{\log(P_{23}/g f_s S_{408})}{\log(23/0.408)}
\end{equation}
where $P_{23}$ is the {\it WMAP} polarization map at 23 GHz, $g$ is a geometrical reduction factor (reflecting depolarization due to magnetic field structure), $f_s$ is the intrinsic polarization fraction from the cosmic rays energy spectrum, and $S_{408}$ is the 408 MHz map of \cite{Haslam}.
As an initial test, we generate our own simple power-law dust model, extrapolating from the predicted 94 GHz map in \cite{Fink99}: $S_{\nu}(p) = S_{94}(p)\left(\frac{\nu}{94}\right)^{\beta_d}$.  We set the dust spectral index to $\beta_d = 1.5$ uniformly over the whole sky.  Although the observed dust emission is known to fit better to a multi-component model, it is reasonable to approximate it with a single-component model at frequencies below 300 GHz where the dust polarization is likely to be dominated by a single component.

We generate diagonal white noise realizations based on the noise levels listed in Table \ref{tab:Planck_specs}, taken from the {\it Planck} Bluebook \citep{Planck06}, and scale the given noise levels at beam-sized pixels to the corresponding noise level at $n_{side}=16$ sized pixels.  Our simplified noise model contains no $1/f$-noise or other correlations, the addition of which would increase effective noise levels.

\begin{table}
\centering
\begin{tabular}{c c c}
\hline
Frequency & FWHM & $\Delta$T (Q and U) \\
GHz & arcmin & $\mu$K (at nside 16) \\
\hline
30 & 33 & 1.15 \\
44 & 24 & 1.16 \\
70 & 14 & 1.16 \\
100 & 9.5 & 0.47 \\
143 & 7.1 & 0.37 \\
217 & 5.0 & 0.61 \\
353 & 5.0 & 1.85 \\
\hline
\end{tabular}
\caption{{\it Planck} specifications.  The sensitivity is specified for a pixel of area $3.66^{\circ}$ (area of pixel of at HEALPix nside=16, npix = 3072).}
\label{tab:Planck_specs}
\end{table}

\section{Results}
\label{sec:results}

We evaluate the results from Gibbs sampling in a variety of ways.  First, we will check that our two Gibbs sampling codes, {\it Commander} and {\it Galclean}, give results that agree both in terms of their estimated maps and also in their estimates on cosmological parameters.  Next, we test the low-$\ell$ pixel likelihood code and foreground cleaning for potential biases.  Through a number of different test cases, we then look at the effect of foreground cleaning on $\tau$ and $r$ estimates.  Finally, we examine the conditional likelihood slices of the $C_{\ell}^{EE}$ and $C_{\ell}^{BB}$ power spectra and compare the power at each multipole in the Gibbs-cleaned CMB map to the template-cleaned CMB map.

\subsection{Comparisons and Testing}

\subsubsection{Comparison between {\it Commander} and {\it Galclean}}

In this section, we compare our two Gibbs sampling codes, {\it Commander} and {\it Galclean}, and show that they produce comparable CMB and foreground amplitude maps and spectral index maps for a single diagonal noise test case.  Then we use the resulting maps from four simulations to place constraints on two cosmological parameters which are relevant to the low-$\ell$ polarization signal: the optical depth to reionization and tensor-to-scalar ratio.  We show that the estimates from {\it Commander} and {\it Galclean} on $\tau$ and $r$ agree to less than $0.1\sigma$ in all cases.  Given the robust agreement between the two Gibbs sampling codes, and since {\it Commander} has the advantages of being highly parallelized and more flexible than {\it Galclean}, the remainder of the results in this section are produced using {\it Commander} only.

\begin{figure*}
   \centering
   \begin{tabular}{c c c}
   {\scriptsize CMB Input Q} & {\scriptsize {\it Commander} Q} & {\scriptsize {\it Galclean} Q} \\
\includegraphics[ width=.20\textwidth, keepaspectratio,angle=90]{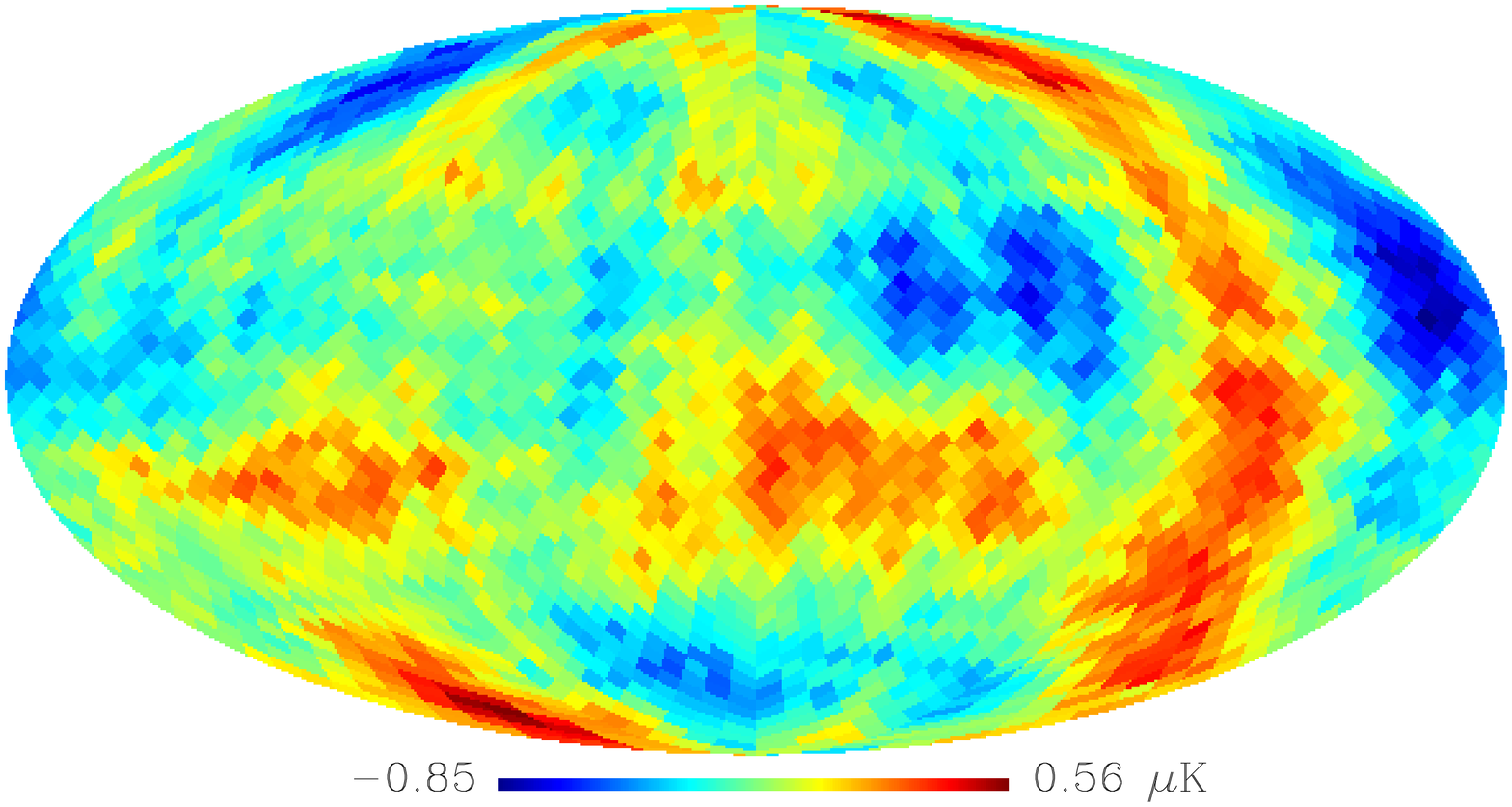} &
\includegraphics[width=.20\textwidth, keepaspectratio,angle=90]{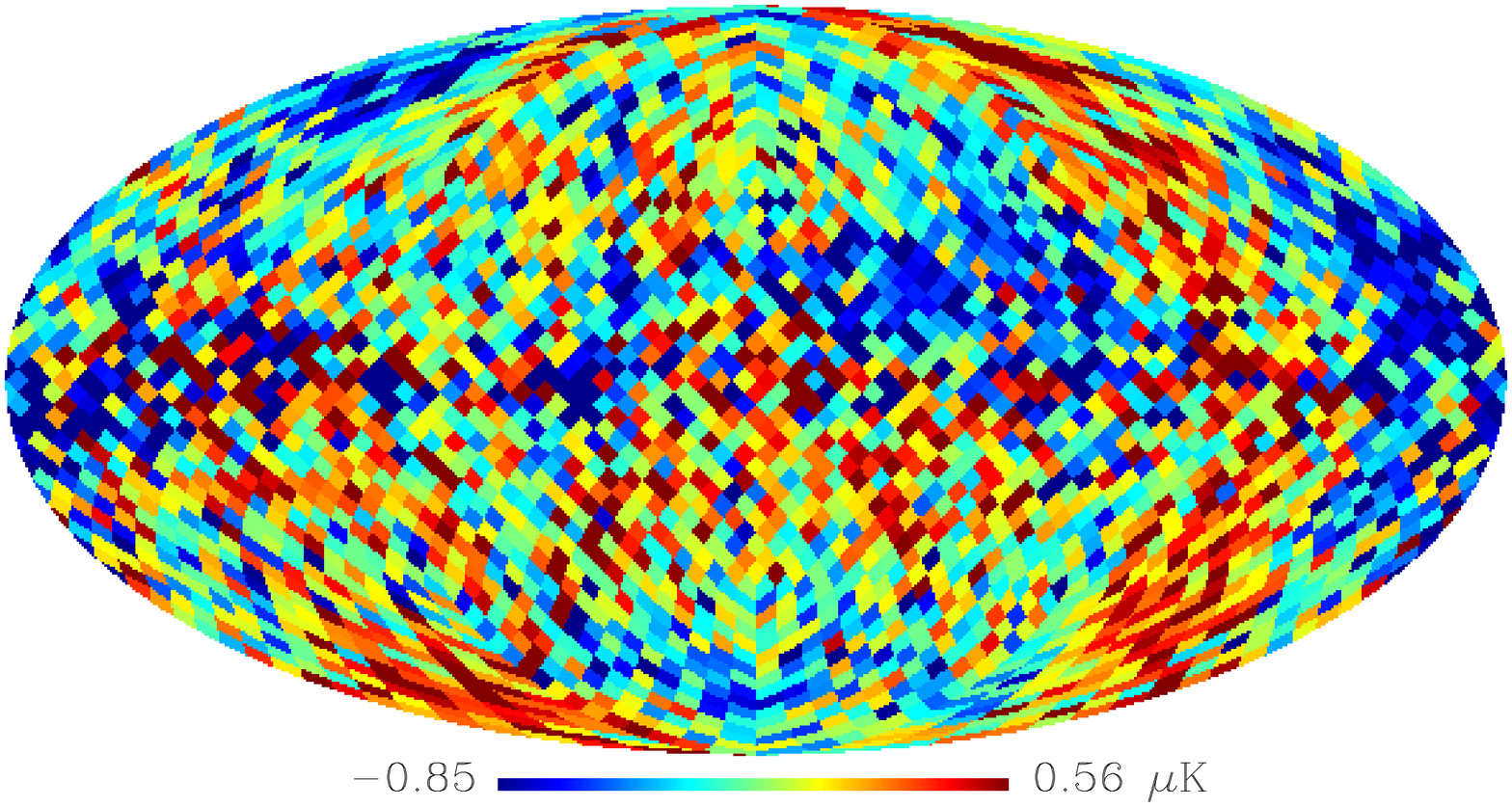} &
\includegraphics[ width=.20\textwidth, keepaspectratio,angle=90]{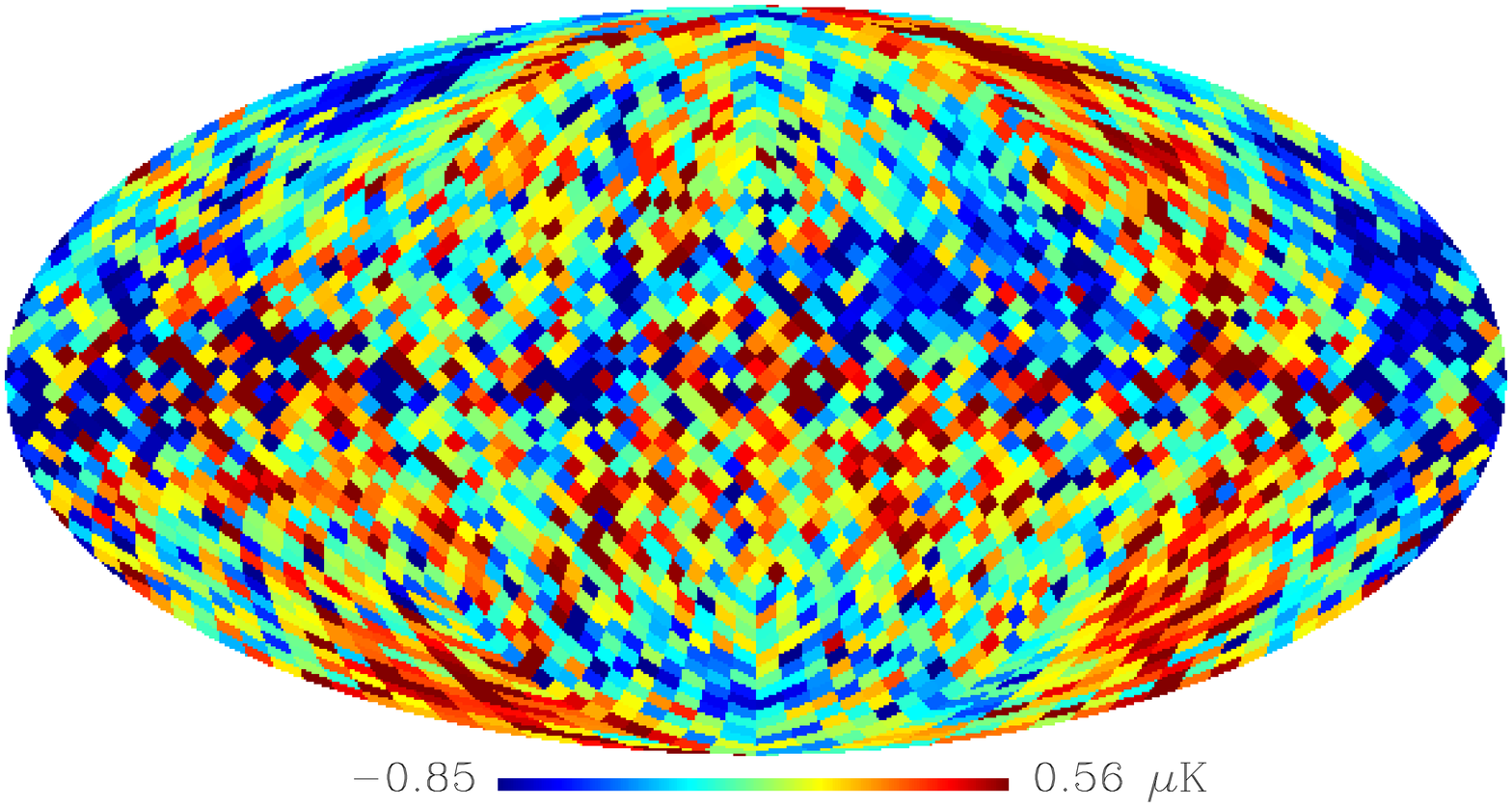} \\
{\scriptsize Error Q} & {\scriptsize {\it Commander} Q deviation} & {\scriptsize {\it Galclean} Q deviation} \\
\includegraphics[width=.20\textwidth, keepaspectratio,angle=90]{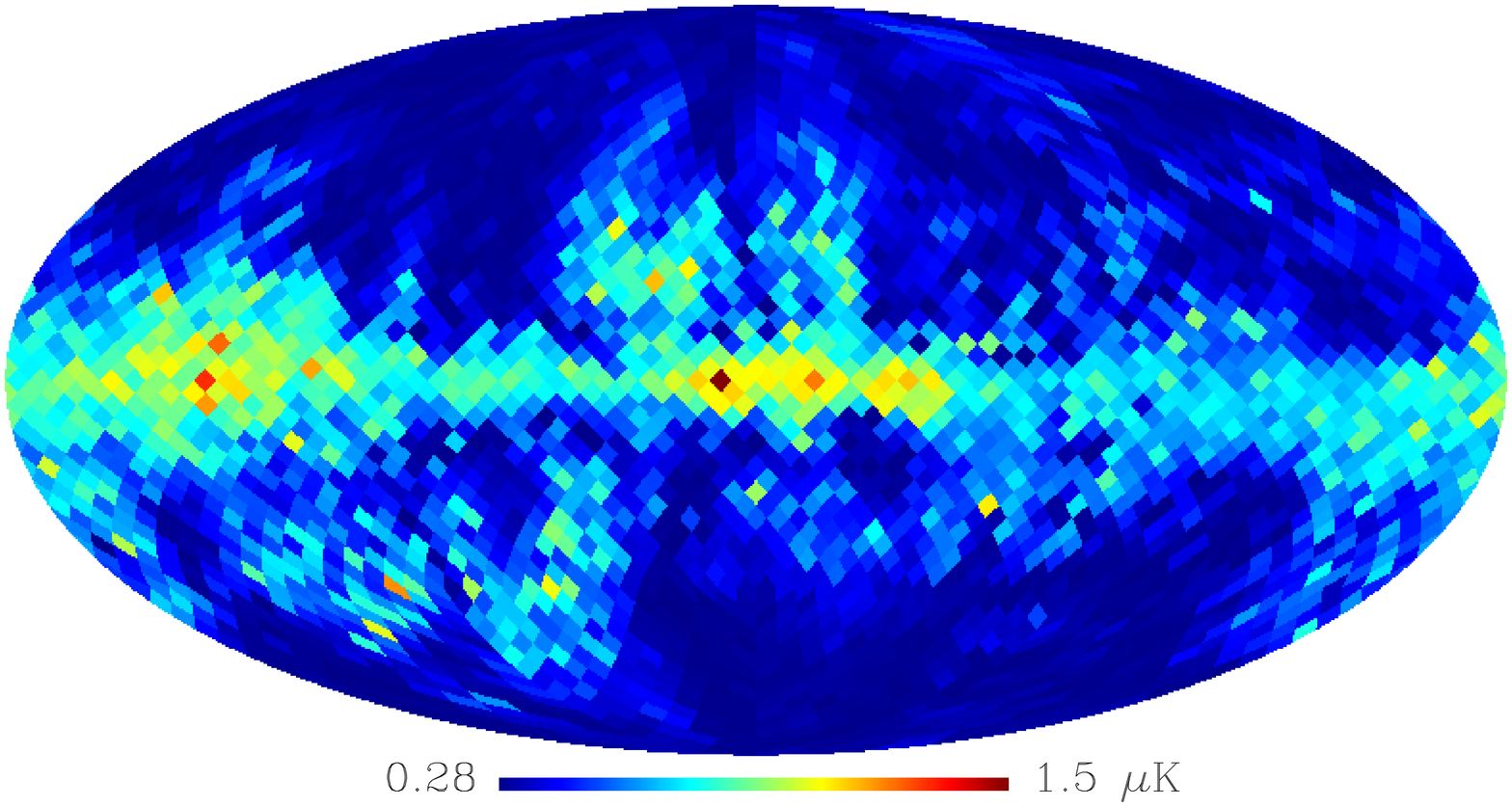}  &
\includegraphics[width=.20\textwidth, keepaspectratio,angle=90]{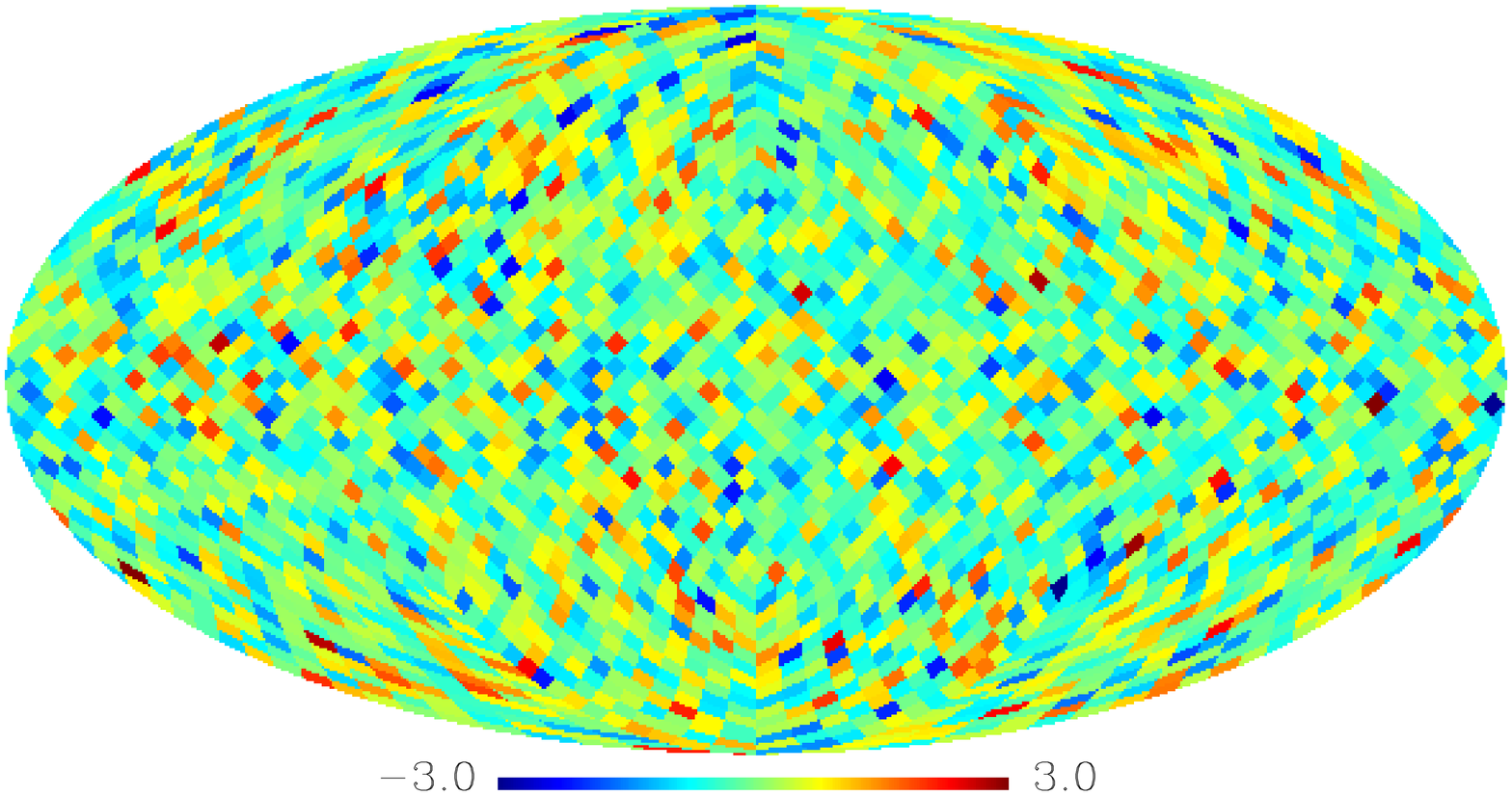}  &
\includegraphics[ width=.20\textwidth, keepaspectratio,angle=90]{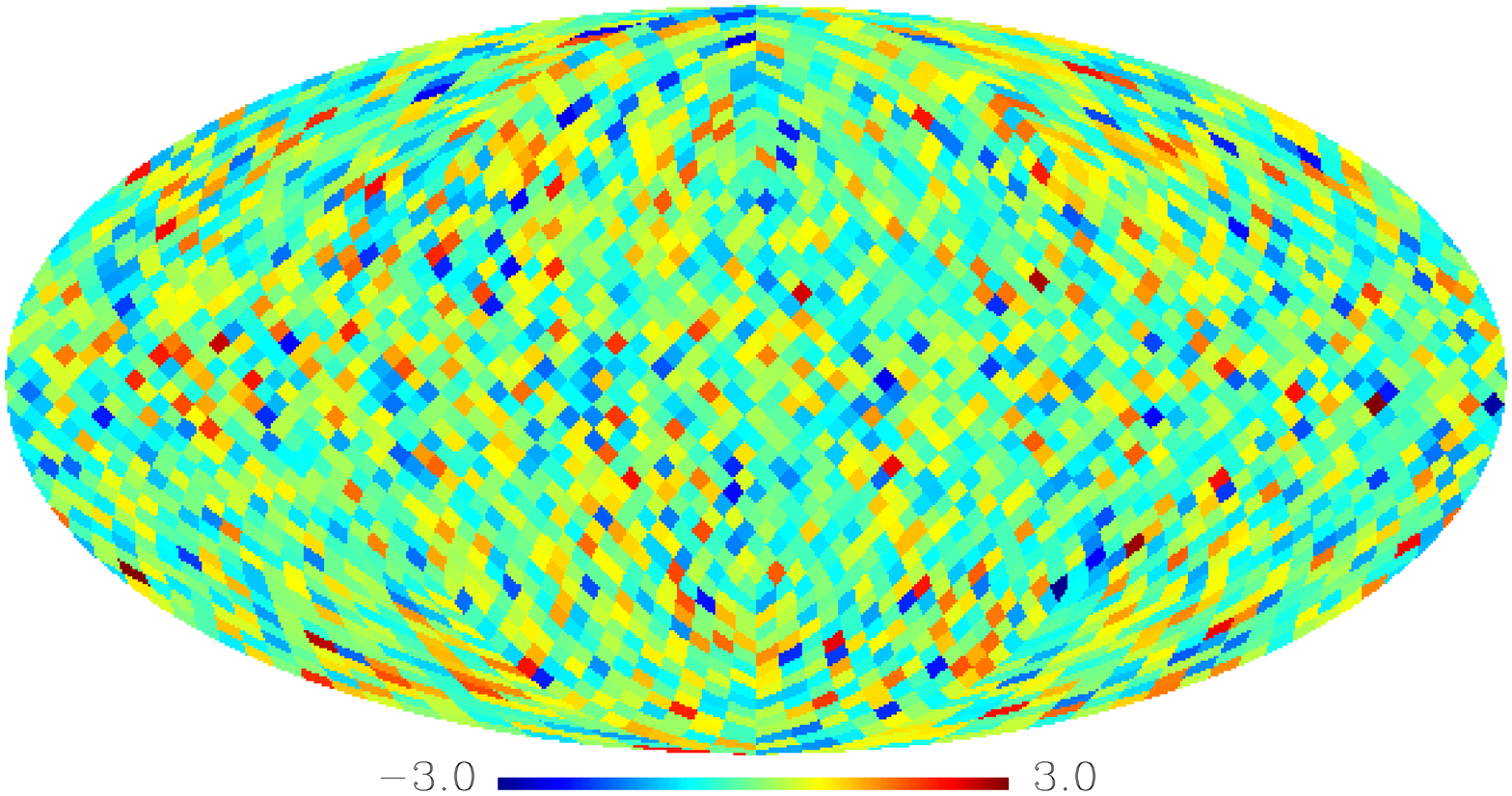} \\
{\scriptsize CMB Input U} & {\scriptsize {\it Commander} U} & {\scriptsize {\it Galclean} U} \\
\includegraphics[ width=.20\textwidth, keepaspectratio,angle=90]{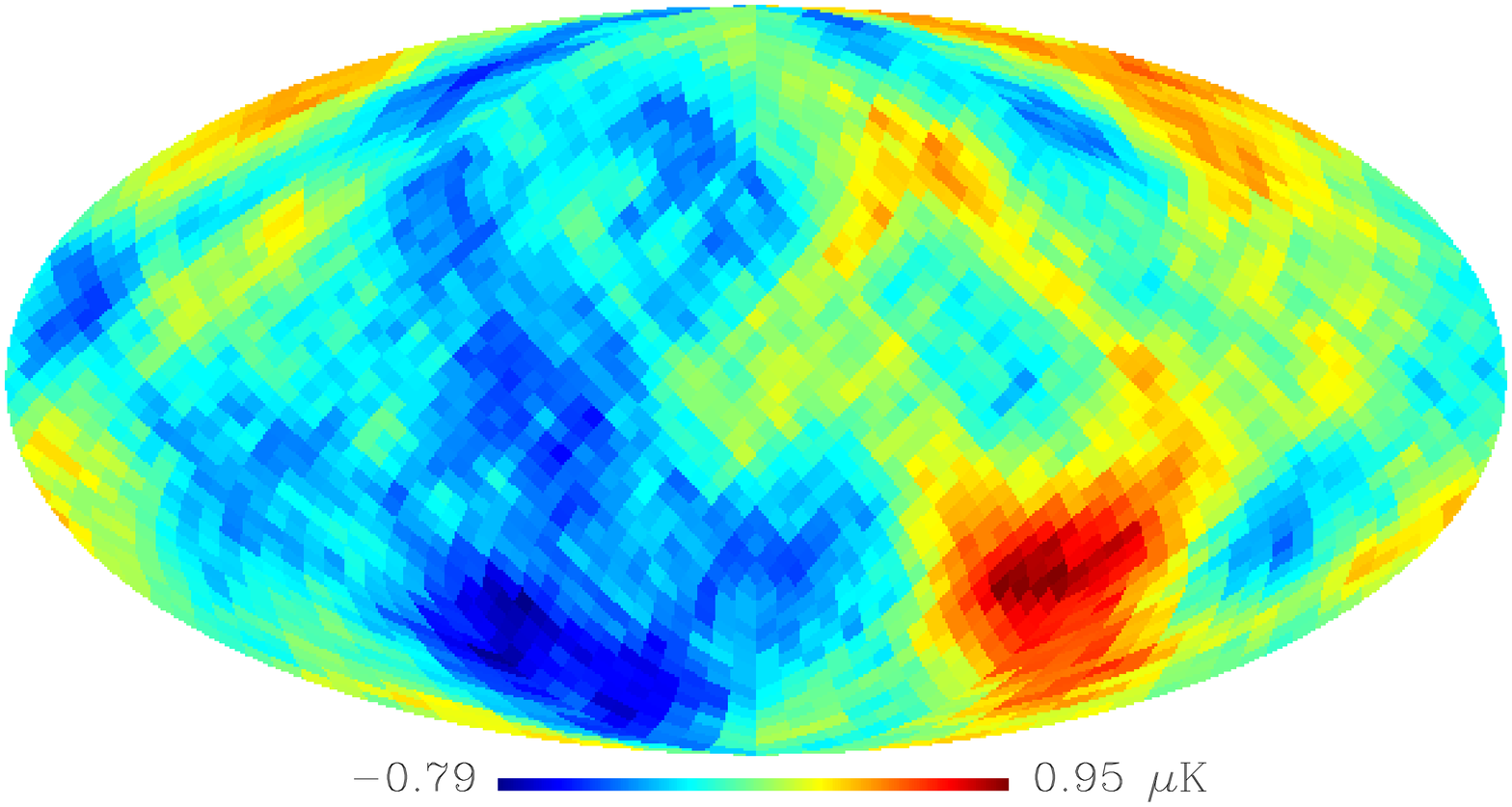} &
\includegraphics[width=.20\textwidth, keepaspectratio,angle=90]{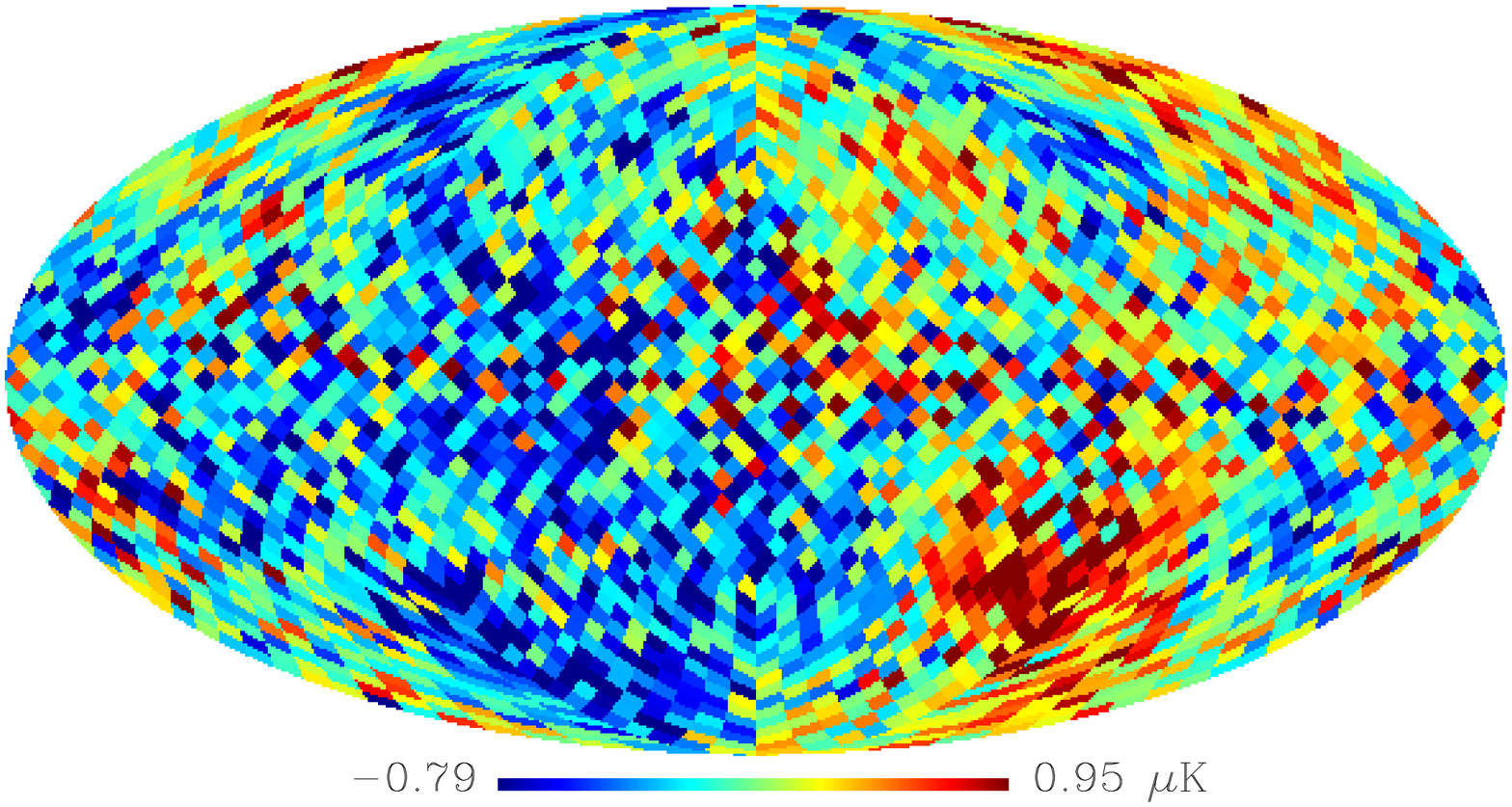} &
\includegraphics[ width=.20\textwidth, keepaspectratio,angle=90]{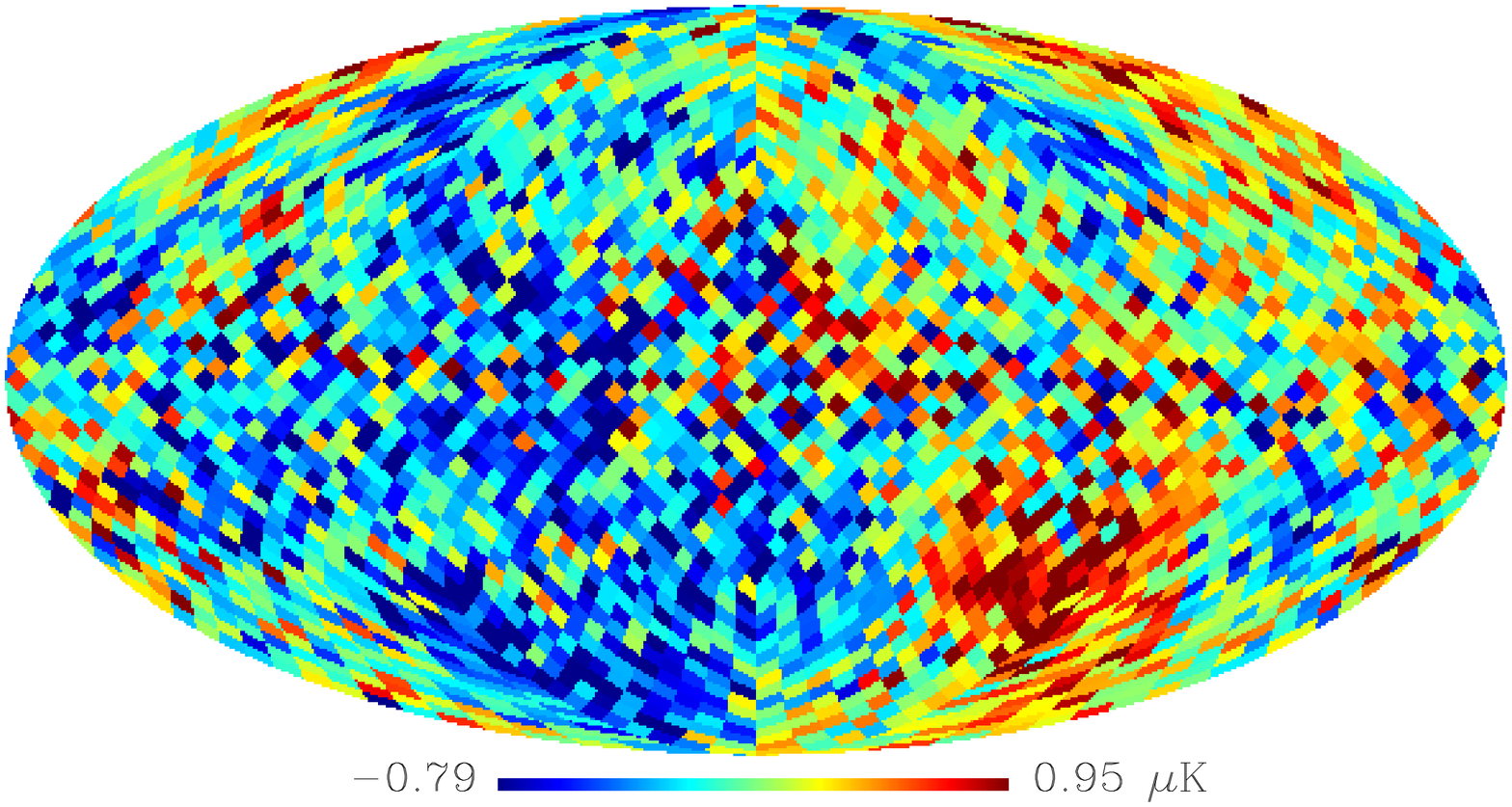}\\
{\scriptsize Error U} & {\scriptsize {\it Commander} U deviation} & {\scriptsize {\it Galclean} U deviation} \\
\includegraphics[width=.20\textwidth, keepaspectratio,angle=90]{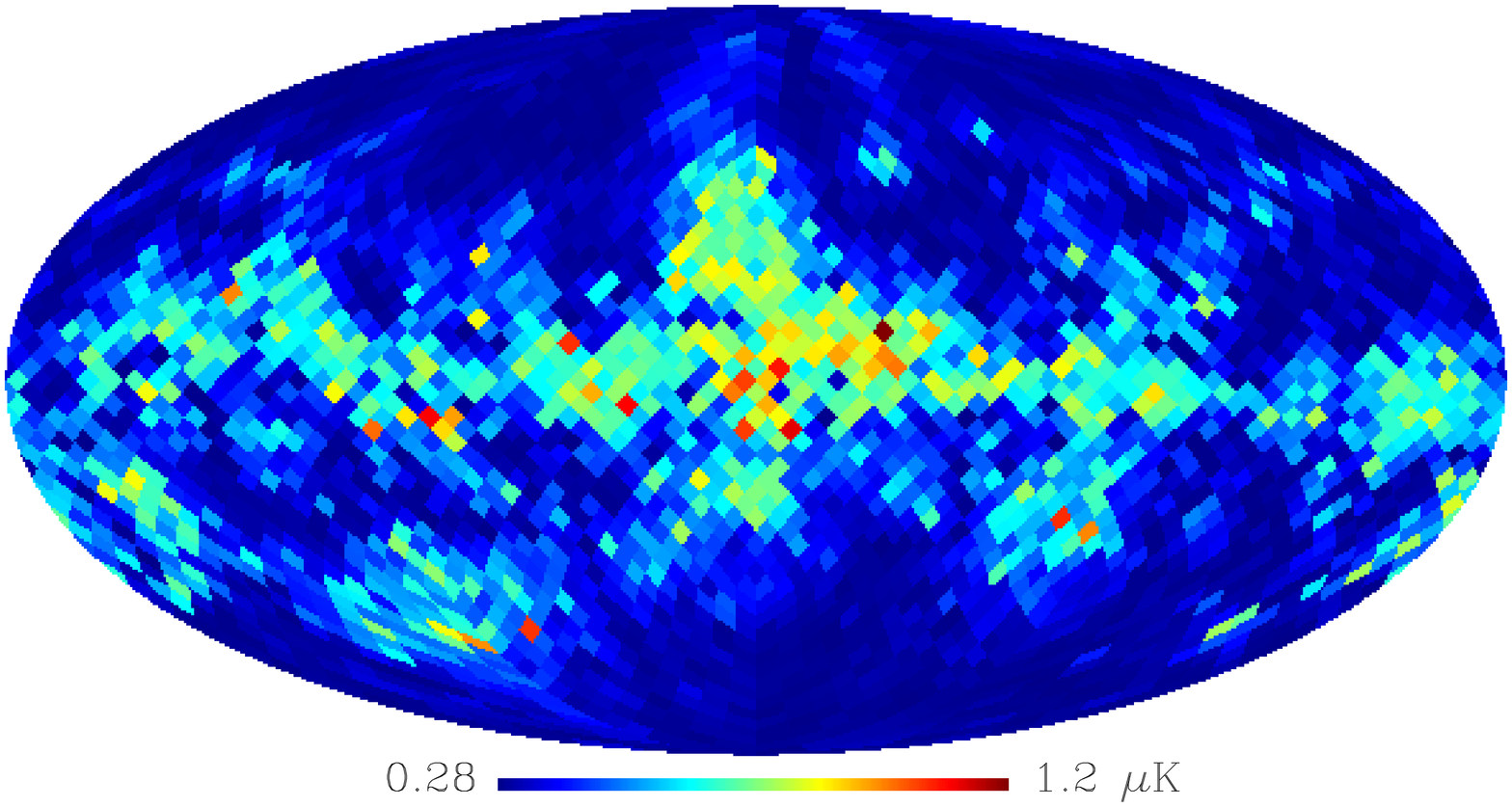} &
\includegraphics[width=.20\textwidth, keepaspectratio,angle=90]{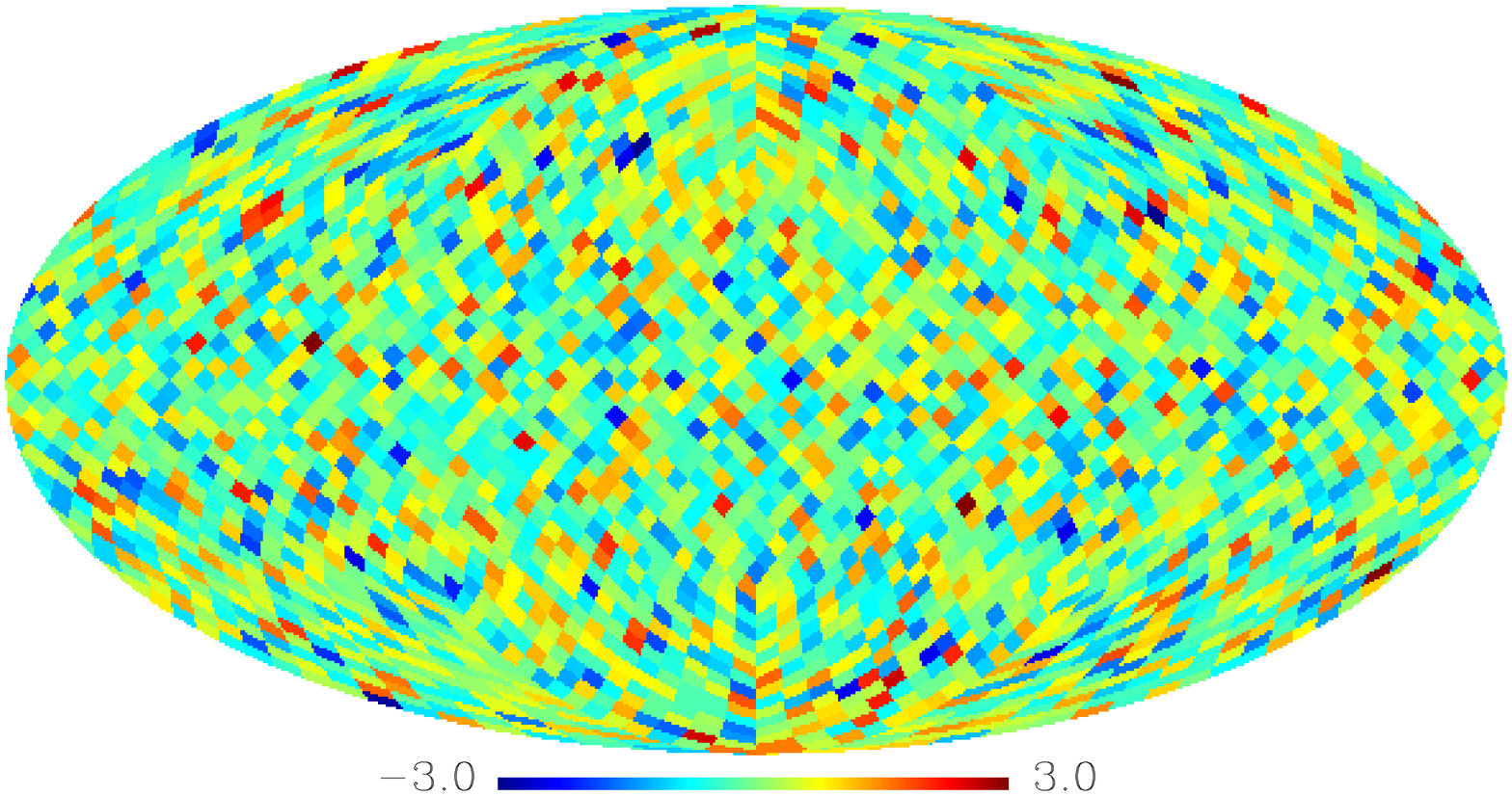} &
\includegraphics[ width=.20\textwidth, keepaspectratio,angle=90]{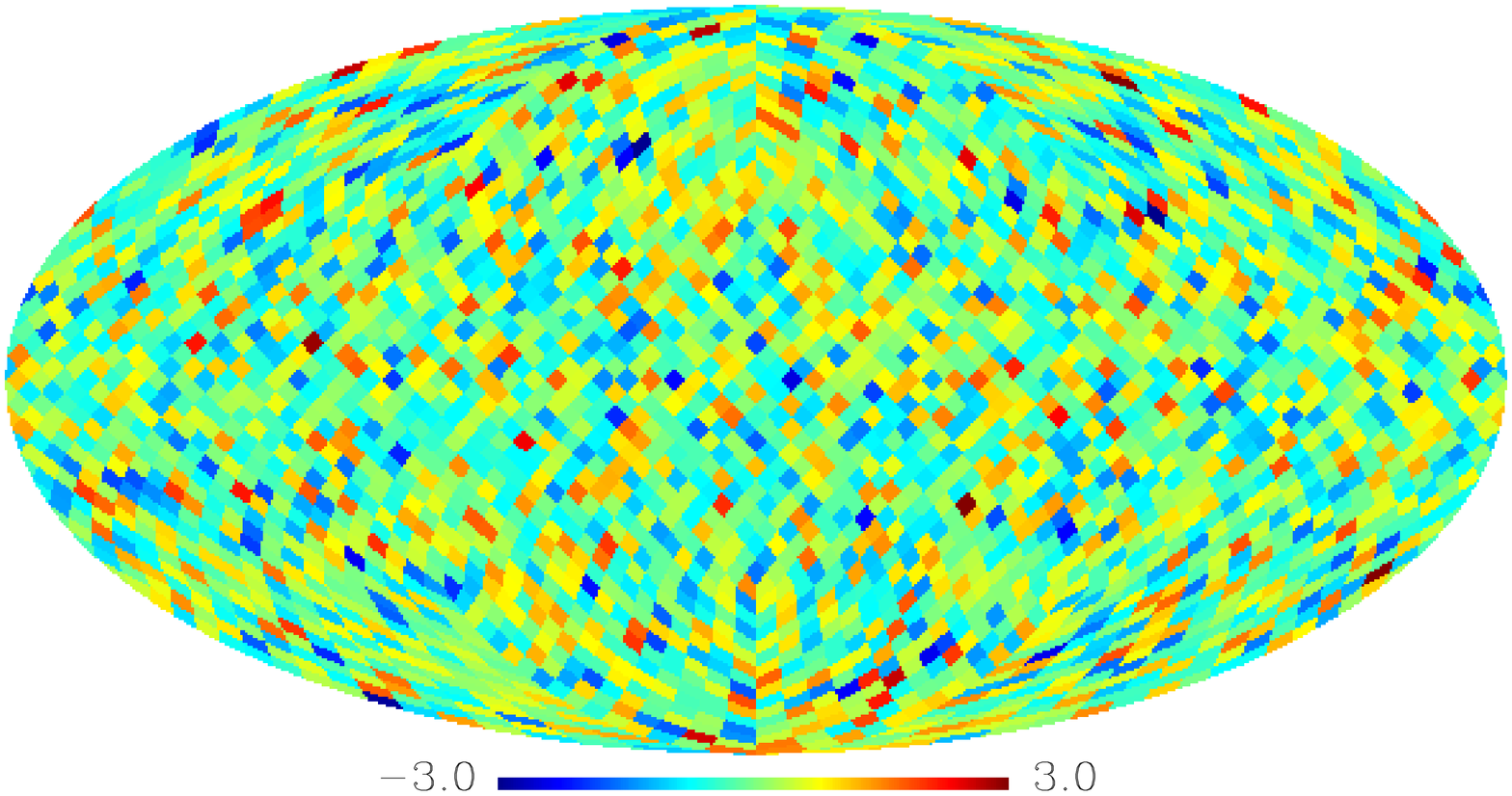}\\
\end{tabular}

   \caption{First row: input Q CMB map (left column), {\it Commander} posterior mean output Q map (middle column), and {\it Galclean} posterior mean output Q map (right column).  Second row: marginalized error, {\it Commander} difference in standard deviations per pixel (middle column), and {\it Galclean} difference in standard deviations per pixel (right column) for the Q component.  Third row: input U CMB map (left column), {\it Commander} posterior mean output U map (middle column), and {\it Galclean} posterior mean output U map (right column).  Fourth row: As in second row but for the U component.}
   \label{fig:CMBmaps}
\end{figure*}

In Fig.~\ref{fig:CMBmaps}, we show the input and marginalized output Q and U polarization maps for the CMB, the difference between the input and output maps in standard deviations per pixel for the Q component, $\delta = (Q_{in} - Q_{out})/\sigma_Q$, and similarly for the U component.  The marginalized amplitude maps are scaled to the same scale as the input CMB map.  The marginalized output CMB maps are found to look correct (with the addition of noise) compared to the input CMB maps and the normalized deviations are Gaussian distributed with a standard deviation of one.  We find that the {\it Commander} and {\it Galclean} marginalized output CMB maps agree to better than $0.2\sigma$ over almost 99\% of the sky.  The small differences that we see between the marginalized output maps for the two codes can be attributed to differences in sampling since we find differences of the same magnitude between output of the same code initialized with different seeds.  We find that the difference between the input and output is greatest in the galactic plane, where the foreground signal is high.  However, these effects are captured in the marginalized errors in the maps (also shown in Fig.~\ref{fig:CMBmaps}) so the deviation maps do not have strong spatial dependence.  The marginalized error maps are a useful product of parametric map estimation and could potentially be used to define a mask threshold level.  Assuming these white noise properties, we find that the signal-to-noise ratio in {\it Planck} should be good enough to clearly see features in the {\it Planck} CMB polarization maps.

\begin{figure*}
   \centering
   \begin{tabular}{c c c}
Synchrotron Input P & {\it Commander} Sync P & {\it Galclean} Sync P \\
\includegraphics[ width=.20\textwidth, keepaspectratio,angle=90]{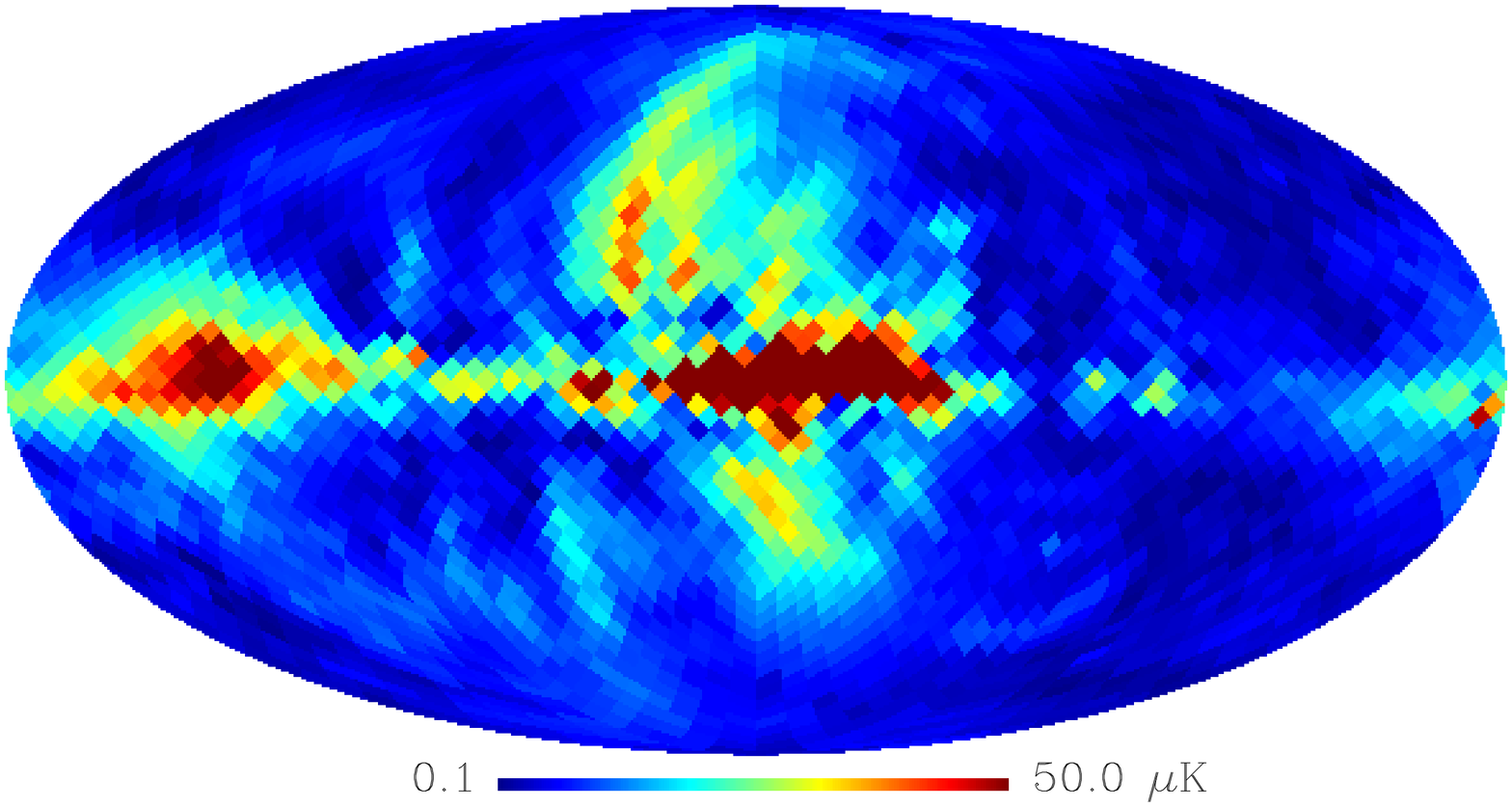} &
\includegraphics[width=.20\textwidth, keepaspectratio,angle=90]{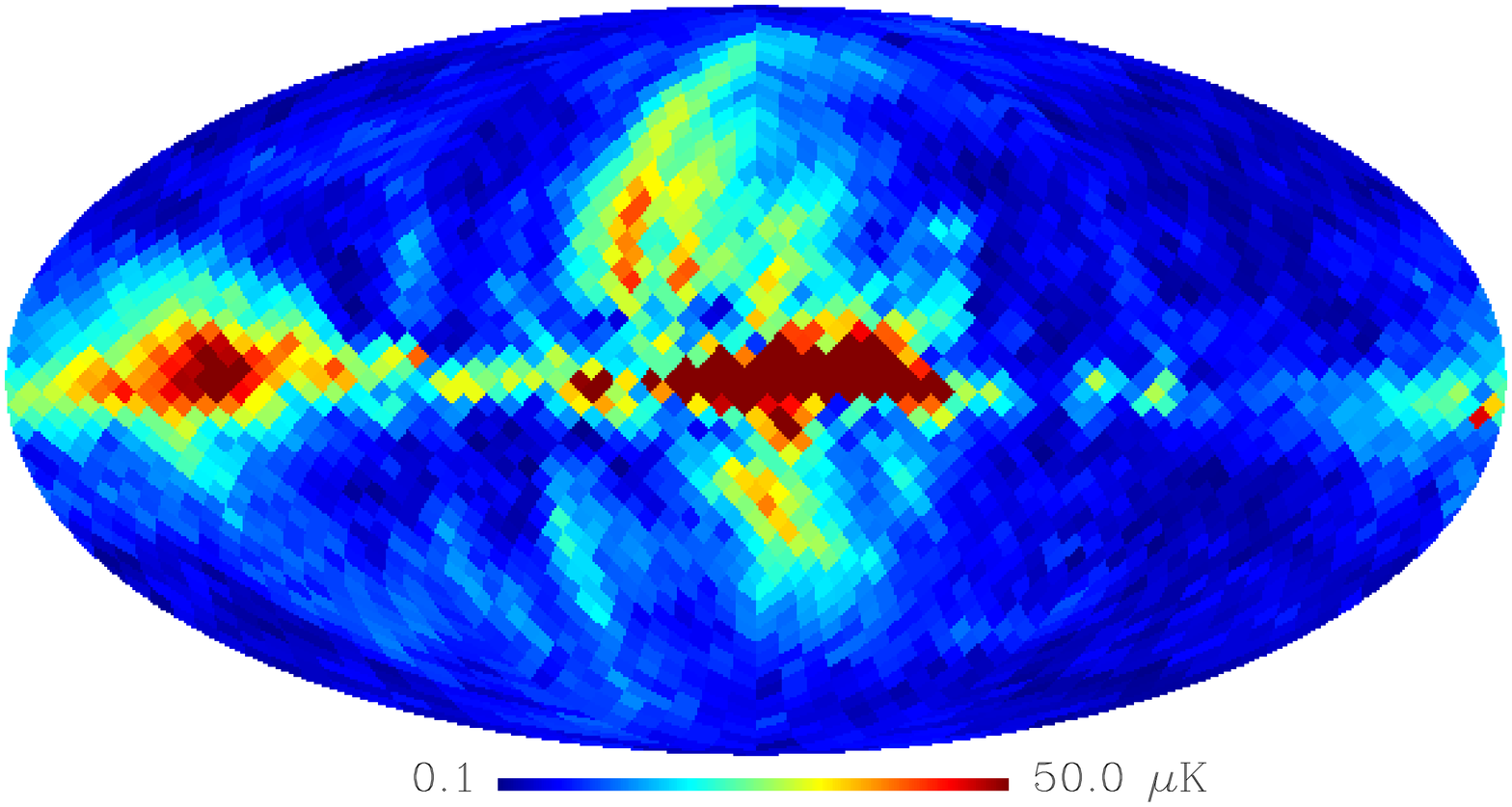} &
\includegraphics[width=.20\textwidth, keepaspectratio,angle=90]{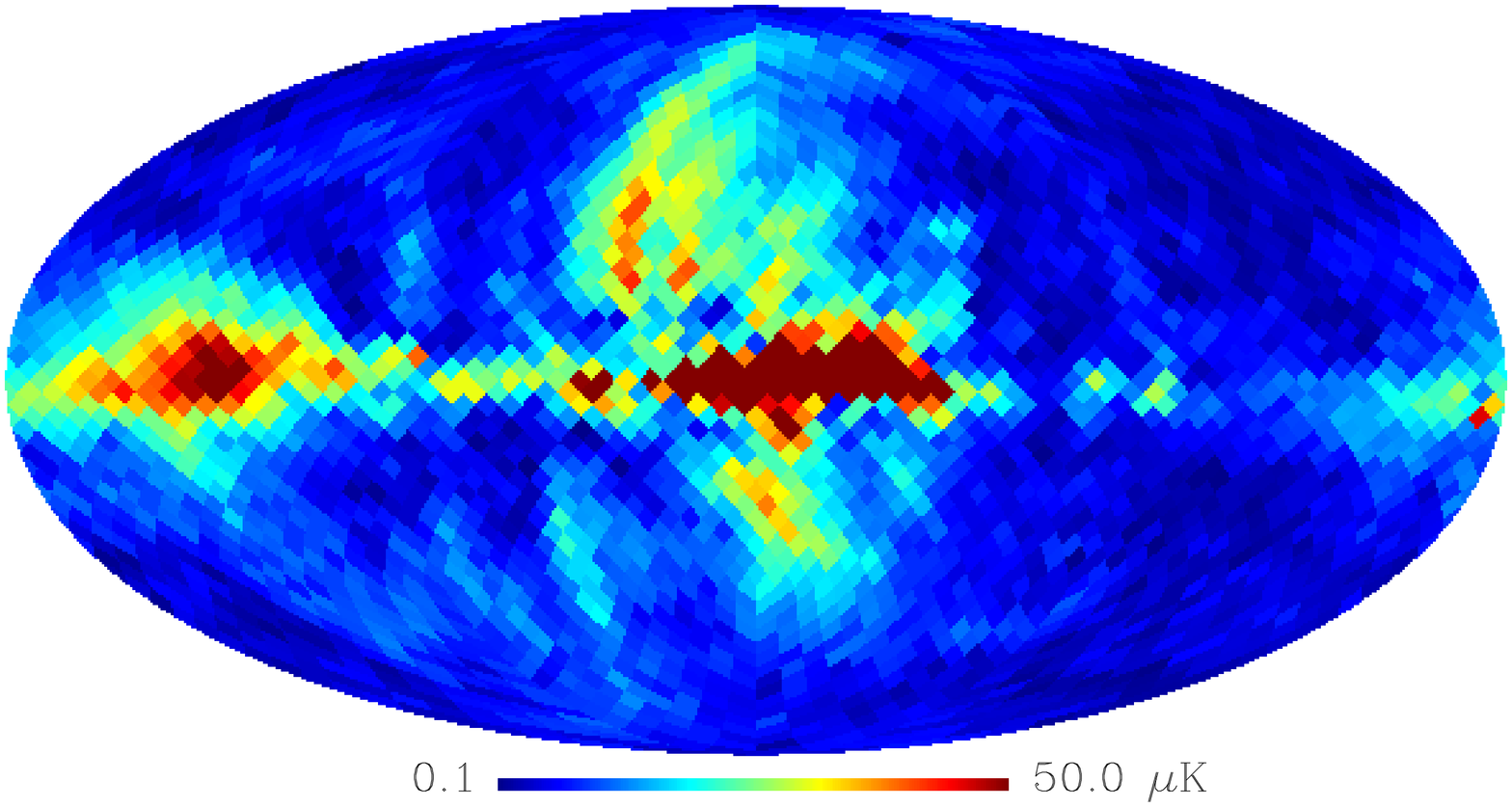} \\

Error Q & {\it Commander} Q deviation & {\it Galclean} Q deviation \\
\includegraphics[ width=.20\textwidth, keepaspectratio,angle=90]{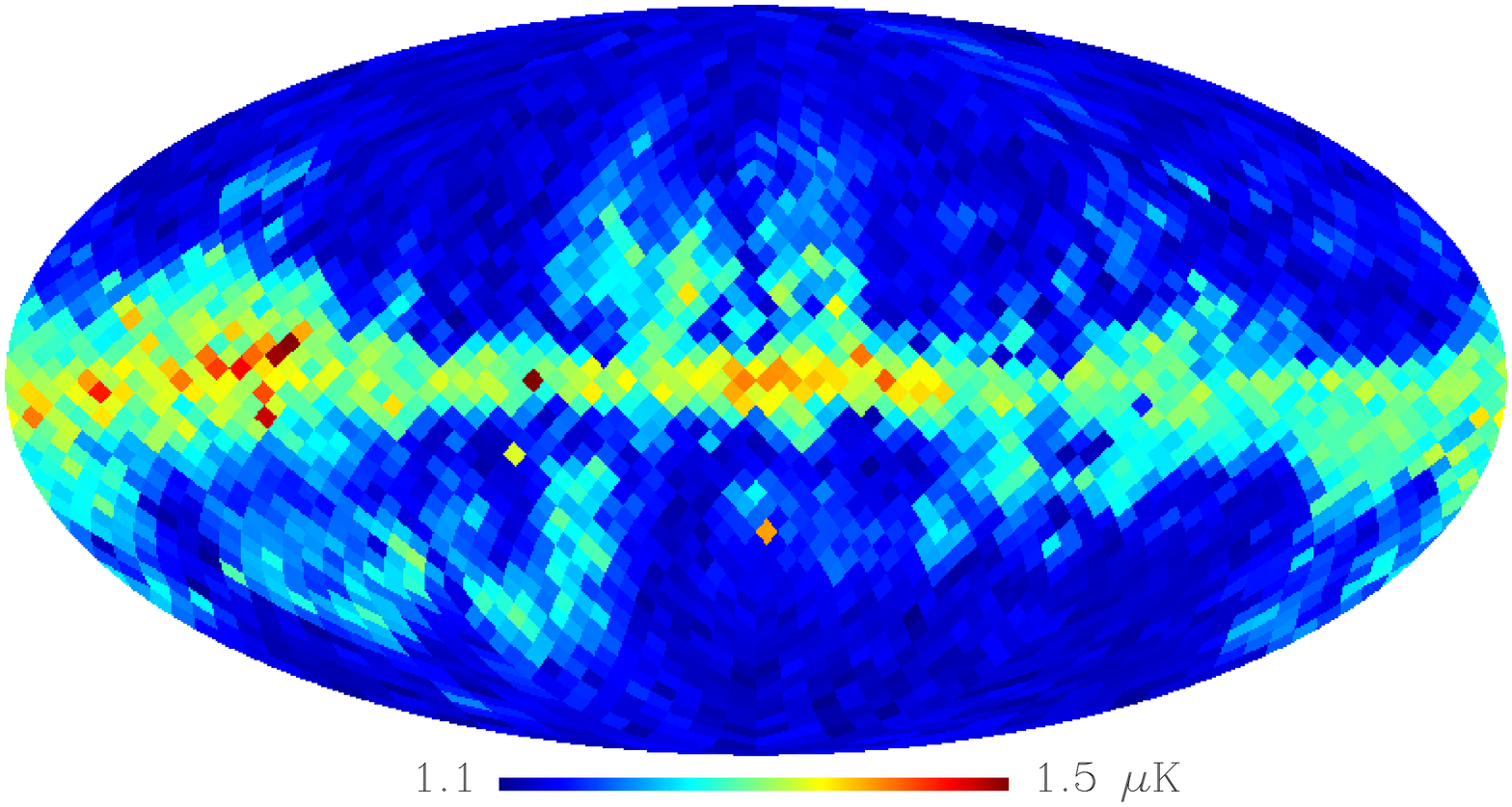} &
\includegraphics[ width=.20\textwidth, keepaspectratio,angle=90]{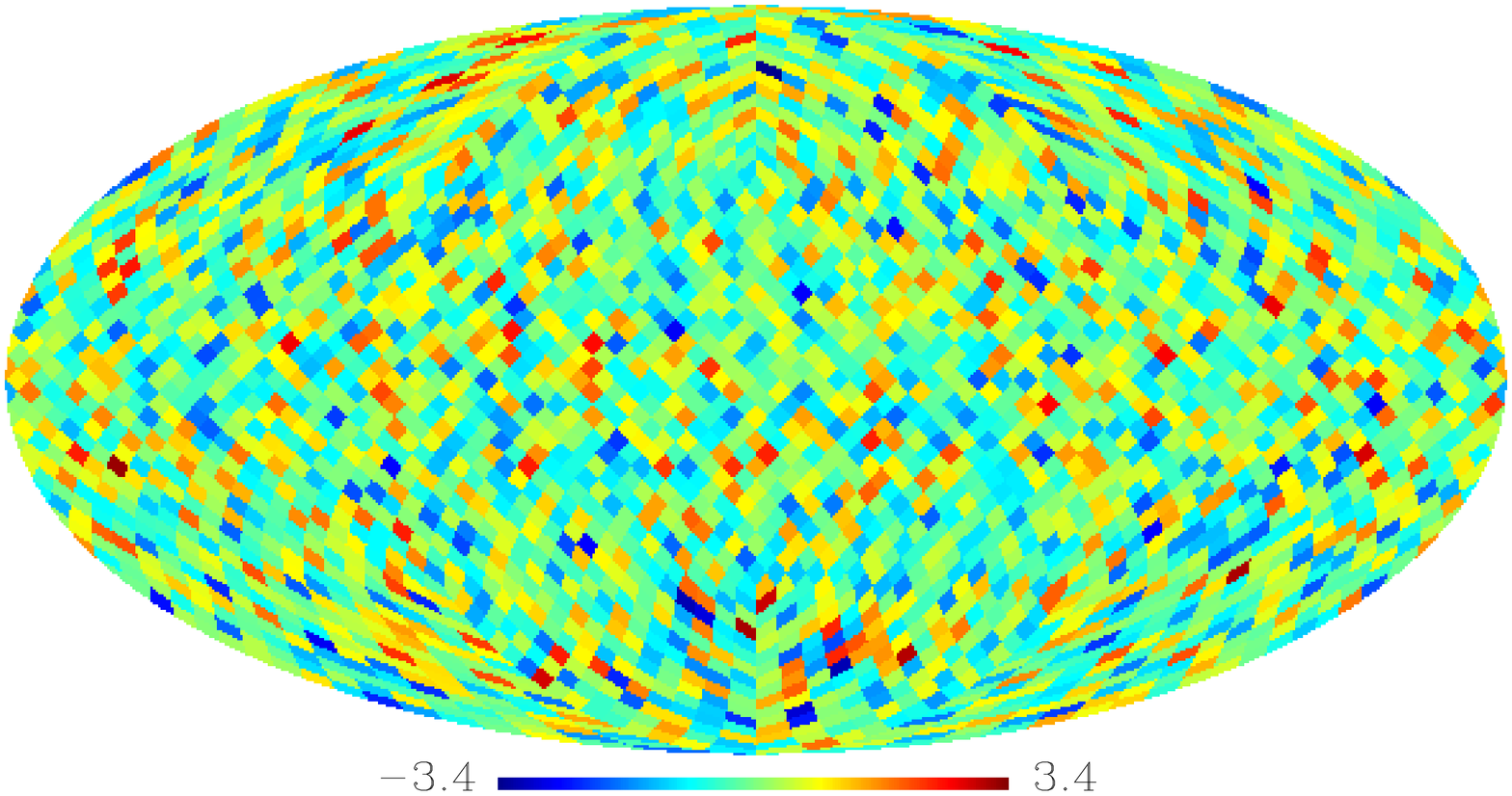} &
\includegraphics[ width=.20\textwidth, keepaspectratio,angle=90]{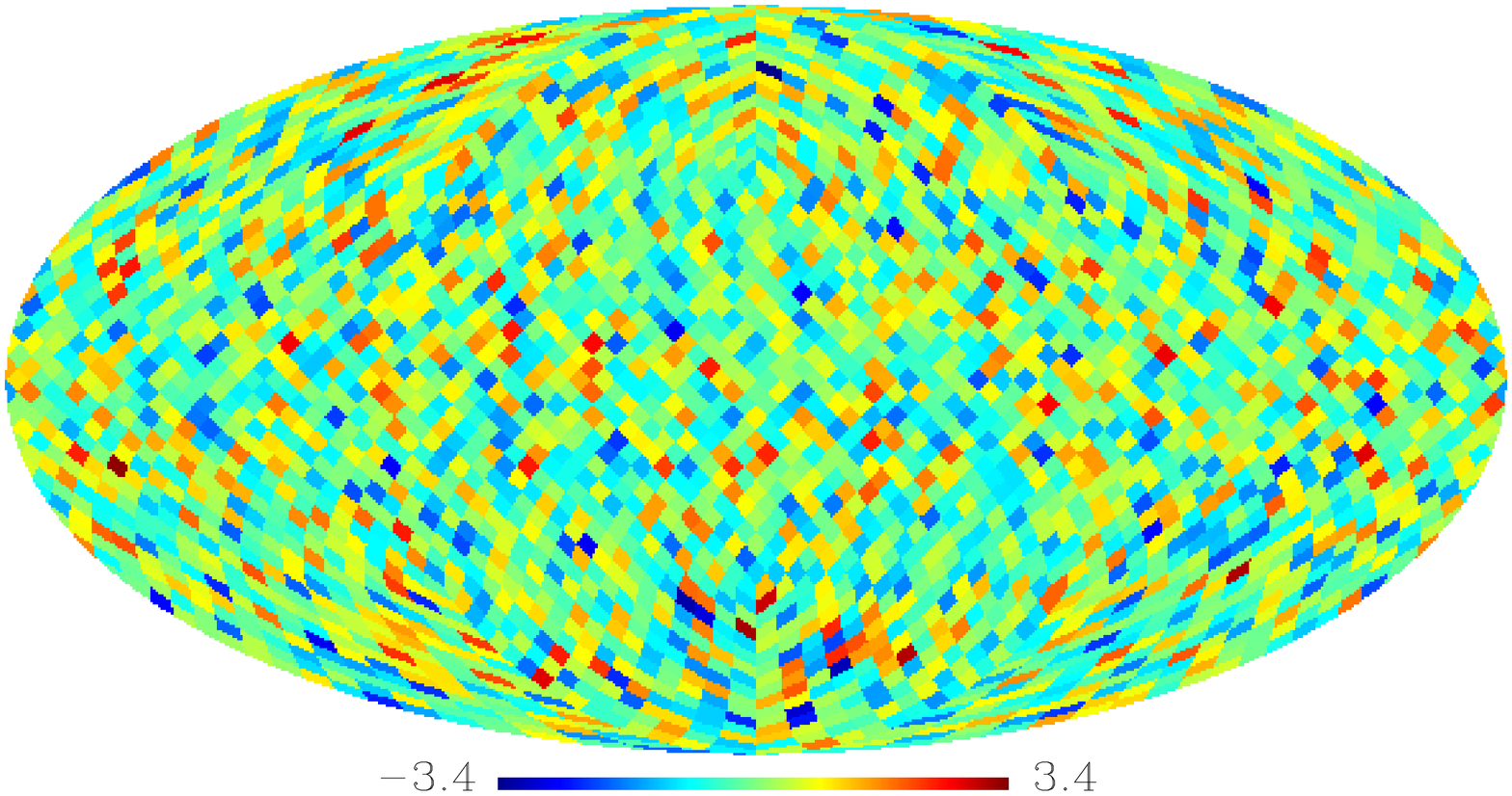} \\

Dust Input P & {\it Commander} Dust P & {\it Galclean} Dust P \\
\includegraphics[ width=.20\textwidth, keepaspectratio,angle=90]{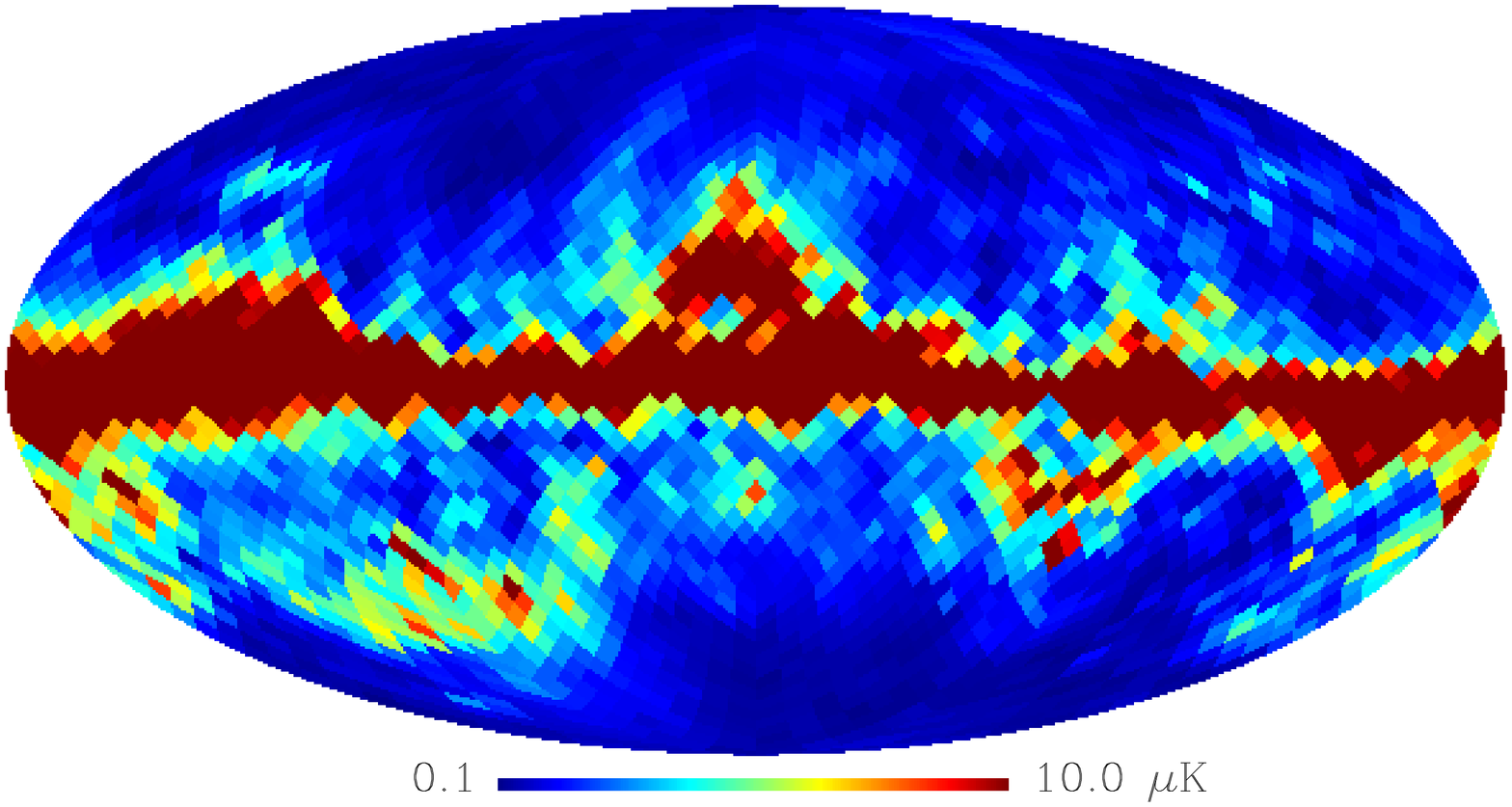} &
\includegraphics[width=.20\textwidth, keepaspectratio,angle=90]{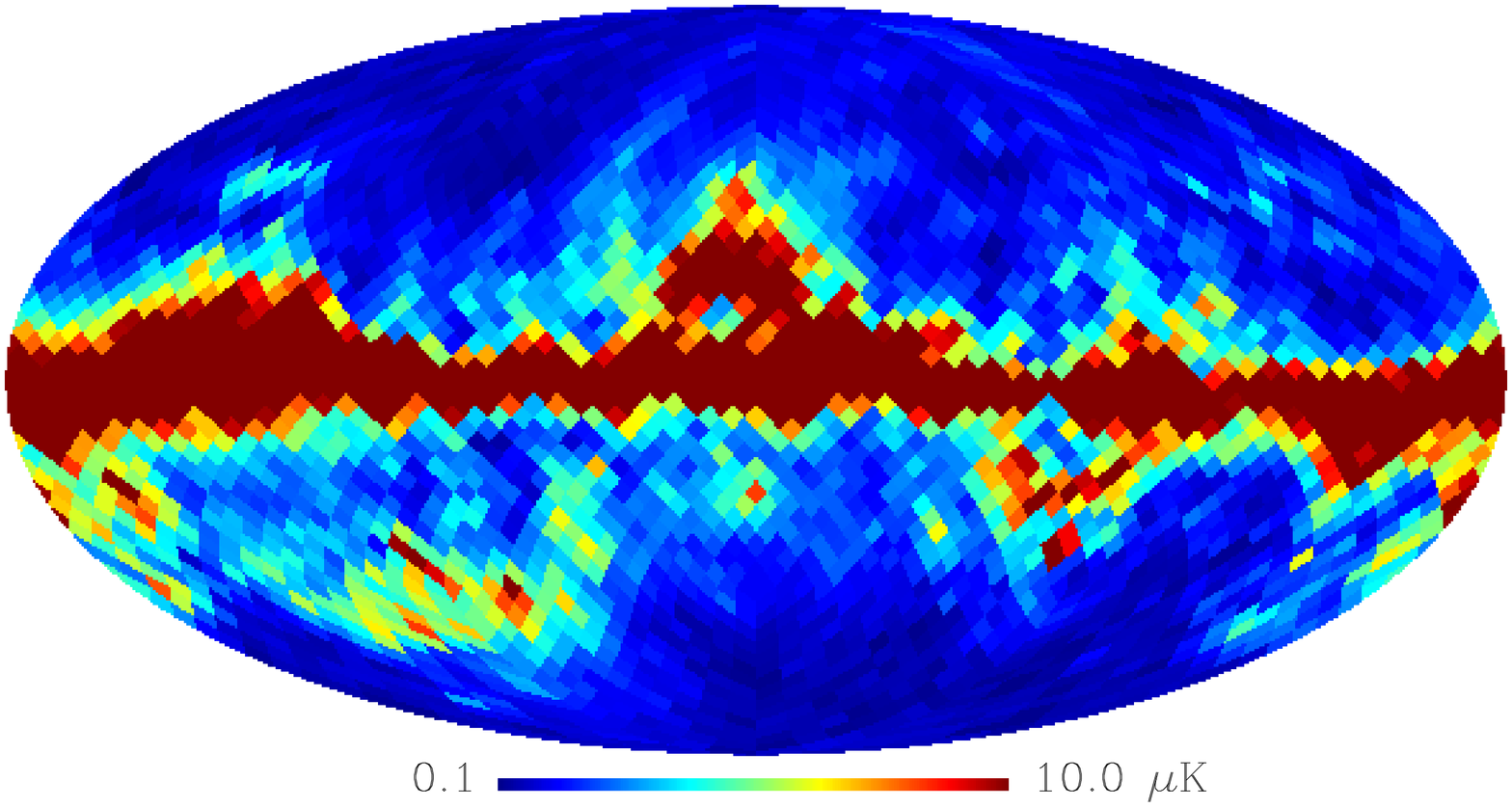} &
\includegraphics[width=.20\textwidth, keepaspectratio,angle=90]{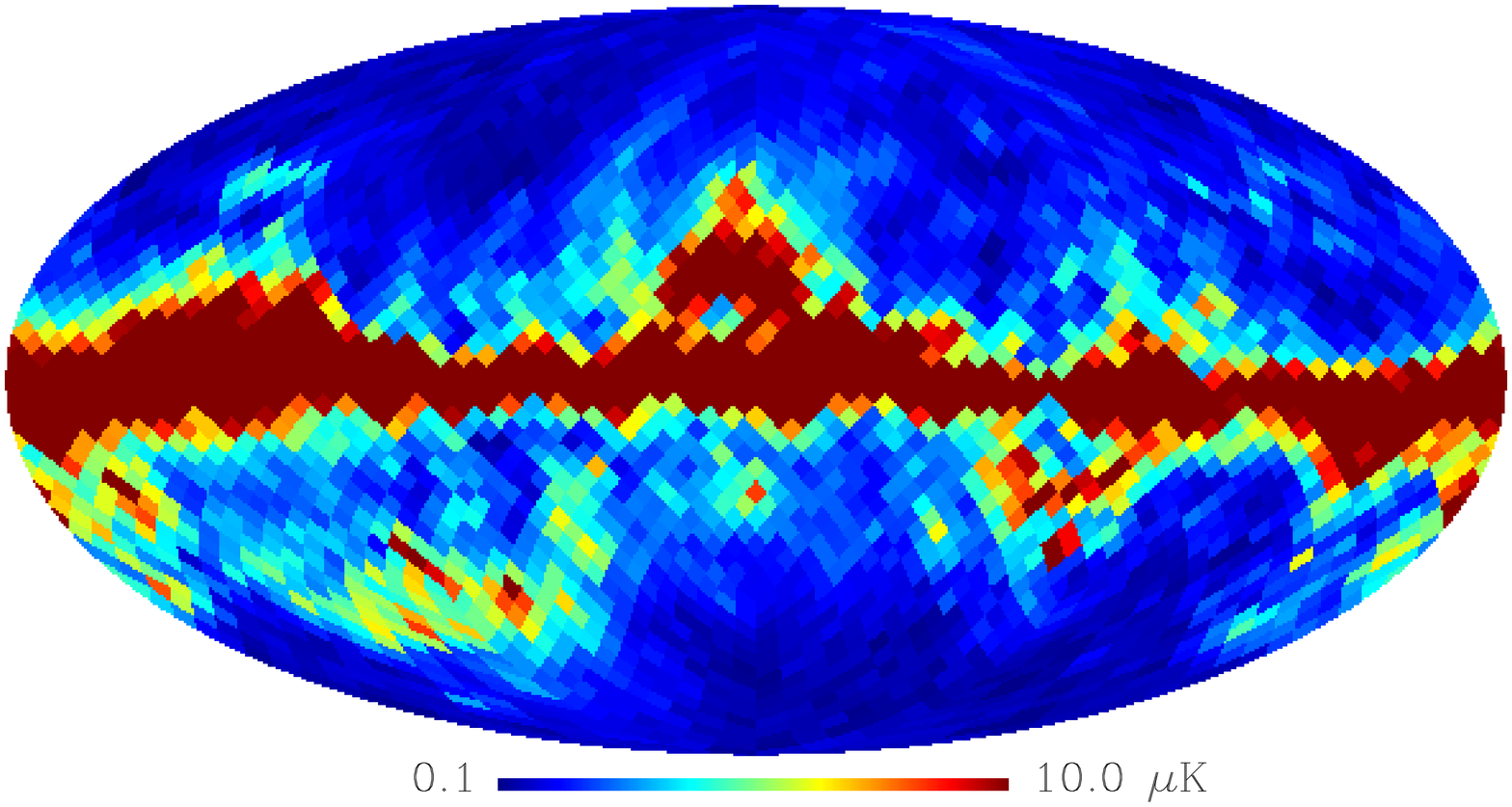}\\

Error Q & {\it Commander} Q deviation & {\it Galclean} Q deviation \\
\includegraphics[ width=.20\textwidth, keepaspectratio,angle=90]{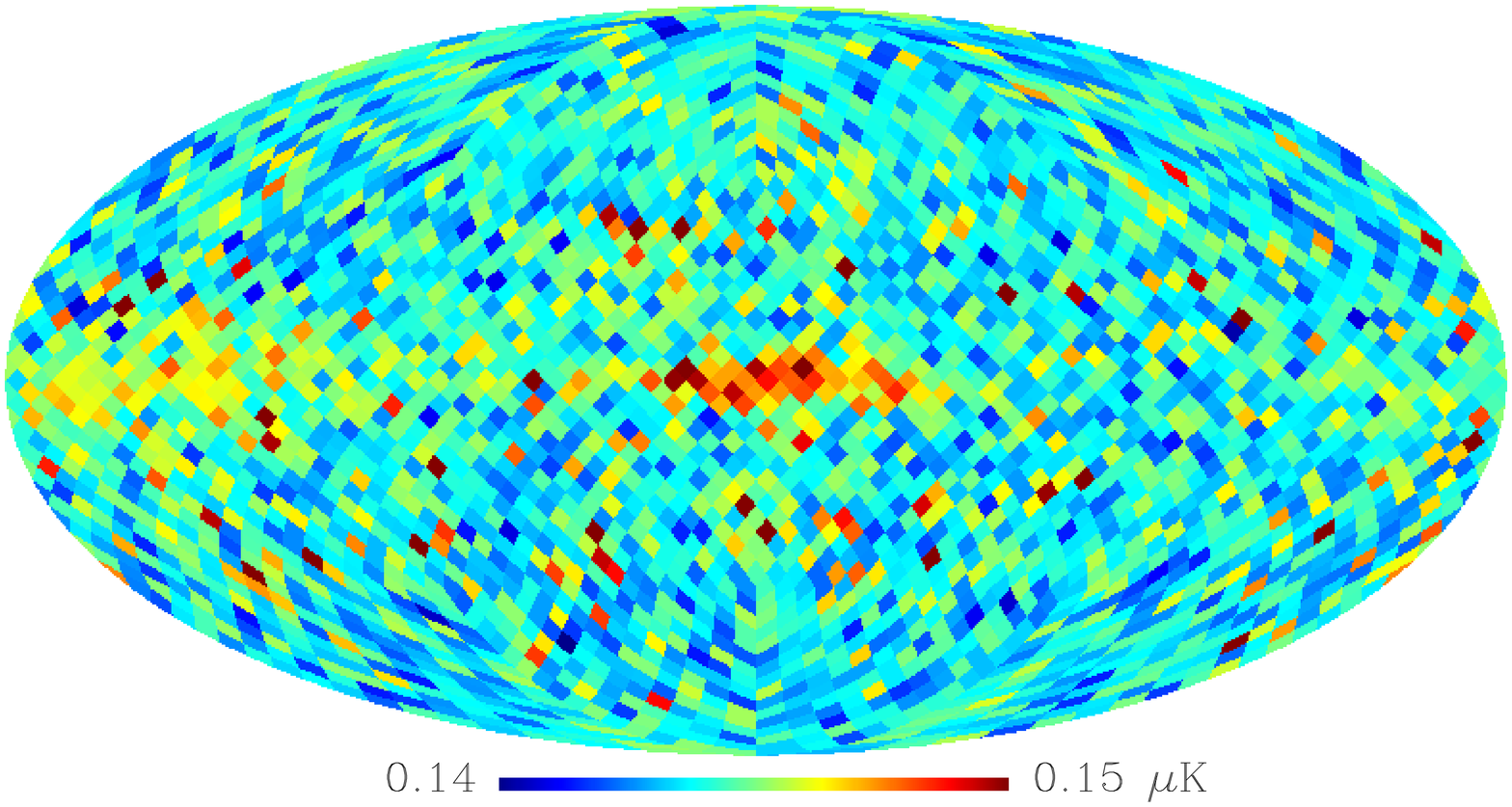} &
\includegraphics[ width=.20\textwidth, keepaspectratio,angle=90]{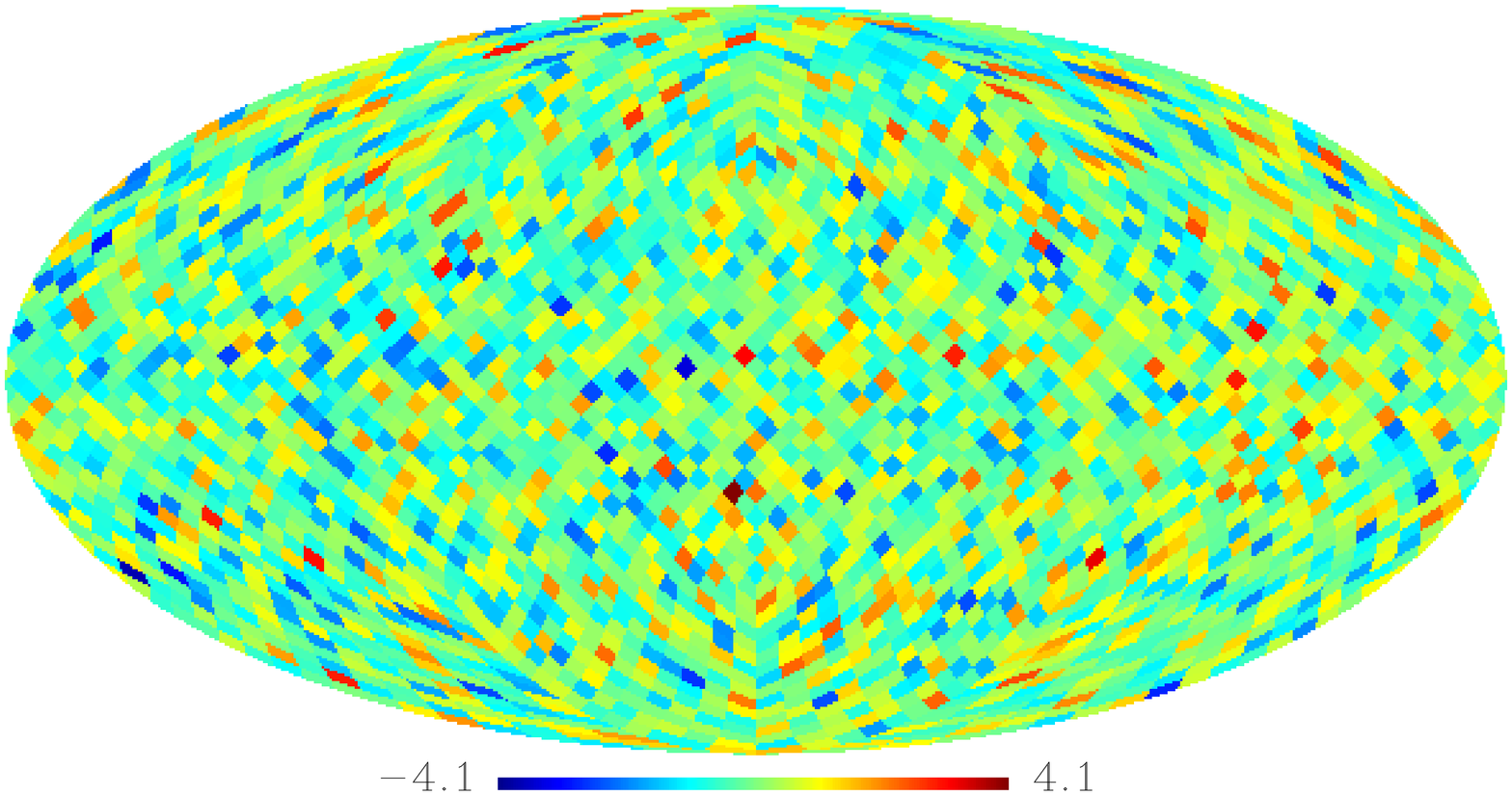} &
\includegraphics[ width=.20\textwidth, keepaspectratio,angle=90]{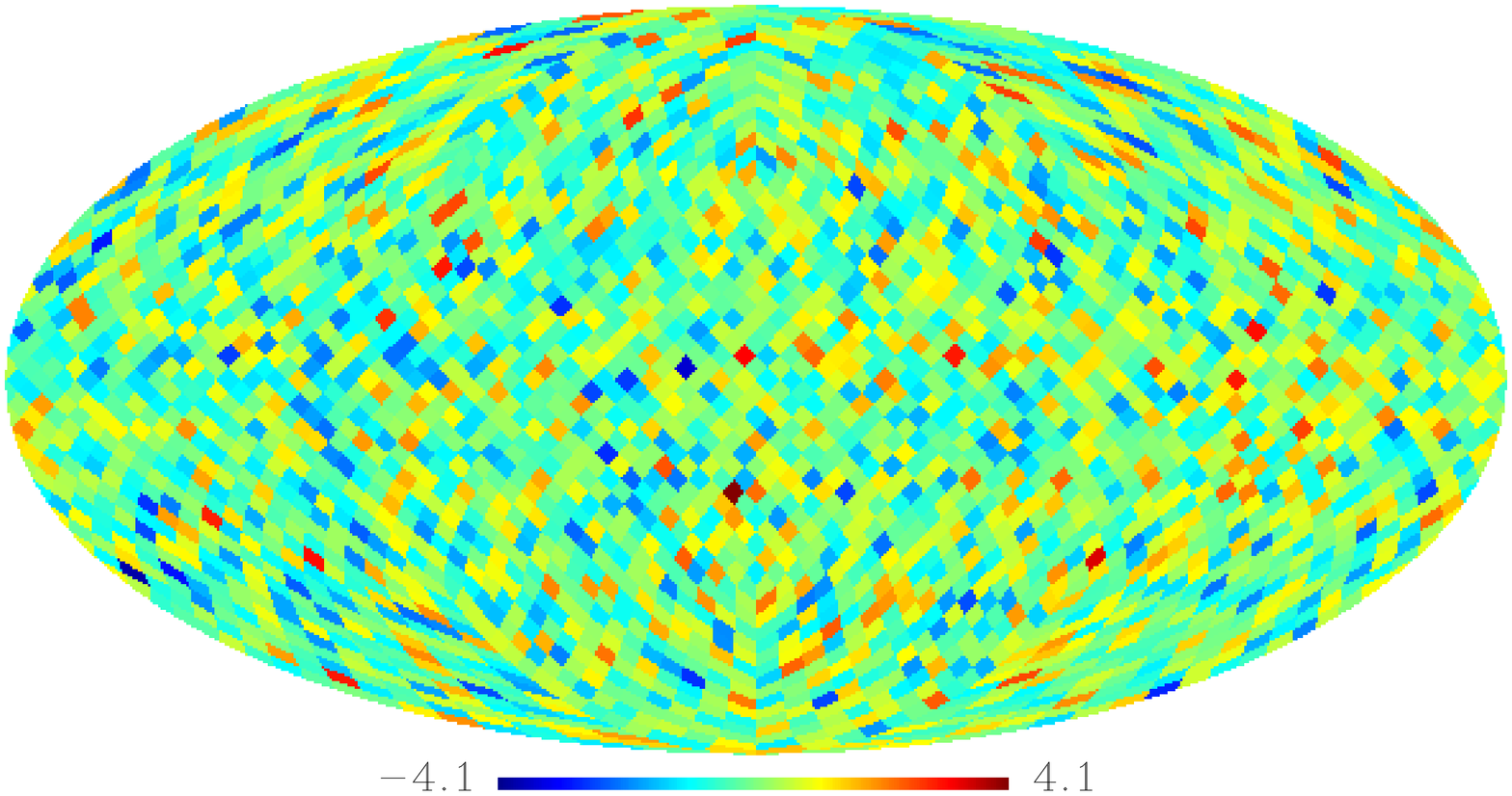}\\

\end{tabular}
   \caption{Maps of the polarization amplitude $P=\sqrt{Q^2+U^2}$ for the synchrotron at 30 GHz (first row) and dust at 353 GHz (third row).  The difference in standard deviations per pixel for the Q component (second and fourth rows) indicate that the synchrotron and dust maps have been recovered to the expected statistical result.}
   \label{fig:foreground_maps}
\end{figure*}

Fig.~\ref{fig:foreground_maps} shows the input and output polarization amplitude $P=\sqrt{Q^2+U^2}$ maps for the synchrotron and dust maps, and the difference in standard deviations per pixel for the Q component.   For the synchrotron and dust amplitude maps, we find that the {\it Commander} and {\it Galclean} marginalized output maps agree to better than $0.1\sigma$ over almost 99\% of the sky. 

\begin{figure*}
   \centering
   \begin{tabular}{c c}
   Dust index & Synchrotron index \\
   \includegraphics[width=0.30\textwidth,keepaspectratio,angle=90]{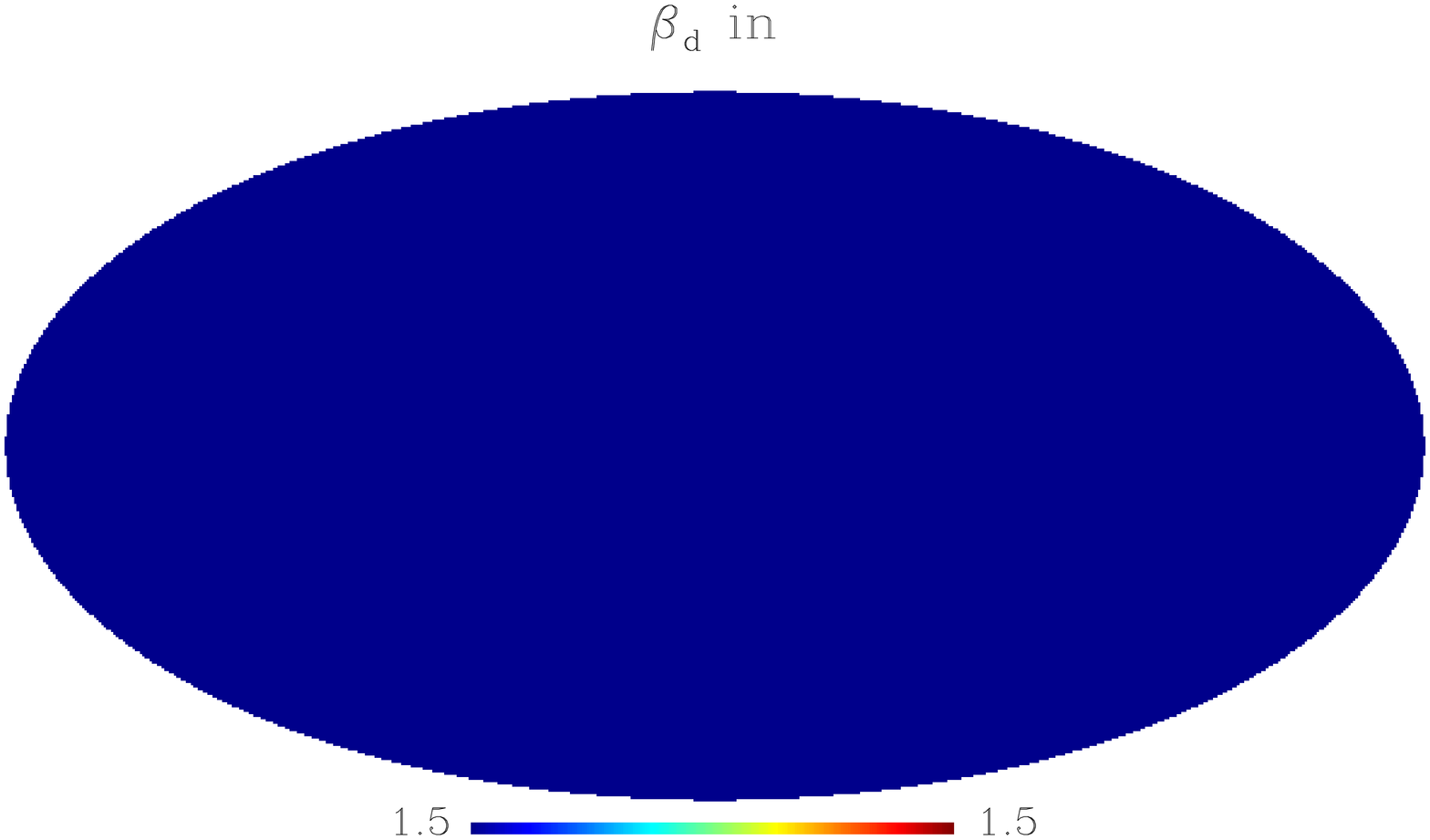} &   
   \includegraphics[width=0.30\textwidth,keepaspectratio,angle=90]{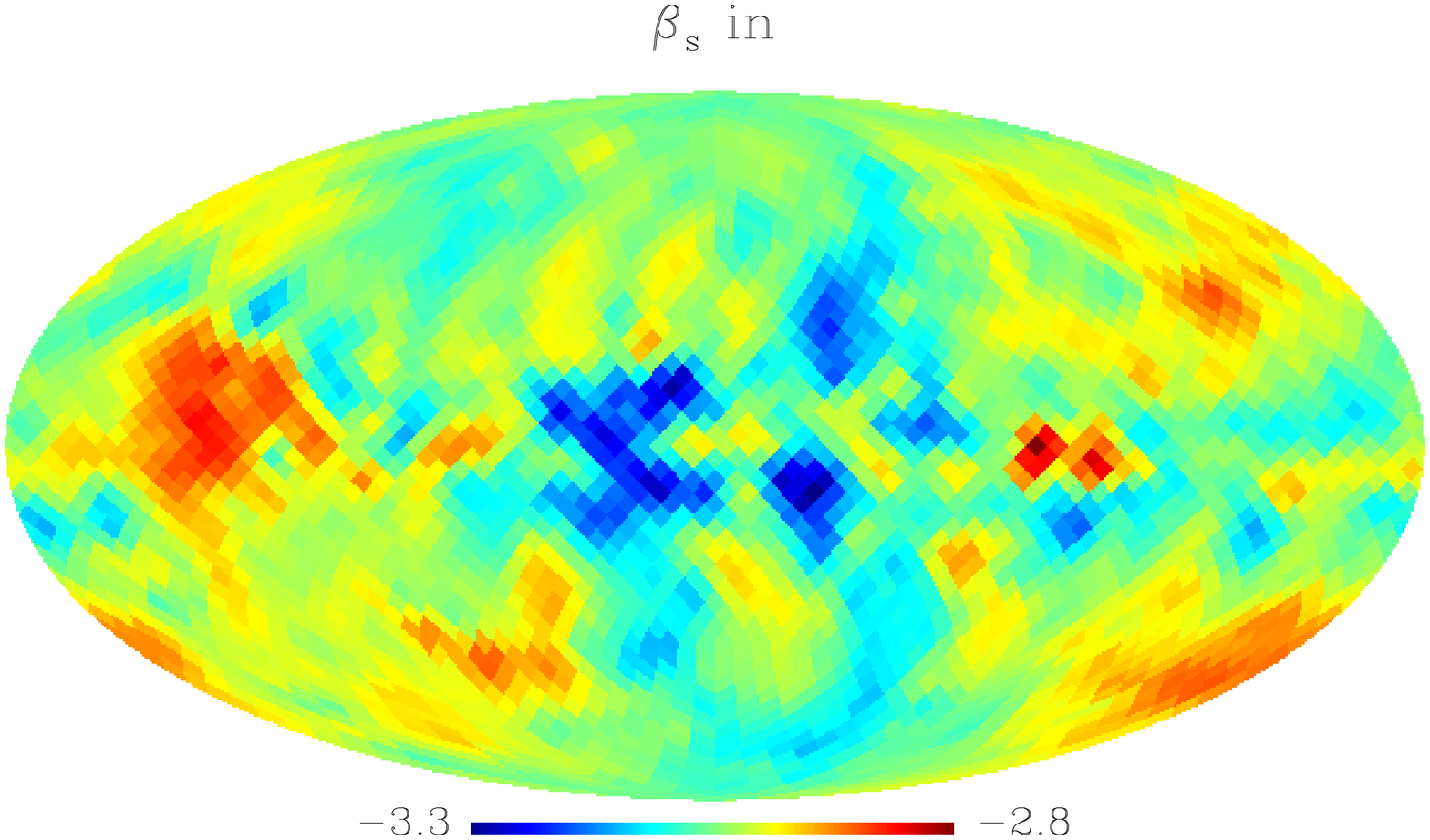} \\   
   \includegraphics[width=0.30\textwidth,keepaspectratio,angle=90]{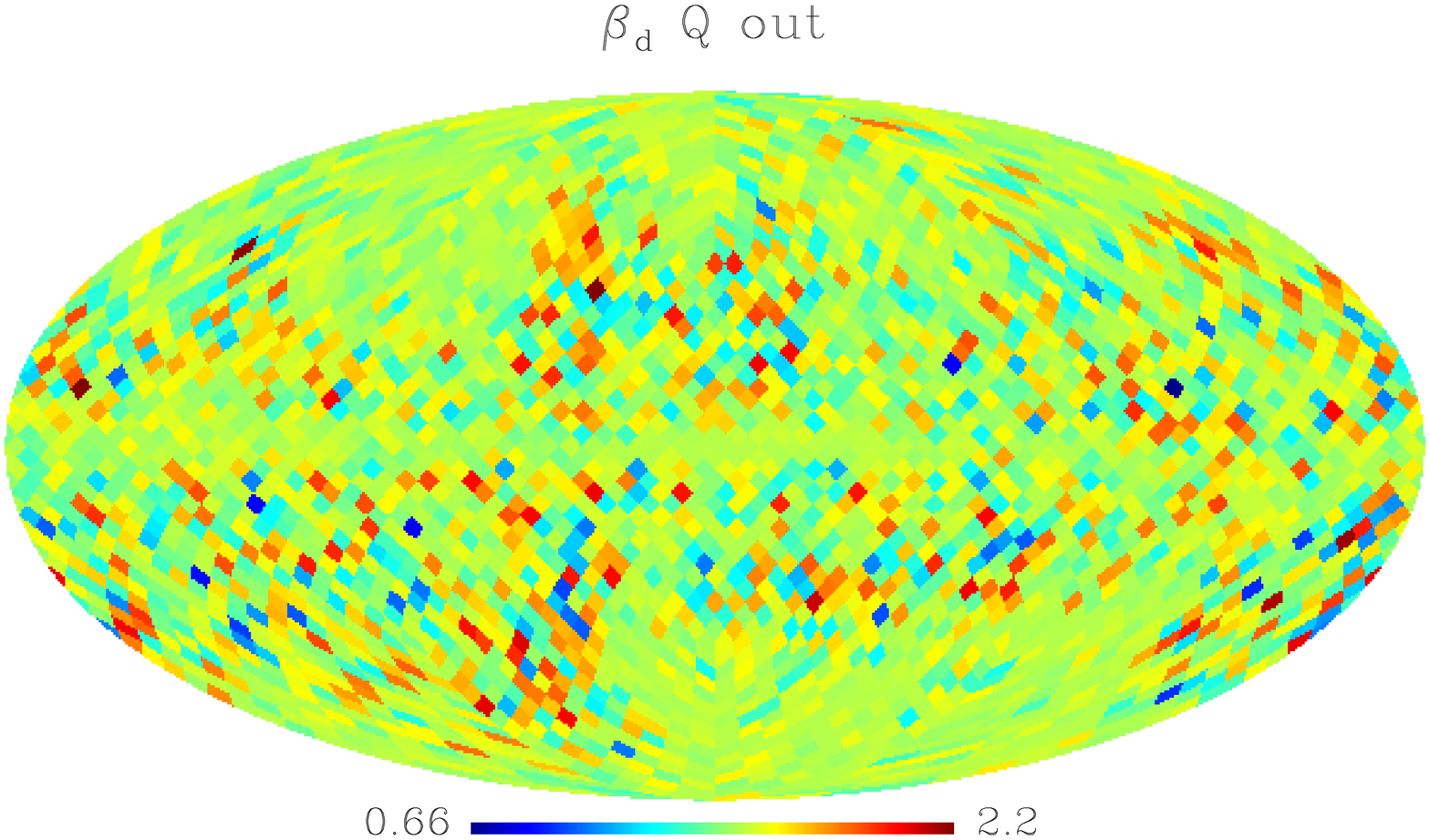} &   
   \includegraphics[width=0.30\textwidth,keepaspectratio,angle=90]{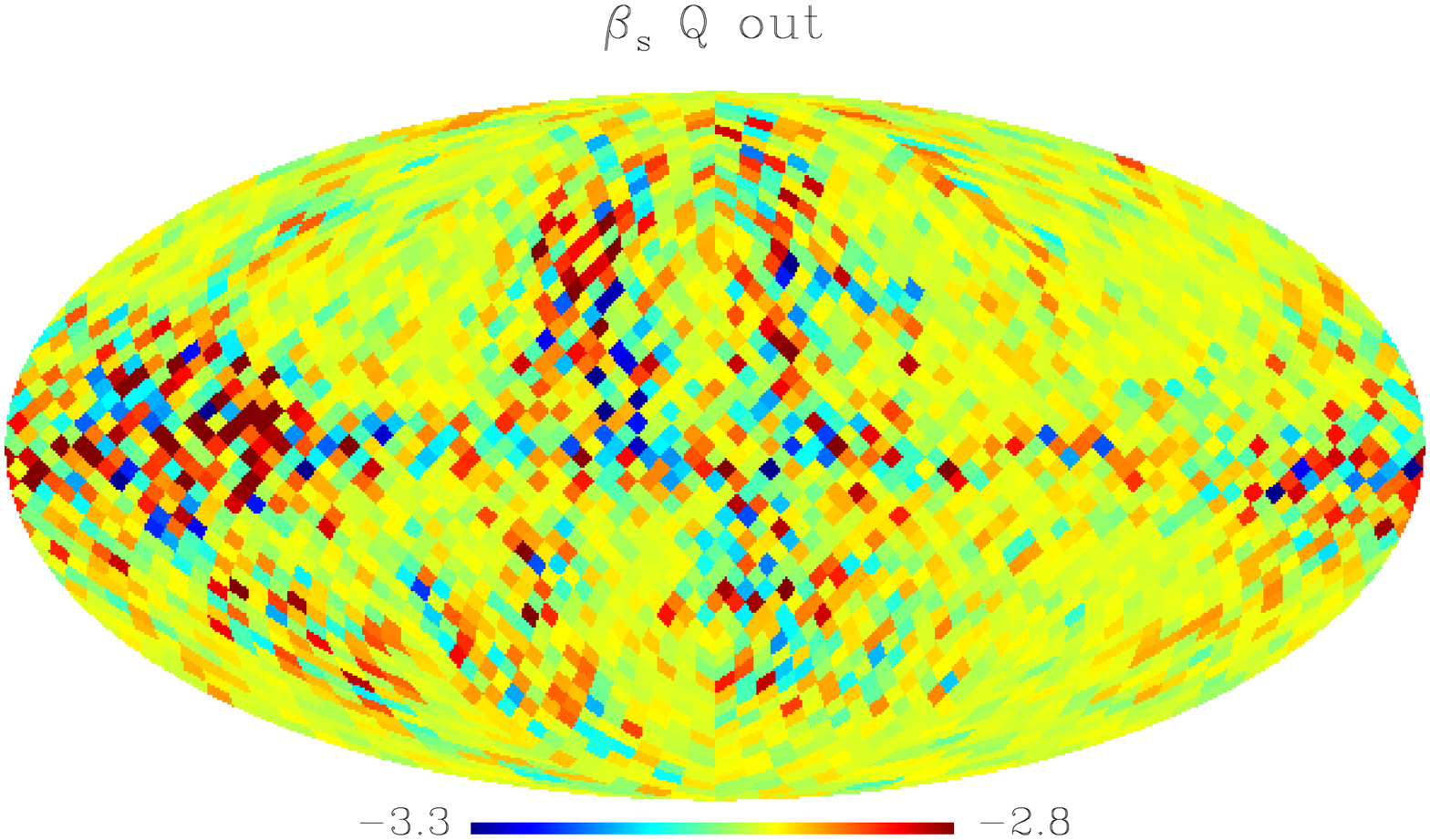} \\   
   \includegraphics[width=0.30\textwidth,keepaspectratio,angle=90]{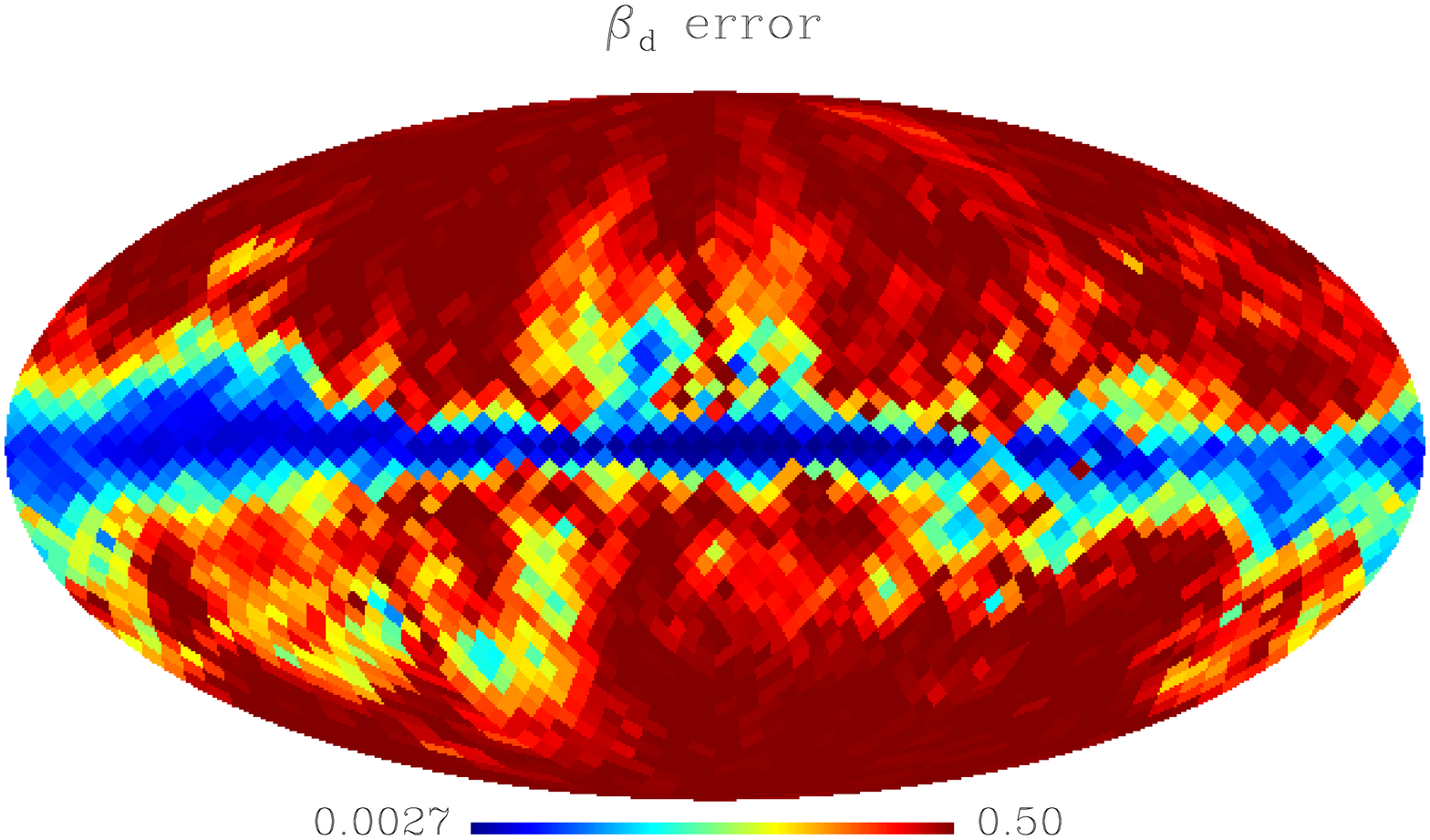}&
   \includegraphics[width=0.30\textwidth,keepaspectratio,angle=90]{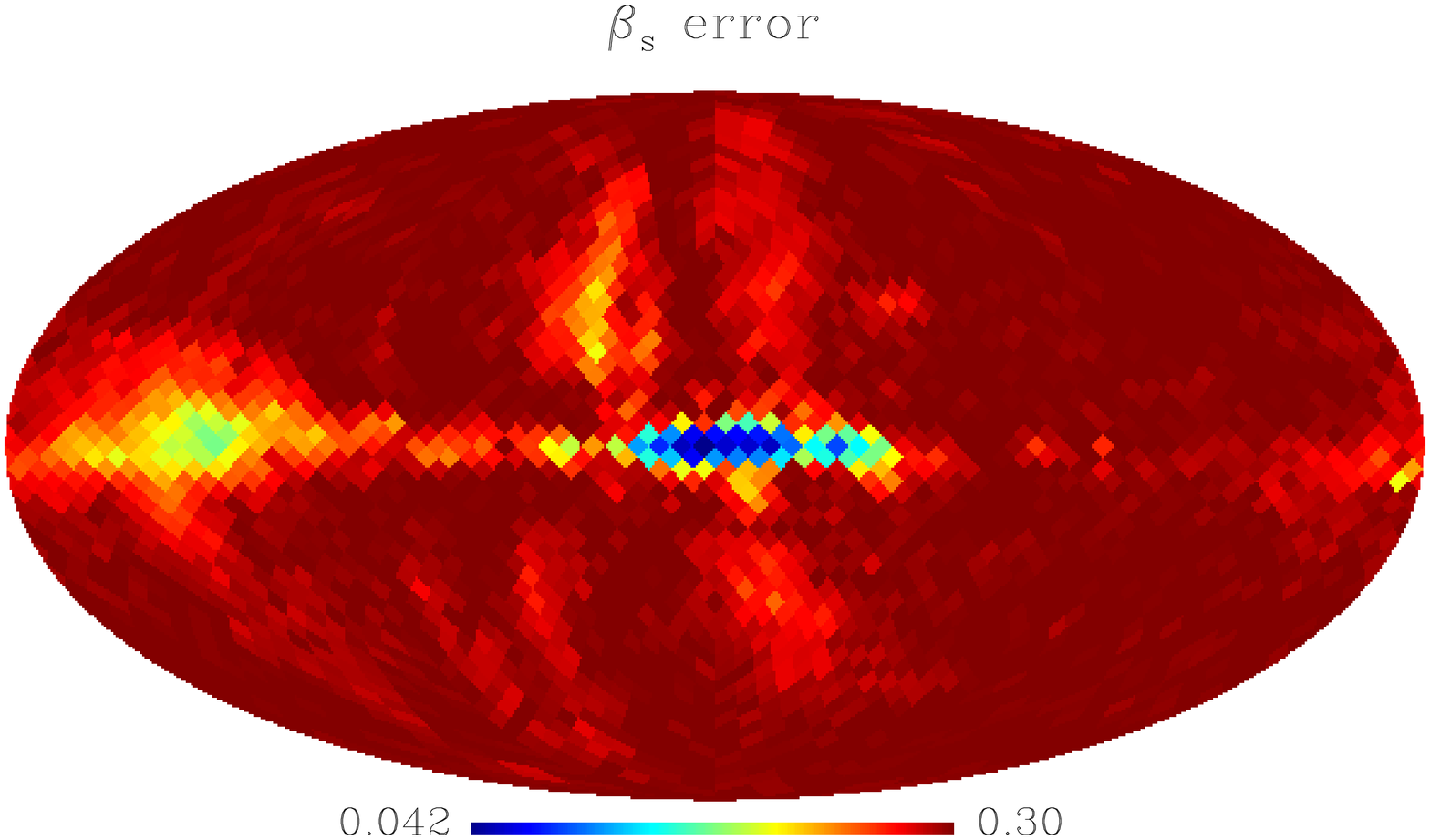}\\
  \includegraphics[width=0.30\textwidth,keepaspectratio,angle=90]{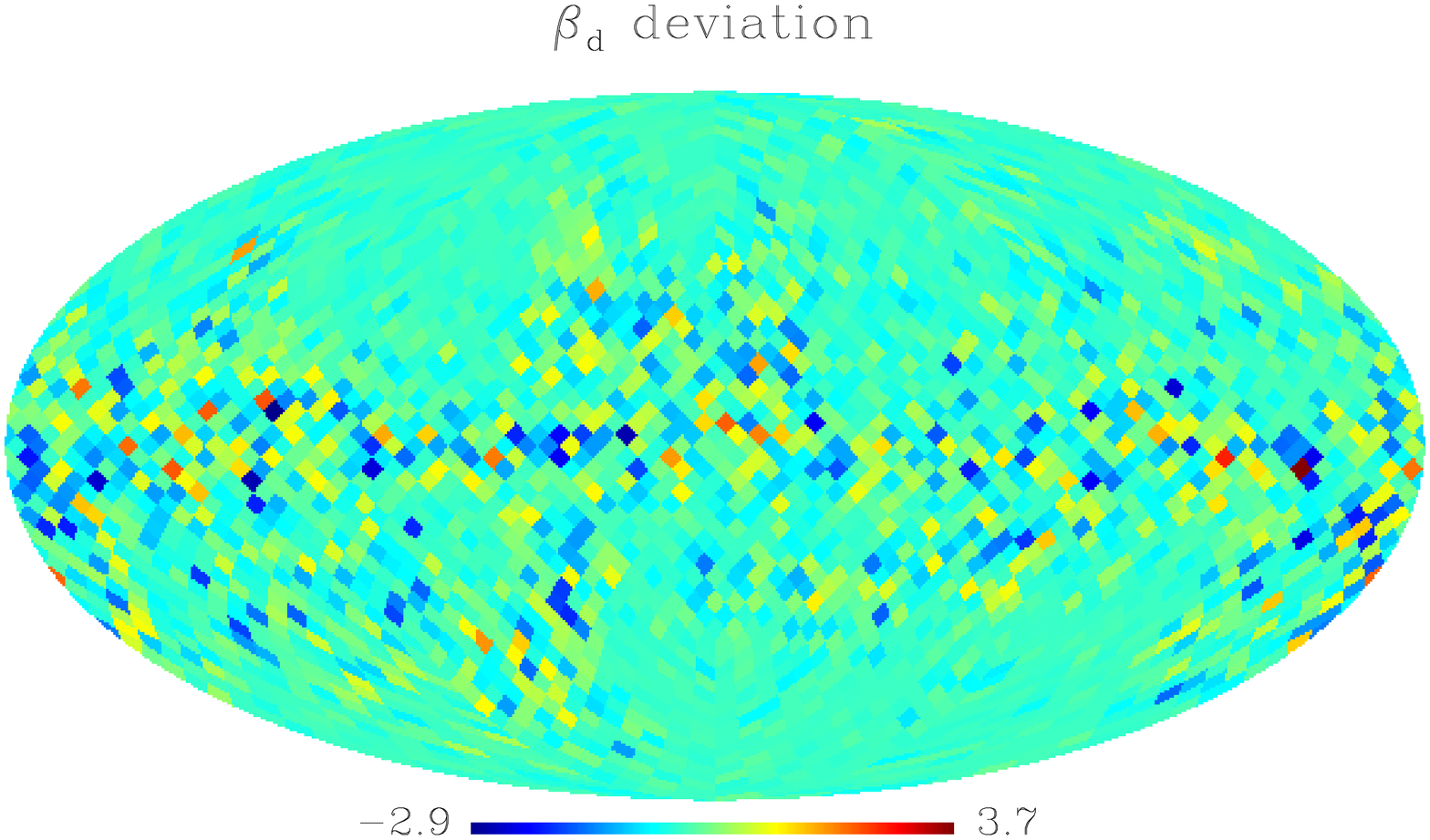}&
  \includegraphics[width=0.30\textwidth,keepaspectratio,angle=90]{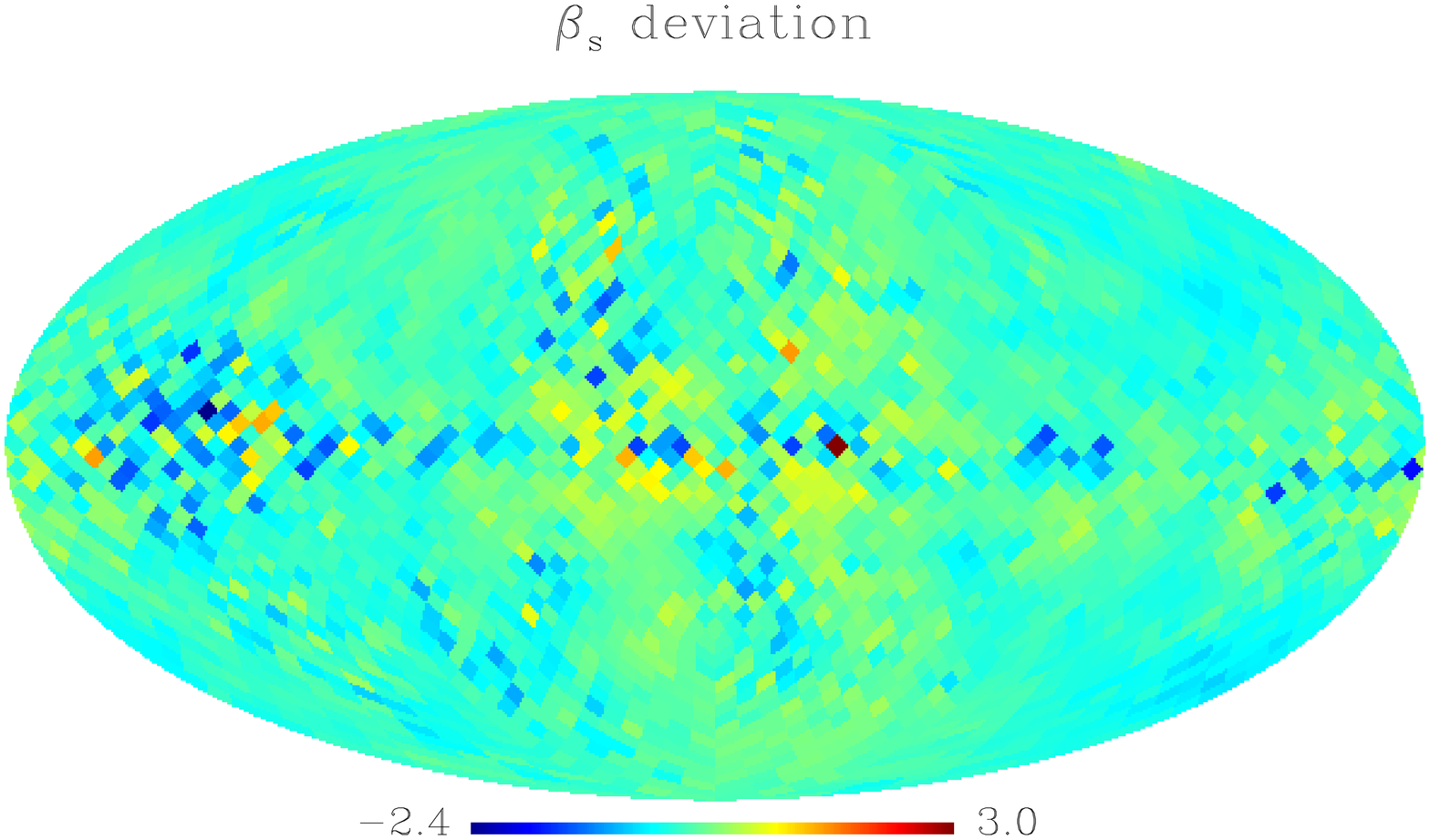}\\
   \end{tabular}
   \caption{Spectral index, $\beta$, input map (top), output map (second row), error map (third row), and deviation map (bottom row) for dust (left) and synchrotron (right).  Note that in areas of low signal-to-noise, the error is driven to the prior value of 0.5 for dust and 0.3 for synchrotron in areas of low signal-to-noise.}
   \label{fig:betad}
\end{figure*}

The results from the spectral index sampling for the thermal dust and synchrotron components are shown in Fig.~\ref{fig:betad}.  We show the input spectral index maps (which is a uniform 1.5 for $\beta_d$) and the marginalized output Q index.  We also plot the error map and difference in standard deviations per pixel.  The bottom panels of Fig.~\ref{fig:betad} show how the effect of our Gaussian priors of $\beta_d = 1.5 \pm 0.5$ and $\beta_s = -3.0 \pm 0.3$ varies over the sky.   In regions of low signal-to-noise, away from the Galactic plane, the prior is stronger than the likelihood and the error is driven to the prior value.  In contrast, the likelihood dominates over the prior in regions of high signal-to-noise.  In the dust error map, we find the error to be as small as 0.0026 in the Galactic plane where the likelihood dominates by a factor of nearly 200.  Thus, for data containing a spatially varying $\beta_d$, we would expect to constrain those variations in the high signal-to-noise regions.  

\begin{figure*}
   \centering
   \includegraphics[width=0.45\textwidth,keepaspectratio,angle=0]{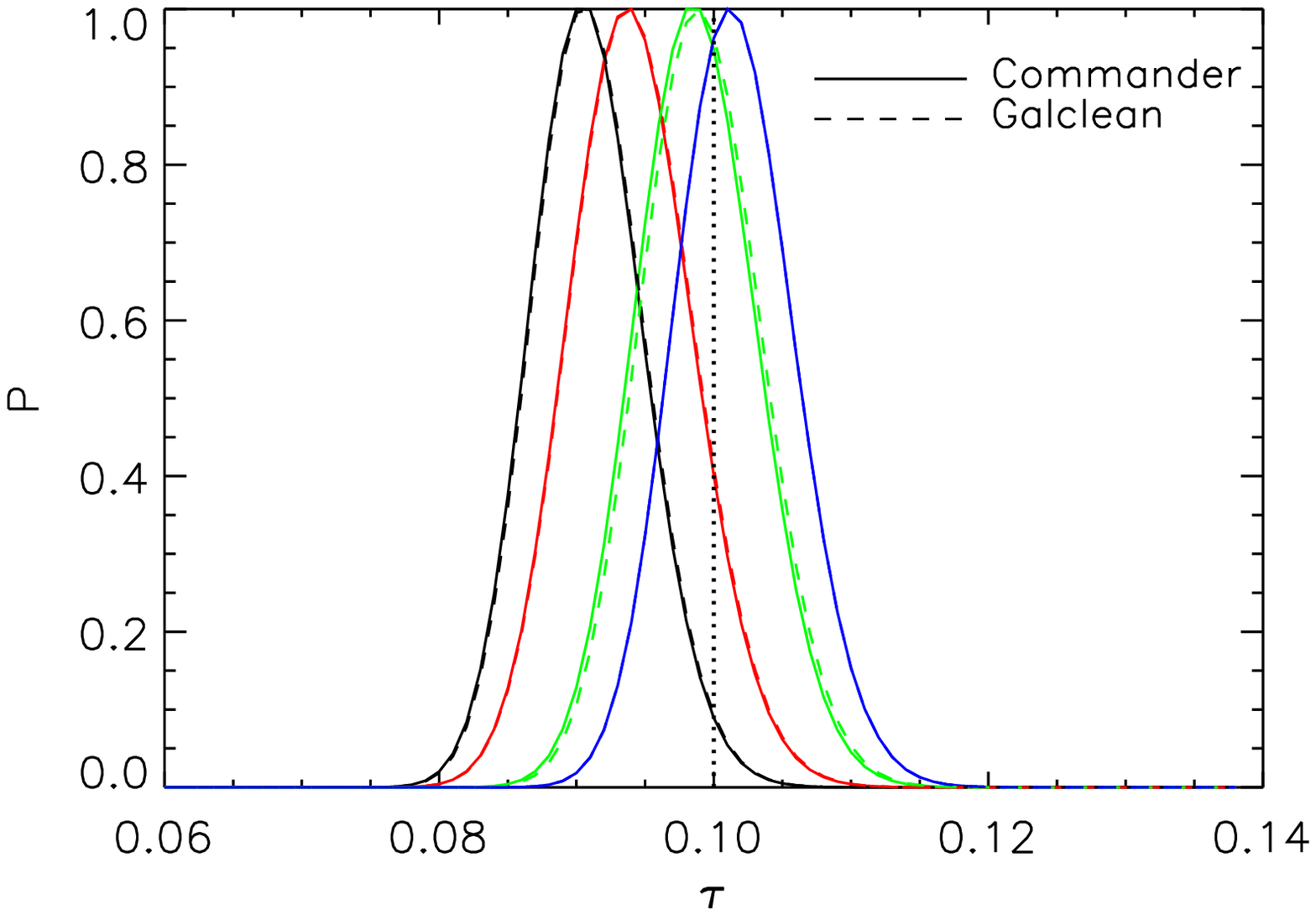} 
    \includegraphics[width=0.45\textwidth,keepaspectratio,angle=0]{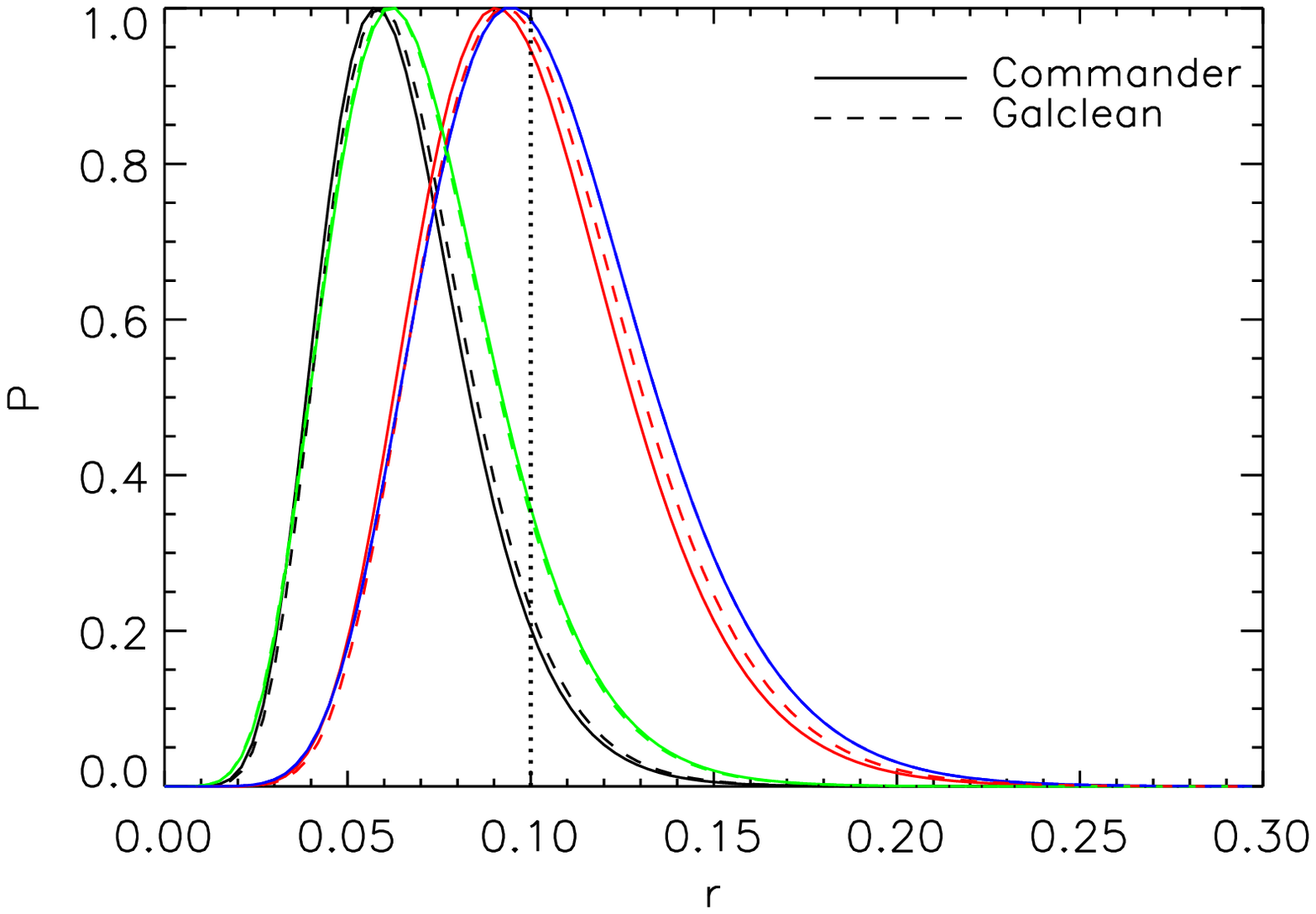} 
   \caption{Likelihood distributions for $\tau$ (left plot) and $r$ (right plot) for four simulations of CMB+foregrounds with $\tau=0.1$ and $r=0.1$.  The four different simulations are represented by the black, blue, green, and red curves.  Results from {\it Commander} are shown with a solid line and results from {\it Galclean} are shown with a dashed line.  Note that the {\it Galclean} curve for the red simulation is completely overlaid by the {\it Commander} curve for the red simulation (left plot), and that the {\it Galclean} curve for the green simulation is completely overlaid by the {\it Commander} curve for the green simulation (right plot).  We find $\sigma(\tau)\approx 0.004$ and $\sigma(r)\approx 0.03$.}
   \label{fig:tau_r_comm_gal}
\end{figure*}

Next, we run our low-$\ell$ pixel likelihood code on the full CMB+foregrounds case, using the resulting CMB maps and covariance matrices from both {\it Commander} and {\it Galclean}.  The $\tau$ likelihoods for four simulations are plotted in Fig.~\ref{fig:tau_r_comm_gal} and we find that three of the four results are consistent with $\tau=0.1$ at $< 2\sigma$ (the fourth is consistent at $\sim2.2\sigma$).  We find constraints on $r$ for the $r=0.1$ CMB+foreground case with maps estimated by {\it Commander} and {\it Galclean}.  These likelihood curves are also shown in Fig.~\ref{fig:tau_r_comm_gal}.  As with the $\tau$ estimates, we find that the $r$ results are consistent with $r=0.1$ at $<2\sigma$ for all of the simulations.  The estimates on both $\tau$ and $r$ for {\it Commander} and {\it Galclean} show that the two codes agree to better than $0.1\sigma$.  We will discuss the estimated errors on $\tau$ and $r$ in \S\ref{sec:parameters}.

\subsubsection{Testing the likelihood code}

To check for potential biases, we have tested the low-$\ell$ pixel likelihood code on foreground-free simulations.  We obtain these simulations by co-adding the 100, 143, and 217 GHz simulated maps, $M_i$, with inverse-noise weightings 
\begin{equation}
M = \sum\limits_i \frac{M_i}{\sigma_i^2}\sigma^2
\end{equation}
where the co-added error, $\sigma^2$ is given by
\begin{equation}
\sigma^2 = \frac{1}{\sum\limits_i 1/\sigma_i^2}
\end{equation}

\begin{figure*}
   \centering
   \includegraphics[width=0.45\textwidth,keepaspectratio,angle=0]{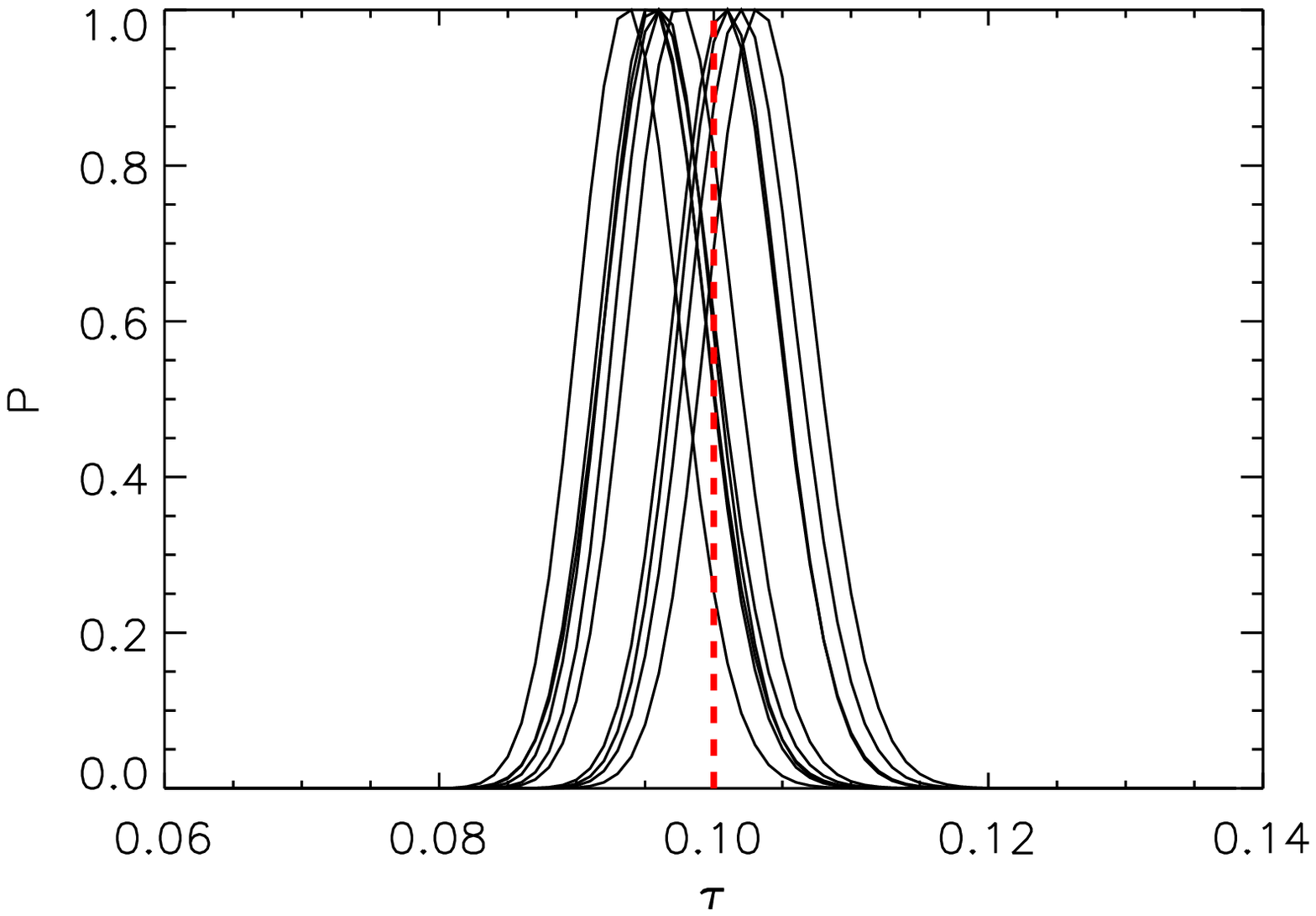} 
   \includegraphics[width=0.45\textwidth,keepaspectratio,angle=0]{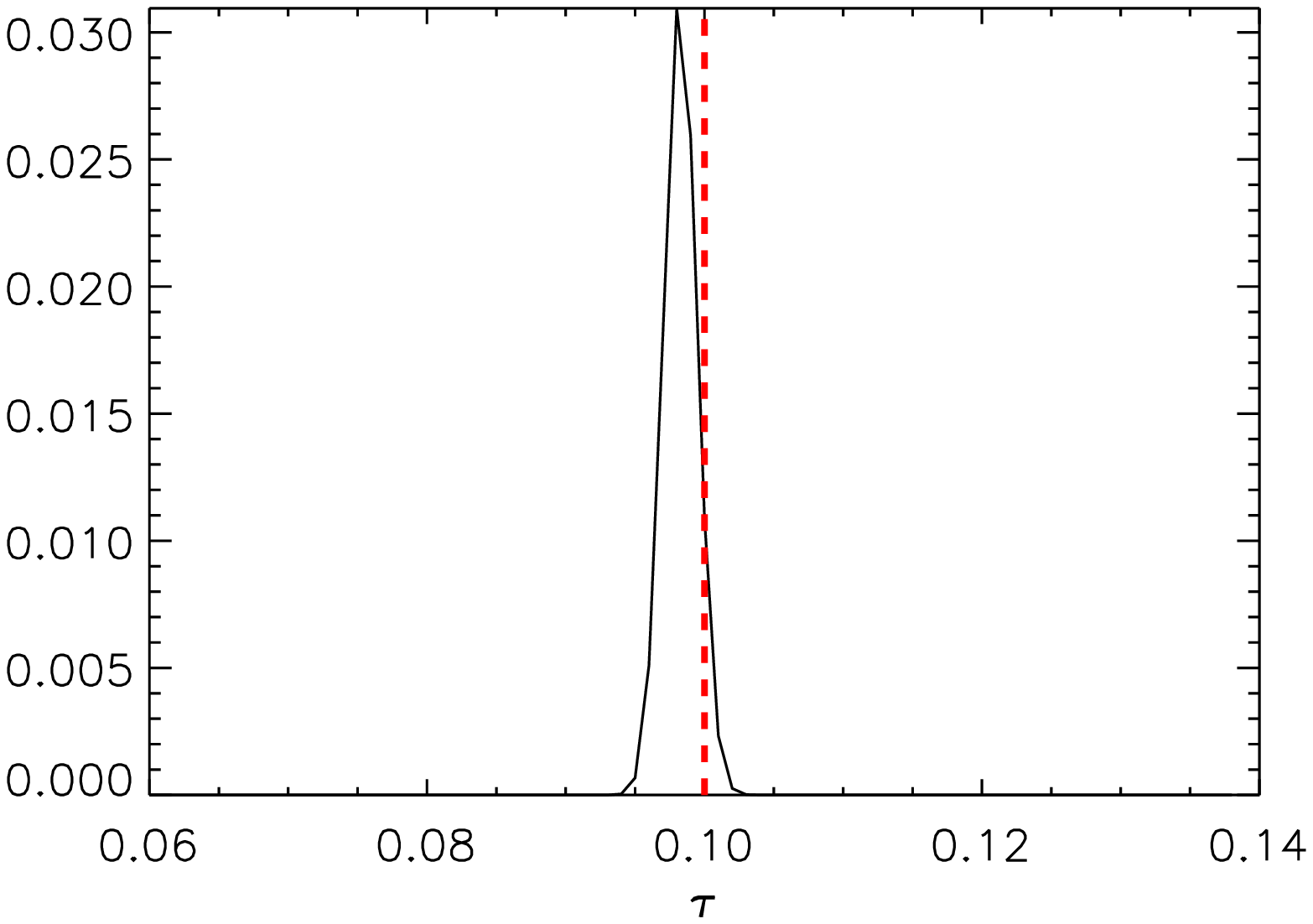}    
   \caption{$\tau=0.1$ foreground-free case for 10 simulations.  The left-hand plot shows the likelihood distributions for each of the 10 simulations, while the plot on the right-hand side is the sum of the log-likelihoods of the 10 distributions.}
   \label{fig:tau_forefree}
\end{figure*}

\begin{figure*}
   \centering
   \includegraphics[width=0.45\textwidth,keepaspectratio,angle=0]{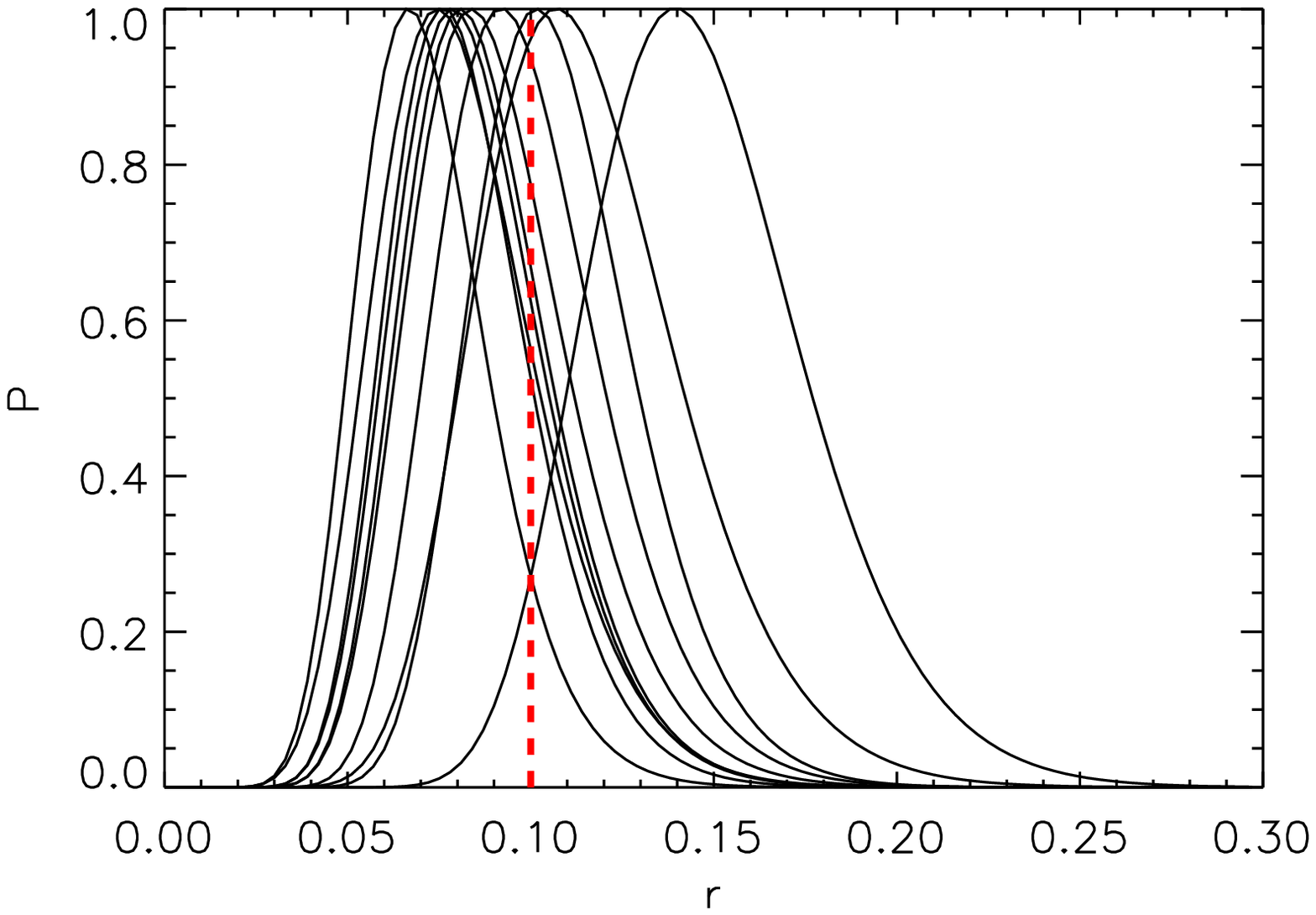} 
   \includegraphics[width=0.45\textwidth,keepaspectratio,angle=0]{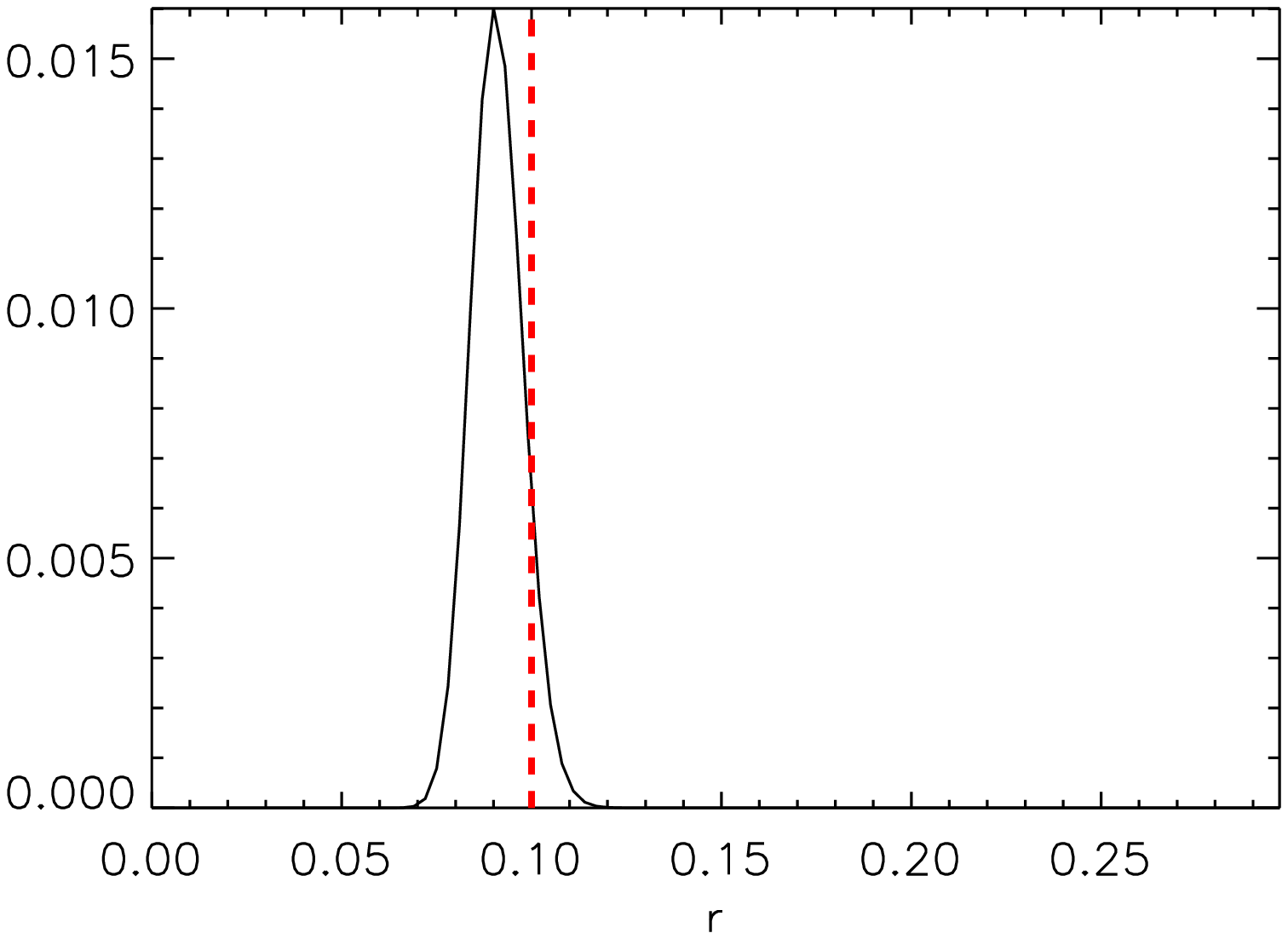}    
   \caption{$r=0.1$ foreground-free case for 10 simulations.  The left-hand plot shows the likelihood distributions for each of the 10 simulations, while the plot on the right-hand side is the sum of the log-likelihoods for the 10 simuations.}
   \label{fig:r_forefree}
\end{figure*}

The likelihood curves for the $\tau=0.1$ foreground-free case are shown in Fig.~\ref{fig:tau_forefree}.  We calculate the likelihoods for 10 simulations and take the sum of the log-likelihoods ($\sum\limits_{i=1}^{N_{sim}}-2 ln L_i$) of the 10 simulations to obtain an average distribution, shown in the right-hand plot of Fig.~\ref{fig:tau_forefree}.  We find that the foreground-free case is consistent with $\tau=0.1$ at $1.4\sigma$.  
We repeat the foreground-free analysis for $r=0.1$ and 10 simulations, as shown in Fig.~\ref{fig:r_forefree}, and take the sum of the log-likelihoods to obtain an average likelihood.  The average likelihood is consistent with $r=0.1$ at 1.3$\sigma$.  A thorough test for biases should combine results from thousands of simulations, however, for our small set of simulations, the deviations that we see from $\tau=0.1$ and $r=0.1$ are not statistically significant.     

\subsection{Effect of foregrounds on parameter estimation}
\label{sec:parameters}
\begin{figure*}
  \centering
   \includegraphics[width=0.45\textwidth]{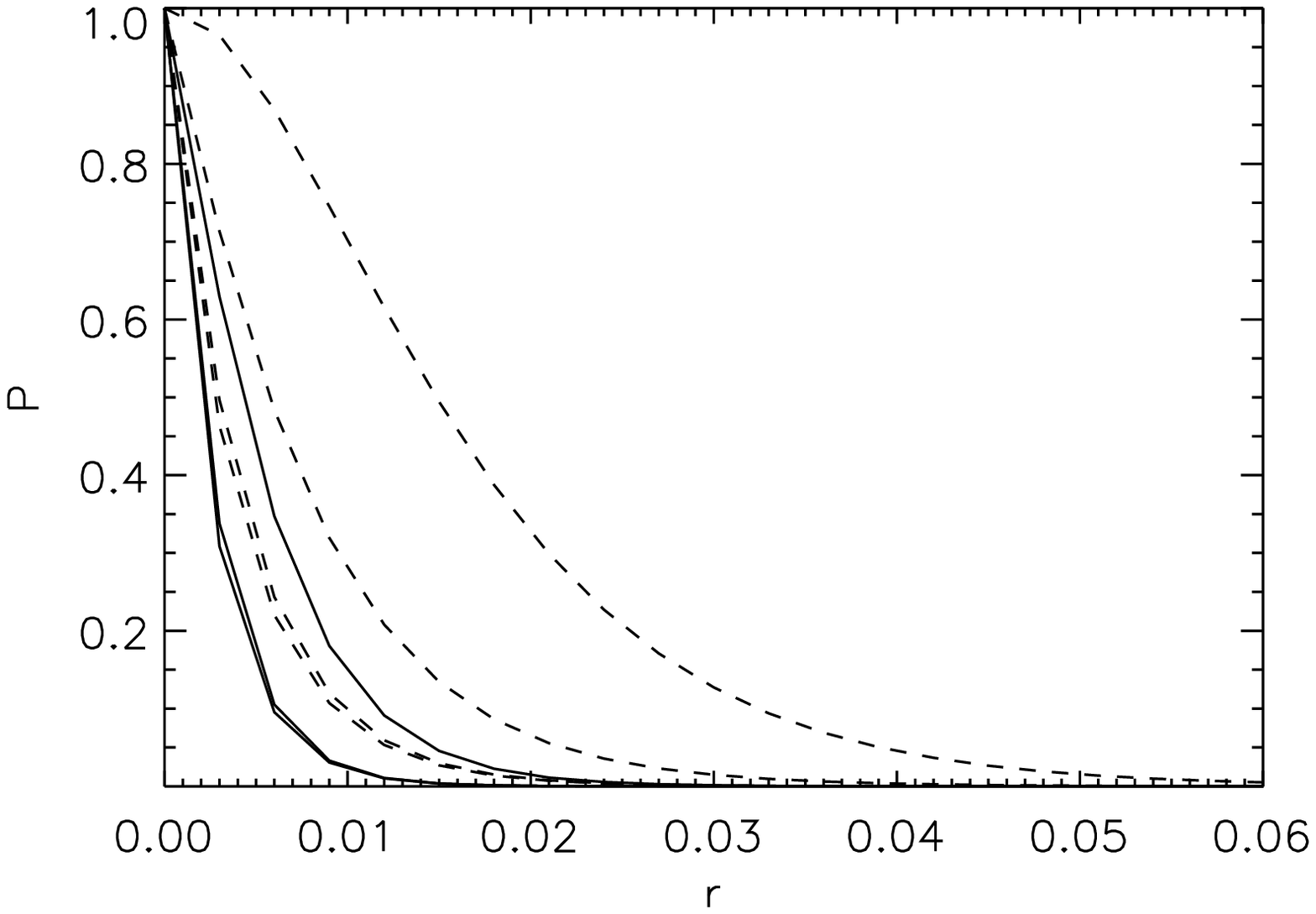}     
   \caption{The solid curves show the likelihood distributions for $r$ for the $r=0.0$ foreground-free simulations while the dashed curves show the likelihood distributions from maps estimated by {\it Commander}.  The widening of the likelihood curves from solid to dashed line is commensurate with the increased uncertainty in our estimate of $r$ due to the presence of foregrounds.}
   \label{fig:r0}
\end{figure*}

The likelihoods for four simulations of the $r=0.0$ case are shown in Fig.~\ref{fig:r0} for the two cases, foreground-free and CMB+foregrounds.  The foreground-free simulations all show likelihoods which are sharply peaked at $r=0.0$ while the CMB+foreground simulations show widening of the likelihoods commensurate with the increased uncertainty due to the presence of the foregrounds.  This effect is summarized in Table \ref{tab:95cutoff} which gives the average upper 95\% cut-off limits on estimates of $r$ for $r=0.0$ and the average estimates on $\sigma(r=0.1)$ and $\sigma(\tau=0.1)$ with and without foregrounds.  We also apply the standard {\it WMAP} P06 mask \citep{Page07},  which masks about 26\% of the sky, and calculate the likelihood distributions for the masked case. 

For our chosen simulations and modeling, we find minimal error inflation in $\sigma_{\tau}$ and $\sigma_r$.  $\sigma_{\tau}$ remains nearly constant at $\sim 0.005$ in the absence or presence of foregrounds.  $\sigma_r$ increases from $\sim 0.02$ to $\sim 0.03$ with the addition of foregrounds.  Our limits on $r=0$ show that it is more sensitive to the presence of foregrounds than an estimation of an $r=0.1$ signal.  We find $\sigma_r/r = 0.32$ for $r=0.1$ and $\sigma_{\tau}/\tau = 0.05$ for $\tau=0.1$ using Commander.  Using a Fisher matrix approach, \cite{Betoule09} find values of $\sigma_r/r$ similar to ours: $\sigma_r/r = 0.34$ with foregrounds and $\sigma_r/r = 0.25$ with noise only.  In another Fisher matrix forecast for {\it Planck}, \cite{Baumann} finds $\sigma_r = 0.011$ for $r=0.01$ without foregrounds.

\begin{table*}
\centering
\begin{tabular}{| c | c | c | c |}
\hline
 & Foreground-Free & With Foregrounds (Unmasked) & With Foregrounds (Masked)\\
\hline
$r=0.0$ & $< 0.008$ & $< 0.017$ & $< 0.023$\\
\hline
$\sigma(r=0.1)$ & 0.023 & 0.027 & 0.032\\
\hline
$\sigma(\tau=0.1)$ & 0.004 & 0.004 & 0.005 \\
\hline
\end{tabular}
\caption{Average upper 95\% cut-off limits on estimates of $r=0$, and average estimates on $\sigma(r=0.1)$ and $\sigma(\tau=0.1)$.  We find a foreground-free error on $r$ that matches the size of errors found in analogous Fisher matrix forecasts for {\it Planck} \citep{Betoule09,Baumann}.  The effect of foregrounds is seen to inflate the error bar in the case of $r$ but not $\tau$.  The error on $r$ for the $r=0.1$ model is amplified by a factor of $\sim 1.4$, and the 95\% cut-off limit on $r$ for the $r=0.0$ model is amplified by a factor of $\sim 3$, when foregrounds are included.}
\label{tab:95cutoff}
\end{table*}

\begin{figure*}
   \centering
   \includegraphics[width=0.85\textwidth,keepaspectratio,angle=0]{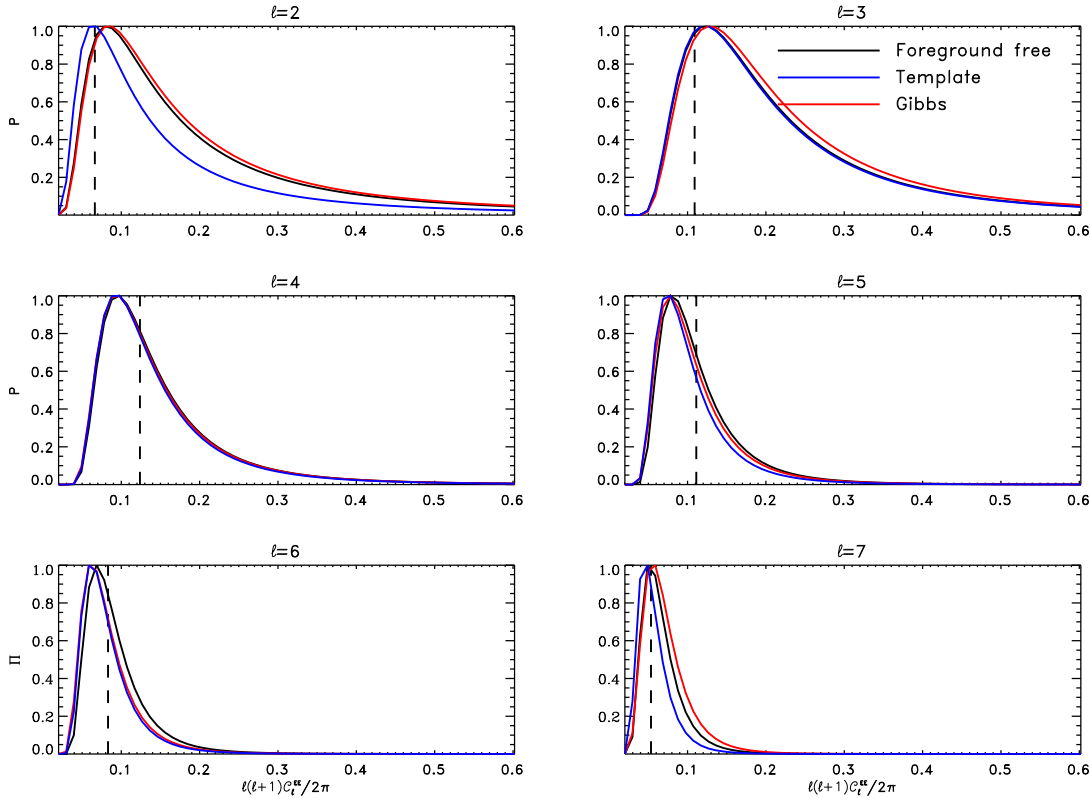} 
   \caption{Conditional likelihood slices for CMB $C_{\ell}^{EE}$ for $\ell = 2-7$, estimated from the polarization CMB maps cleaned using Gibbs sampling (red), compared to the template cleaned maps (blue) and to the foreground-free case (black).  The dashed vertical line represents the true value of $C_{\ell}^{EE}$.}
   \label{fig:clee}
\end{figure*}

\begin{figure*}
   \centering
   \includegraphics[width=0.85\textwidth,keepaspectratio,angle=0]{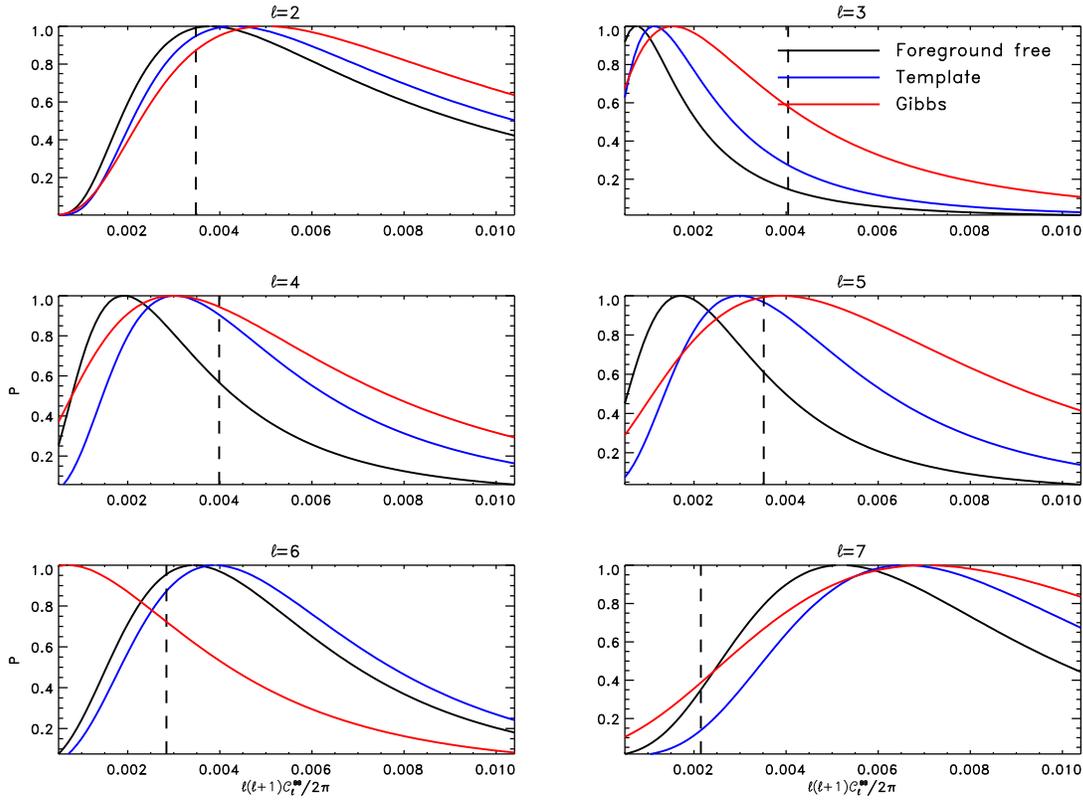} 
   \caption{Conditional likelihood slices for CMB $C_{\ell}^{BB}$ for the $r=0.1$ simulation, estimated from the polarization CMB maps cleaned using Gibbs sampling (red), compared to the template cleaned maps (blue) and to the foreground-free case (black).  The dashed vertical line represents the true value of $C_{\ell}^{BB}$.}
   \label{fig:clbb}
\end{figure*}

\begin{figure*}
   \centering
   \includegraphics[width=0.85\textwidth,keepaspectratio,angle=0]{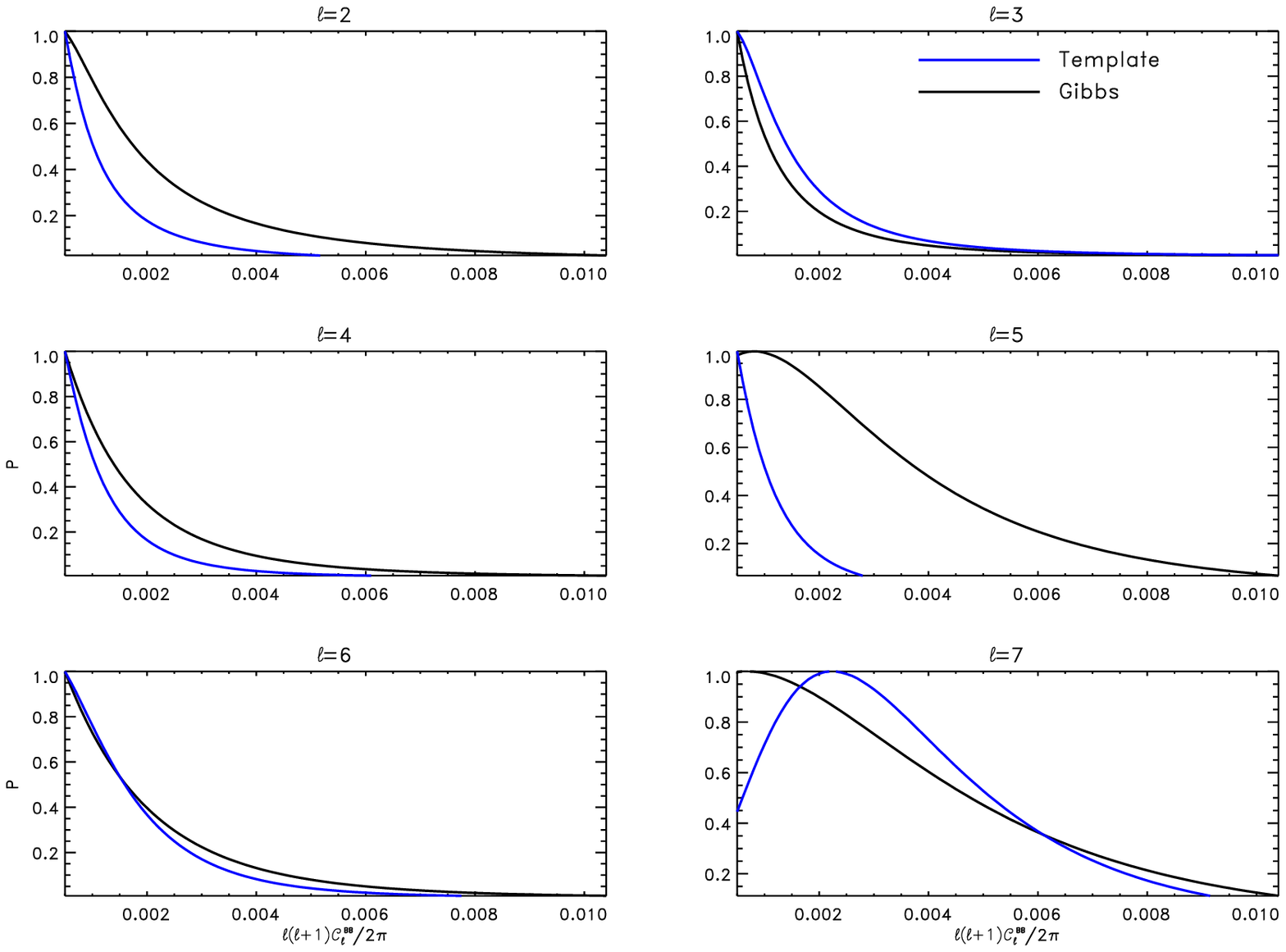} 
   \caption{Conditional likelihood slices for CMB $C_{\ell}^{BB}$ for the $r=0.0$ simulation, estimated from the polarization CMB maps cleaned using Gibbs sampling (black), and compared to the template cleaned maps (blue).  We expect that Gibbs sampling is needed over template cleaning to provide appropriate $r$ limits and find that Gibbs sampling does tend to inflate the distributions, particularly for $\ell=2, 4$ and 5, in this case.}
   \label{fig:clbb_r0}
\end{figure*}

Our pixel likelihood code can be used not only to constrain parameters, but also to find the power at each multipole in the polarized power spectra.  At each multipole, we compute the conditional likelihood as a function of $C_{\ell}^{EE}$ and $C_{\ell}^{BB}$ for $\ell = 2-7$ with all other multipoles held fixed at the fiducial $\Lambda$CDM values, using the method described in \cite{Nolta08}.  For example, the conditional likelihood of $C_{4}^{EE}$ is $f(x) \propto L(d|...,C_3^{EE},C_4^{EE}=x,C_5^{EE}...)$.  We compare the power at each multipole in the Gibbs CMB map to the template-cleaned case (described in \S\ref{sec:template_cleaning}) and to the foreground-free case, shown in Fig.~\ref{fig:clee} and Fig.~\ref{fig:clbb} for the $r=0.1$ simulation.  For $C_{\ell}^{EE}$, the results are consistent between the three cases.  For the $C_{\ell}^{BB}$ spectra, we find that the template-cleaned conditional slices agree with the foreground-free curves as well, or better than, the Gibbs slices, indicating that the more economical template-cleaning method is an effective (and fast) option for foreground removal in the case of low spectral index variation established in our data model.  However, we argue that Gibbs sampling should be used instead of, or in addtition to, template cleaning, in order to benefit from the Gibbs feature that the inclusion of foreground uncertainties in the covariance matrix can be propagated to the limits on $r$.  This effect appears as the inflation in the Gibbs $C_{\ell}^{BB}$ distributions over the template distributions for the $r=0.0$ simulation, shown, in particular for $\ell=2,4$ and 5 in Fig.~\ref{fig:clbb_r0}. 


\begin{figure*}
   \centering
   \includegraphics[width=0.85\textwidth,keepaspectratio,angle=0]{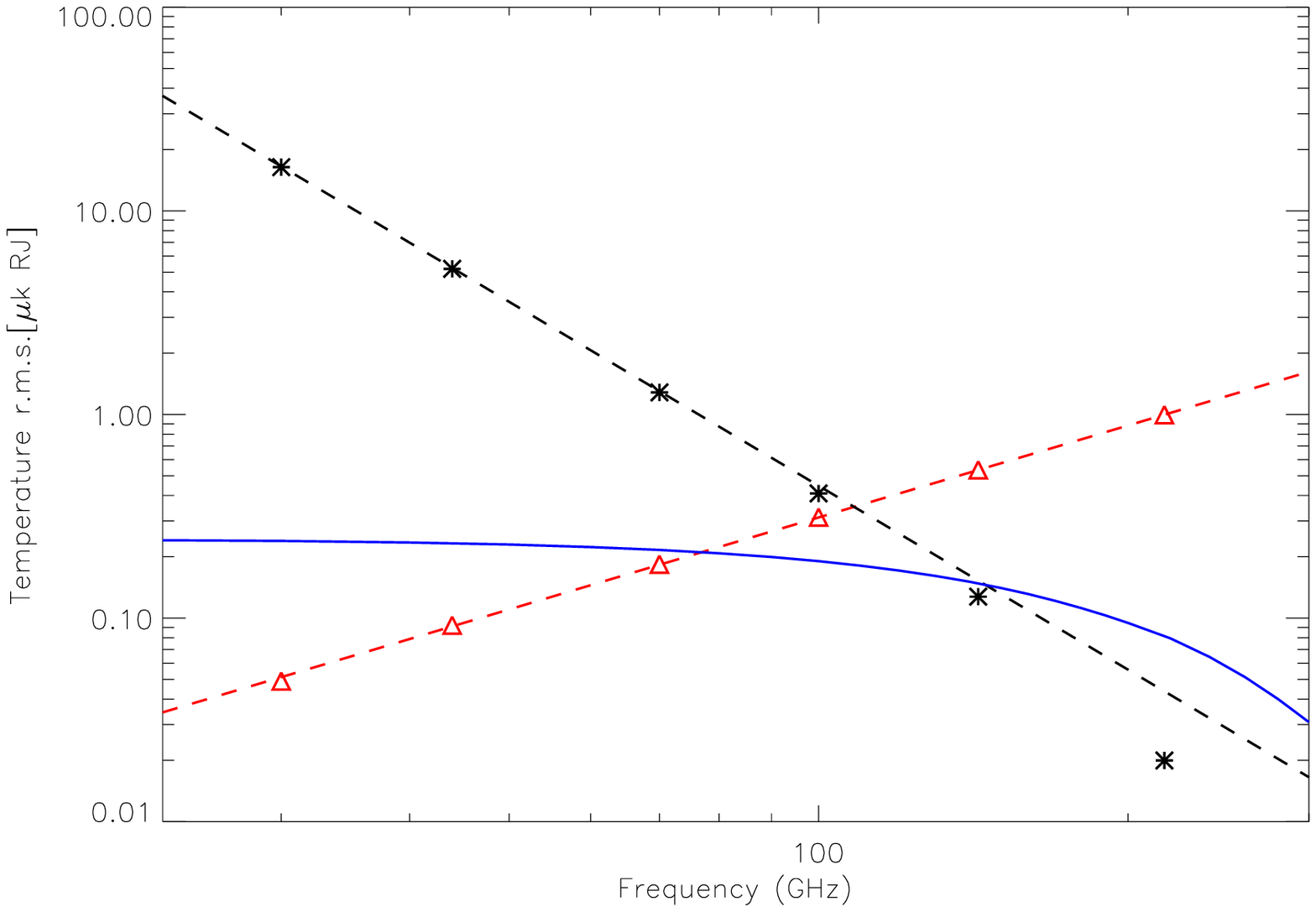} 
   \caption{R.m.s. fluctuations spectrum (antenna temperature units) of the polarized dust and synchrotron components.  The symbols are the best-fit template coefficients for dust (red triangles) and synchrotron (black stars).  The dashed curves represent dust for $\beta_d = 1.5$ (red dashed line) and synchrotron for $\beta_s = -3$ (black dashed line).  The CMB fluctuations (blue line) are normalized to the r.m.s. value, 0.24 $\mu K$ (thermodynamic), of the simulated Q and U components of the CMB map.}
   \label{fig:template_coeff}
\end{figure*}

In Fig.~\ref{fig:template_coeff}, we plot the results from the {\it Commander} template fitting.  The data points are the best-fit template coefficients for the dust and synchrotron emission at 30, 44, 70, 100, 143, and 217 GHz .  The dashed curves show the emission, in antenna units, of the thermal dust for $\beta_d = 1.5$ and of the synchrotron for $\beta_s = -3$.  The curves are normalized to the r.m.s. values of the 23 GHz and 353 GHz template maps for synchrotron and dust, respectively.

\subsection{Comparison between $C_{\ell}$ estimates from Gibbs sampling and pixel likelihood code}
\label{sec:cl_comparison}

In \S\ref{sec:likelihood} we discussed a potential issue with our standard pixel likelihood code in the case that the marginalized distributions $p(\mathbf{A}|\mathbf{d})$ contain non-Gaussianities.  We investigate our CMB marginal posteriors and do find a small level of non-Gaussianity particularly in regions where the foreground signal is large.  We proposed several options for addressing this issue in \S\ref{sec:likelihood}, and in this section we show a comparison between our standard pixel-likelihood and Gibbs sampled $C_{\ell}$ estimates in order to assess the level of non-Gaussianity seen in the CMB marginal posteriors. 

\begin{figure*}
\centering
  \includegraphics[width=0.45\textwidth,keepaspectratio,angle=0]{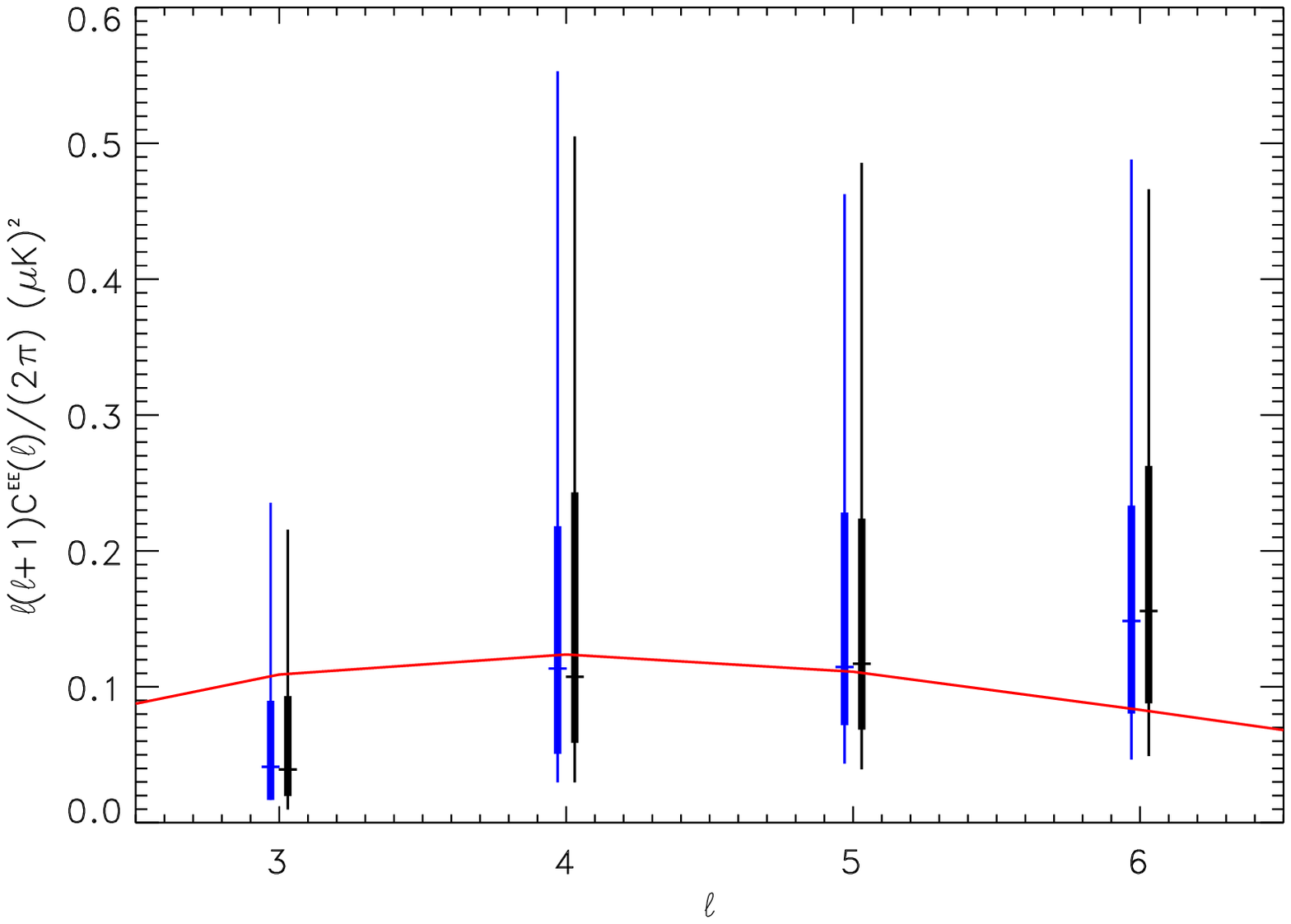} 
  \includegraphics[width=0.45\textwidth,keepaspectratio,angle=0]{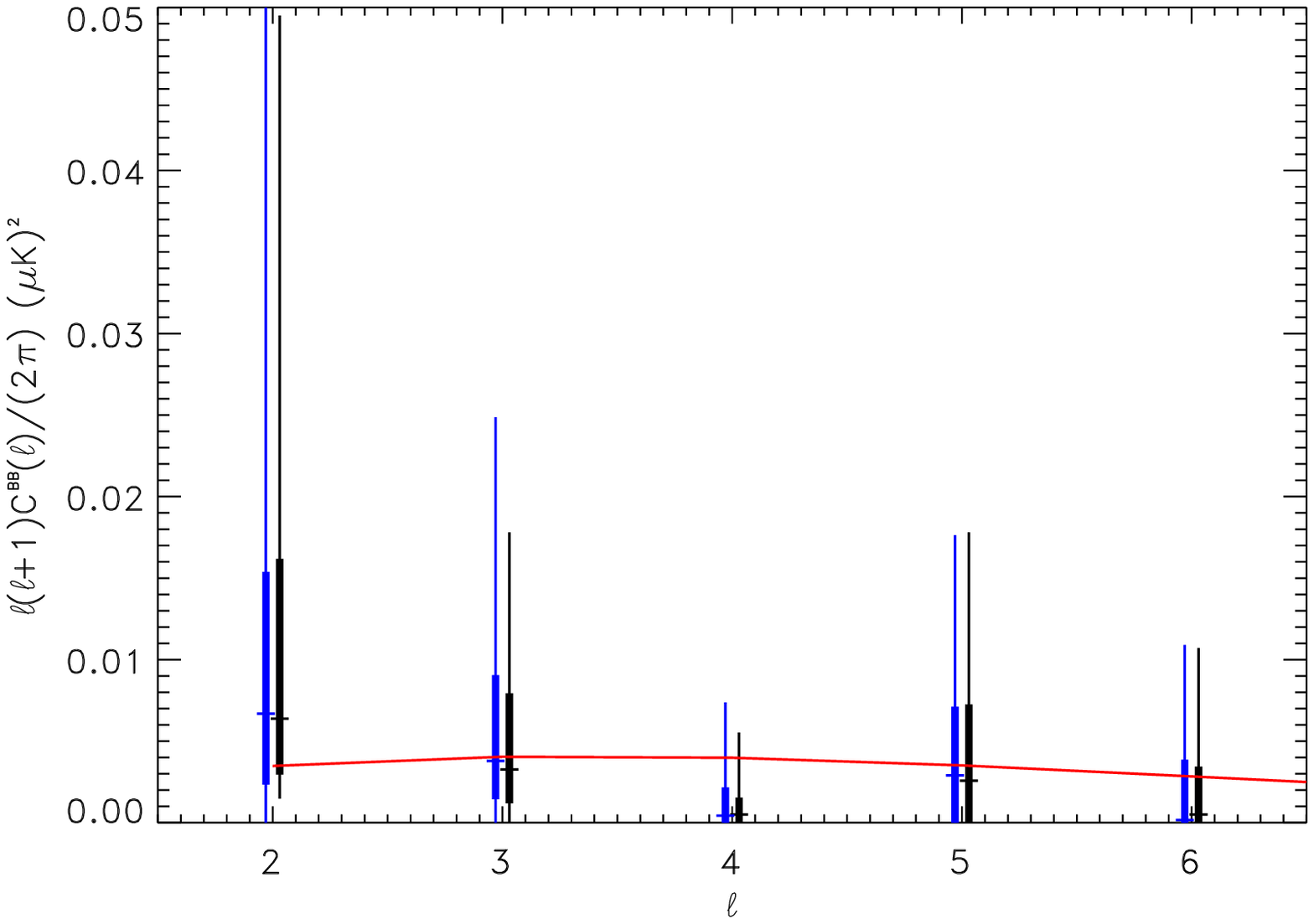} 
\caption{Estimates of $C_{\ell}^{EE}$ (left plot) and $C_{\ell}^{BB}$ (right plot) computed with two different techniques.  At each $\ell$ value, we plot the maximum likelihood value (tic mark), the region where the likelihood is greater than 50\% of the peak value (thick line) and the region where the likelihood is greater than 95\% of the peak value (thin line).  The black lines (right side of each pair) are estimated with a pixel-based likelihood code with $N_{side}=8$.  The blue lines (left side of each pair) are estimated by Gibbs sampling the maps and $C_{\ell}$s simultaneously at $N_{side}=16$.  Note that we do not show results for $C_2^{EE}$ as the comparison between conditional and marginal distributions for $\ell=2$ is not feasible using this method.}
\label{fig:clslices}
\end{figure*}

We run the pixel-likelihood code to compute the conditional likelihood as a function of $C_{\ell}^{EE}$ and $C_{\ell}^{BB}$ for $\ell=2-6$ with all other multipoles held fixed at the fiducial $\Lambda$CDM values.  We additionally marginalize over $C_{\ell}^{TT}$ and $C_{\ell}^{TE}$ when computing the $C_{\ell}^{EE}$ likelihood in order to account for correlations between the TT, TE, and EE components.  We neglect correlations between $\ell$ values.  We run the Gibbs sampler ({\it Commander}) in the mode in which the CMB power spectra is sampled simultaneously with the foreground components, as described in \S\ref{sec:likelihood}.  In Fig.~\ref{fig:clslices} we show slices through the $C_{\ell}$ distribution obtained from the Gibbs estimator compared to the pixel likelihood.  We find the estimates from the two methods to be equivalent up to small differences.  The small discrepancies between the Gibbs and pixel likelihood estimates are due to the pixel likelihood code using $N_{side} = 8$, compared to the higher resolution $N_{side}=16$ used for the Gibbs code.  Another source of differences may be from $\ell-\ell'$ correlations present in the Gibbs samples but not in the pixel likelihood which estimates slices of $C_{\ell}$ for all other multipoles fixed.
These results indicate that it is reasonable to approximate the foreground-marginalized CMB pixel amplitudes as Gaussian in the pixel-based likelihood.

\section{Conclusions}

We have investigated two independent Gibbs sampling codes for polarized CMB foreground separation in the case of diagonal noise, and power law dust and synchrotron models.  We have constructed the large-scale posterior CMB and foreground amplitude maps as well as the dust and synchrotron spectral index maps using the {\it Planck} sky model and noise levels, and without masking the Galactic plane.  We explored constraints on $\tau$ and $r$ for our {\it Planck} model and found that our Bayesian algorithms produced results consistent with $\tau=0.1$ and $r=0.1$ at 1- and 2$\sigma$.  We find $\sigma(\tau=0.1)\approx0.004$ and $\sigma(r=0.1)\approx0.03$.  We find a 95\% cut-off limit on an $r=0$ detection at $r<0.02$.  While our specific predictions on $\sigma_{\tau}$ and $\sigma_r$ are limited by our simplified noise and data models, we have shown that the Gibbs-estimated CMB maps and errors capture the additional uncertainty due to the presence and removal of foregrounds, which is then translated into an error inflation on $\tau$ and $r$.

There are many interesting extensions to this work that can be further explored using the tools described in this paper.  More realistic data models can be considered, including two-component dust and a synchrotron curvature model.  Other modelling effects can be considered, such as polarized free-free emission and polarized spinning dust.  Mismatches between data and model should be investigated in terms of the amplification factor or biases that they impart to estimates on $\tau$ and $r$.  The noise model can be extended to include the full Q/U noise covariances and also $1/f$ noise.  The tools adapted and developed in this work can also be used to estimate parameters for small-scale versus large-scale experiments (e.~g.~COrE)

We acknowledge the use of the Planck Sky Model, developed by the Component Separation Working
Group (WG2) of the {\it Planck} Collaboration.  We thank the {\it WMAP} team for the use of their polarization code.  This work was performed using the Darwin Supercomputer of the University of Cambridge High Performance Computing Service (http://www.hpc.cam.ac.uk/), provided by Dell Inc. using Strategic Research Infrastructure Funding from the Higher Education Funding Council for England.  CD acknowledges an STFC Advanced Fellowship and ERC grant under the FP7.


\begin{thebibliography}{}

\bibitem[Baumann et al.(2009)]{Baumann} Baumann D. et al. 2009, AIP Conf.~Proc.~1141:10-120
\bibitem[Bersanelli et al.(2010)]{Bersanelli10} Bersanelli, M. et al. 2010, A\&A, available on line, arXiv:1001.3321
\bibitem[Betoule et al.(2009)]{Betoule09} Betoule, M. et al. 2009, A\&A, 503, 691
\bibitem[Delabrouille \& Cardoso(2009)]{DC09} Delabrouille, J. \& Cardoso, J.-F. 2009, LNP, 665, 159
\bibitem[Dunkley et al.(2005)]{Dunkley05} Dunkley, J. 2005, MNRAS, 356, 925
\bibitem[Dunkley et al.(2009a)]{Dunkley-WMAP} Dunkley, J. et al. 2009, ApJ, 701, 1804
\bibitem[Dunkley et al.(2009b)]{Dunkley-cmbpol} Dunkley, J. et al. 2009, AIP Conference Proceedings, 1141, 222
\bibitem[Efstathiou et al.(2009)]{EGP09} Efstathiou, G., Gratton, S., \& Paci, F. 2009,
        MNRAS, 397, 1355
\bibitem[Eriksen et al.(2004)]{E04} Eriksen, H.K. et al. 2004,
        ApJS, 155, 227
\bibitem[Eriksen et al.(2006)]{E06} Eriksen, H.K. et al. 2006, New Astron.~Rev.~ 50, 861
\bibitem[Eriksen et al.(2008)]{E08} Eriksen, H.K. et al. 2008,
        ApJ, 676, 10
\bibitem[Finkbeiner et al.(1999)]{Fink99} Finkbeiner, D.~P., Davis, M. \& Schlegel, D.~J. 1999, ApJ, 524, 876
\bibitem[Fraisse et al.(2008)]{Fraisse08} Fraisse, A.~.A., et al. 2008, arXiv:0811.3920v1
\bibitem[Gold et al.(2009)]{Gold09} Gold, B. et al. 2009, ApJS, 180, 265
\bibitem[Haslam et al.(1982)]{Haslam} Haslam et al. 1982, A\&AS, 47, 1
\bibitem[G\'orski et al.(2005)]{Healpix} G\'orski, K. et al. 2005, "The HEALPix Primer" (Version 2.15), available at {\tt http://healpix.jpl.nasa.gov}
\bibitem[Jewell et al.(2004)]{J04} Jewell, J., Levin S., \& Anderson, C.H. 2004,
        ApJ, 609, 1
\bibitem[Knox et al.(2001)]{Knox01} Knox, L., Christensen, N., \& Skordis, C. 2001, ApJ, 563, L95

\bibitem[Larson et al.(2007)]{L06} Larson, D. et al. 2007,
        ApJ, 656, 653
\bibitem[Leach et al.(2008)]{Leach08} Leach, S. et al. 2008, A\&A, 491, 597
\bibitem[Lewis \& Bridle(2002)]{Lewis02} Lewis, A. \& Bridle, S. 2002, Phys. Rev. D, 66, 103511
\bibitem[Mandolesi et al.(2010)]{Mandolesi10} Mandolesi, N. et al. 2010, A\&A, available on line, arXiv:1001.2657
\bibitem[Miville-Deschenes et al.(2008)]{MAMD08} Miville-Deschenes, M.~ -A. et al. 2008, arXiv:0802.3345v1
\bibitem[Nolta et al.(2009)]{Nolta08} Nolta, M.~R. et al. 2009, ApJS, 180, 296
\bibitem[Page et al.(2007)]{Page07} Page, L. et al. 2007, ApJS, 170, 335
\bibitem[Planck Collaboration(2006)]{Planck06} Planck Collaboration 2006,
        The Scientific Programme of {\it Planck}, astro-ph/0604069 
        %also available at {\tt http://www.rssd.esa.int/SA/PLANCK/docs/Bluebook-ESA-SCI(2005)1\_V2.pdf}
\bibitem[Ricciardi et al.(2010)]{Ricciardi10} Ricciardi, S. et al. 2010, arXiv:1006.2326v1
\bibitem[Rosset et al.(2010)]{Rosset10} Rosset, C. et al. 2010, A\&A, available on line, arXiv:1004.2595
\bibitem[Planck Collaboration(2011)]{Tauber11} Planck Collaboration, 2011, arXiv:1101.2022
 \bibitem[Wandelt et al.(2004)]{W04} Wandelt, B. et al. 2004,
        Phys. Rev. D, 70, 083511
       
        

\end{thebibliography}
\end{document}